\documentclass[twocolumn,preprintnumbers,amsmath,amssymb,pra]{revtex4}

\usepackage{graphicx}
\usepackage{amssymb}
\usepackage{esvect}
\usepackage{enumerate}
\usepackage{mathptmx}      
\usepackage{latexsym}

\begin{document}

\newtheorem{defi}{Definition}
\newtheorem{assu}{Assumption}
\newtheorem{state}{Statement}
\newtheorem{hypo}{Hypothesis}

\title{Quantum mechanics from an epistemic state space}

\author{Per \"{O}stborn}
\affiliation{Division of Mathematical Physics, Lund University, S--221 00 Lund, Sweden}

\begin{abstract}
We derive the Hilbert space formalism of quantum mechanics from epistemic principles. A key assumption is that a physical theory that relies on entities or distinctions that are unknowable in principle gives rise to wrong predictions. An epistemic formalism is developed, where concepts like individual and collective knowledge are used, and knowledge may be actual or potential. The physical state $S$ corresponds to the collective potential knowledge. The state $S$ is a subset of a state space $\mathcal{S}=\{Z\}$, such that $S$ always contains several elements $Z$, which correspond to unattainable states of complete potential knowledge of the world. The evolution of $S$ cannot be determined in terms of the individual evolution of the elements $Z$, unlike the evolution of an ensemble in classical phase space. The evolution of $S$ is described in terms of sequential time $n\in \mathbf{\mathbb{N}}$, which is updated according to $n\rightarrow n+1$ each time potential knowledge changes. In certain experimental contexts $C$, there is knowledge at the start of the experiment at time $n$ that a given series of properties $P,P',\ldots$ will be observed within a given time frame, meaning that a series of values $p,p',\ldots$ of these properties will become known. At time $n$, it is just known that these values belong to predefined, finite sets $\{p\},\{p'\},\ldots$. In such a context $C$, it is possible to define a complex Hilbert space $\mathcal{H}_{C}$ on top of $\mathcal{S}$, in which the elements are contextual state vectors $\bar{S}_{C}$. Born's rule to calculate the probabilities to find the values $p,p',\ldots$ is derived as the only generally applicable such rule. Also, we can associate a self-adjoint operator $\bar{P}$ with eigenvalues $\{p\}$ to each property $P$ observed within $C$. These operators obey $[\bar{P},\bar{P}']=0$ if and only if the precise values of $P$ and $P'$ are simultaneoulsy knowable. The existence of properties whose precise values are not simulataneously knowable follows from the hypothesis that collective potential knowledge is always incomplete, corresponding to the abovementioned statement that $S$ always contains several elements $Z$.

\end{abstract}

\maketitle

\tableofcontents

\section{Introduction}
\label{intro}
The meaning of quantum mechanics (QM) has been subject to debate since the birth of the theory almost a hundred years ago. A wide spectrum of interpretations have been proposed, with radically different perspectives \cite{interpretations}.
Some theorists think that QM is incomplete or approximate. The suggested changes may involve the introduction of hidden variables \cite{hidden,leggett}, mechanisms for objective state reduction \cite{GRW,OR}, or other nonlinearities in the evolution \cite{nonlinear}. Another view is that there is redundancy in the standard postulates of QM \cite{neumann}; the most radical example is the 'many-worlds' interpretation of Everett and DeWitt \cite{manyworlds,dewitt}, according to which linear evolution of superposed states is all there is. Zurek tries to derive some postulates from the others \cite{zurek,zurek2}. In recent years, several attempts have been made to derive the Hilbert space formalism of QM from other principles, which are easier to interpret physically \cite{axiomatic,dakic,masanes}. One approach is to use as a foundation the concept of information \cite{infoderivation,goyal,aaronson}.

Already Bohr, Heisenberg and colleagues focused on information, or rather knowledge. The Copenhagen interpretation stresses that the quantum state encapsulates what can be known about a system, and that it is meaningless to ask for anything else. This epistemic perspective has gained renewed interest. Caves, Fuchs and Schack have introduced an interpretation of QM in which the collapse of the wave function is an update of subjective, bayesian probabilities. Fuchs and Schack have given the name Qbism to this approach \cite{fuchs}. In Anton Zeilinger's eyes "the reduction of the wave packet is just a reflection of the fact that the representation of our information has to change whenever the information itself changes" \cite{zeilinger}.

Knowledge has an inevitable subjective side to it: someone has knowledge about something. The association of quantum mechanical states with states of knowledge therefore suggests that knowing subjects play a fundamental role in the modern scientific world view. Nevertheless, the common drive behind many attempts to understand or alter QM has been to explain away or suppress this feature. The aim in this paper is to confront the subjective aspect of knowledge face to face, turn such an epistemology into symbolic form, and show that the formalism that emerges provides a simple and coherent way to understand QM.

To do so, we need to be more clear about the structure of the subjective aspect of the world than is normally the case in epistemic interpretations of QM. For example, we distinguish between actual and potential knowledge, and between individual and collective knowledge. We also need to formulate epistemic principles that constrain the form of physical law. 

The present approach can be seen as an attempt to walk as far as possible in the direction pointed out by the fathers of the Copenhagen interpretation of quantum mechanics. In a certain sense, though, the present approach is opposite to theirs. They wanted to understand a given physical formalism, and arrived at an epistemic interpretation. Here we start with a set of epistemic assumptions and use them to motivate the physical formalism. The advantage of this reverse approach is that the conceptually well-defined starting point enables a better understanding of the components of the formalism, and its domain of validity. Another advantage is that it makes it possible to understand better not only the meaning of quantum mechanics, but also some other physical concepts and principles, such as Pauli's exclusion principle, the gauge principle and entropy \cite{epistemic}.  

In the present paper, the more limited aim is to account for the Hilbert space formalism of QM as the only generally valid algebraic representation of well-defined experimental contexts. This means that the Hilbert space is not seen as the fundamental state space, but as a contextual state space that can be defined in certain states of knowledge.

The paper is organized as follows. In Section \ref{basic} the basic approach is presented in qualitative terms. The fundamental epistemic state space is introduced in Section \ref{knowledgestate}. In Section \ref{guiding} we formulate some epistemic rules that govern the game played on this stage. Physical law is expressed in terms of a general evolution operator that is introduced in Section \ref{evolution}. Empirically, physical law is determined in well-defined experimental contexts. In Section \ref{measurements} we try to define such contexts formally in the language introduced in the previous sections. This definition makes it possible to define the probability of different outcomes of an experiment, as described in Section \ref{probability}. Equipped with these concepts and definitions we try to argue in Section \ref{qmelements} that QM is the only generally valid algebraic representation of experimental contexts that conforms with the epistemic rules of the game formulated in Section \ref{guiding}. In Section \ref{discussion} we put the present approach into perspective and compare it to other approaches currently discussed within the field of quantum foundations.

\section{The basic picture}
\label{basic}

The main idea at the heart of this paper is that the structure of knowledge and the structure of the world reflect each other. A thorough analysis of what we can know about the world and what we can't teaches us a lot about the world itself. This view is similar to that maintained by Kant \cite{kant}.

To arrive at a level of discussion where we can use epistemic principles to derive physics, we first have to express some basic philosophical assumptions. Each of these assmptions can be discussed and debated at great length, but here we merely state them and discuss briefly what we mean by them. The purpose is not to convince the reader that the assumptions are correct or essential for physics, but to make it clear what perspective we choose. We refer the reader to the more detailed, but preliminary discussion in Ref. \cite{epistemic}.

\begin{figure}
\begin{center}
\includegraphics[width=80mm,clip=true]{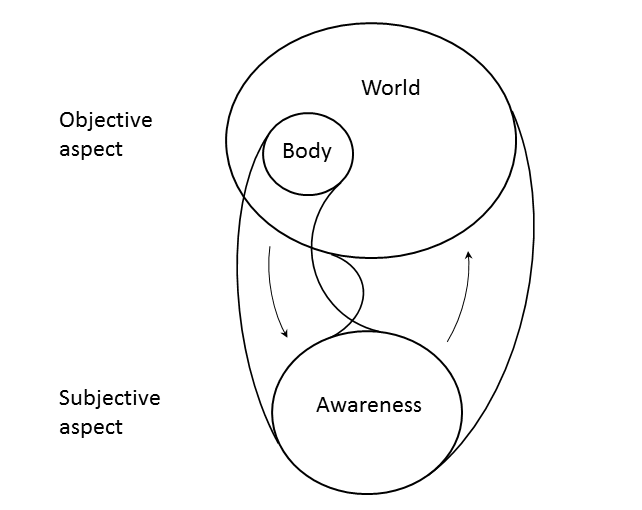}
\end{center}
\caption{Schematic illustration of the world view assumed in this study. There is nothing objective independent of the subjective, and the subjective is completely embedded in the objective. Since the body (including the brain) is regarded as a part of the world, fantasies and illusions are also seen as awareness of the world. However, the \emph{outside} world emerges from a subset of awareness, just as awareness emerges from a subset of the world.}
\label{Figure1}
\end{figure}

It is assumed that the observer and the observed are equally fundamental aspects of the world, being distinct but inseparable. One of them cannot be reduced to the other, explained in terms of the other, or even be properly imagined in isolation from the other. We call this view \emph{intertwined dualism}, and it is expressed schematically in Fig. \ref{Figure1}.

The role of subjective experiences is often discussed in terms of \emph{qualia} - the redness of red. Such qualia are put in opposition to the scientific description in which light of a certain frequency hits the retina and gives rise to the experience of redness \cite{lewis}. The question how qualia may arise is often called the `hard problem of consciousness' \cite{chalmers}. From the present perspective, this question is misguided. It assumes that qualia arise from an underlying physical world and is therefore secondary. In the present description, the physical world and qualia always arise together, at the same footing. Just as it is generally accepted that physics cannot explain why there is something rather than nothing, it has to be accepted that physics cannot explain why there are subjective experiences rather than no such experiences. Existence and experience go hand in hand.

We may even say that everything is qualia. The idea of the light wave with a certain frequency that gives rise to the perception of redness is a mental image. It is therefore a kind of qualia itself. What transcends the subjective perceptions is therefore not an objective world of `things', but objective physical laws that govern the relations between the qualia and their evolution. This does not mean that the world view at the basis of the present study is solipsistic. We assume that there are other subjects whose perceptions transcend our own perceptions. There are other aware beings `out there'.

\begin{figure}[tp]
\begin{center}
\includegraphics[width=80mm,clip=true]{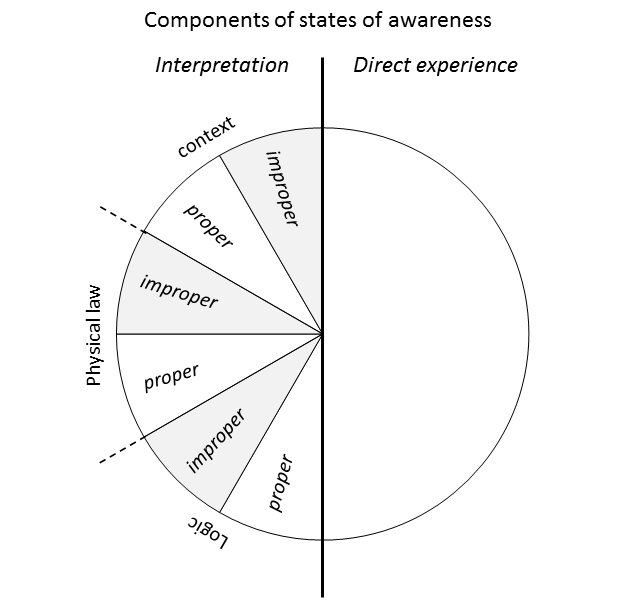}
\end{center}
\caption{Knowledge is direct experience in combination with at least one of the three components of interpretation. There must be no improper such component involved. The component of interpretation called `context' may, for instance, be the distinction between external and internal experiences, or between present and past experiences.}
\label{Figure2}
\end{figure}

To be able to identify objective physical laws, and even to imagine and speak about them properly, we have to be able to interpret and generalize our subjective perceptions. We have to assume a primary distinction between proper and improper interpretations of perceptions (Fig. \ref{Figure2}). This distinction transcends the perceptions themselves. We may say that `the truth is out there'. Properly interpreted perceptions are called knowledge, and physical law is expressed in terms of states of knowledge and the changes they undergo.

The interpretational ability to distinguish between perceptions that originate in the body and in the outside world is necessary in order to formulate the world view expressed in Fig. \ref{Figure1}. An example is the ability of a sane person to distinguish between reality and fantasies, where the latter originate in the brain, which is part of the body. This ability makes it possible to talk about a body of a subject being a proper subset of the world.

It should be noted that the interpretational ability to distinguish the body from the outside world does not pressuppose a transcendent world, independent from subjective perceptions, which consists of objects just like the ones we subjectively perceive. Following Kant, the `thing in itself' must exist, however, but its nature is beyond our knowledge; it cannot be \emph{intuited} or \emph{cognized}. Therefore it is not a concern of physics. Metaphorically speaking, it is the background on which the picture is drawn that expresses intertwined dualism in Fig. \ref{Figure1}. It is not accessible via the forms of perception that we use to do physics, reaching out to the objective aspect of the world via the arrow pointing upwards, and to understand ourselves, via the arrow pointing downwards. We are stuck in this circle.

Nevertheless, the assumed interpretational ability to distinguish the body from the outside world makes it possible to assume what we call \emph{detailed materialism}. According to this assumption, different states of subjective awareness always correspond to different states of perceivable objects within the body, and vice versa. This means that all perceptions are rooted in processes among the perceivable objects that make up the body.

The qualification \emph{detailed} when speaking about materialism refers to the assumption that each detail of a subjective perception corresponds to a detail of a process going on among the objects in the body. We may imagine a weaker form of materialism, in which the existence of an aware subject indeed requires a body consisting of objects, but in which some perceptions of this subject are not rooted in the body, but have a purely spiritual origin, so to say.

In the above discussion the term \emph{object} has been essential. We should define what we mean by that term. An object is simply an element of perception. It is a basic quality of perceptions that they can be subjectively divided into a finite set of objects. The division is not necessarily spatial; we may hear a sound that we are able to divide into two tones with different pitch. 

Apart from the ability to distinguish between objects of internal and external origin, we have to assume the interpretational ability to distinguish objects belonging to the present from those belonging to the past. We have to be able to tell things happening here and now from memories. This is necessary to give meaning to the concepts of time and change.

The basic notion of time in the present picture is that of \emph{sequential time} $n$. The knowledge of a subject at time $n$ consists of the knowledge belonging to the present time $n$, as well as the knowledge referring to the set of past times $\{n-1,n-2,\ldots\}$. Sequential time is updated according to $n\rightarrow n+1$ each time knowledge changes, each time a perceivable event occurs.

All properly interpreted objects are assumed to obey the same laws of physics, so that their states are updated in the same way as $n\rightarrow n+1$. This goes for the objects in the bodies of aware subjects as well. This means the present world view cannot be called creationistic in the usual sense of the word. God-made human beings cannot appear from nowhere, since the history of the objects in our bodies then cannot be properly accounted for by means of physical law. Instead, the present world view is perfectly compatible with Darwinism. However, the first stages of evolution have to be philosophically re-interpreted. The world was created when the first dim awareness appeared in some primitive creature. Everything that happened before that moment must be considered to be an abstract mathematical extrapolation backwards in time using the laws of physics. This perspective also puts cosmology in an unconventional light, since the Big Bang must be seen as nothing more than such an extrapolation taken to its extreme. At that stage, the extrapolated state of the world was not compatible with bodies of aware subjects that formed knowledge about what was going on. Therefore there was no actual physical world at all, according to the assumption of intertwined dualism (Fig. \ref{Figure1}).

To be able to use knowledge as a basis of a well-defined physical world view, the knowledge at any given time must be precisely defined. However, our aware knowledge is quite fuzzy. Our attention shifts, we may have vague ideas, our interpretations of what we perceive may be correct or erroneous, or something in between. Therefore we have to distinguish between the state of aware or \emph{actual knowledge} at time $n$, which may be fuzzy, and the state of \emph{potential knowledge}, which is assumed to be well-defined. The potential knowledge at time $n$ corresponds to all knowledge that may in principle be deduced from all subjective experiences at time $n$. It includes, for example, experiences at time $n$ that we do not become fully aware of until later.

\begin{figure}[tp]
\begin{center}
\includegraphics[width=80mm,clip=true]{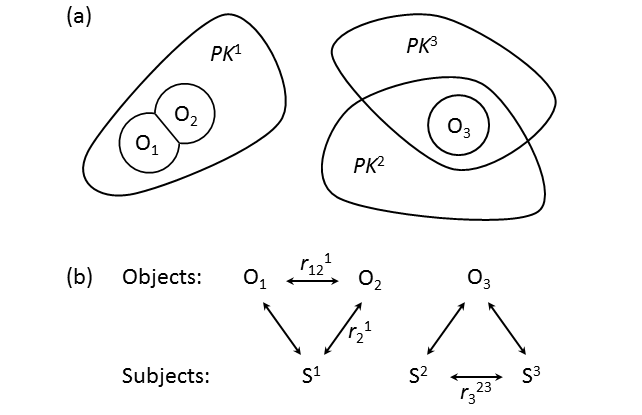}
\end{center}
\caption{a) Potential knowledge $PK$ is the union of the individual potential knowledge $PK^{k}$ of subjects $k$. Individual potential knowledge may overlap, which means knowledge of at least one common identifiable object $O_{l}$. b) The fact that subject $k$ knows about an object $O_{l}$ defines a relation $r_{l}^{k}$ between them. That one subject knows two objects defines a relation $r_{ll'}^{k}$ between the objects. If two subjects know the same object, a relation $r_{l}^{kk'}$ between the subjects is defined.}
\label{Figure3}
\end{figure}

Another question that arises when we try to identify the physical state with a state of knowledge is `whose knowledge?'. The only way to avoid arbitrariness in the definition of such a physical state is to answer `everyone's!'. We have to take the union of the potential knowledge of all perceiving subjects. To define the notion of collective potential knowledge of all these subjects we do not need to define exactly what organisms are aware and which of these have knowledge-forming interpretational abilities. All we need to assume at the conceptual level is that there are several subjects with potential knowledge. The nature of the associated body necessary to uphold the assumed detailed materialism is irrelevant.

To be able to say that the individual states of knowledge overlap we must assume that different subjects may know about the \emph{same} object (Fig. \ref{Figure3}). This possibility can be taken as a definition of the notion that these subjects live in the same world. Just as we have assumed the possibility that there are several subjects living in one world, it is of course possible that there are several objects in this world. Subjective experiences are differentiated. These seemingly trivial relations are expressed in Fig. \ref{Figure3} and constitute a more detailed version of the symmetry between the subjective and objective aspects of the world expressed in Fig. \ref{Figure1}.

It may be noted that the assumption of the existence of several subjects is not merely wishful thinking to avoid the loneliness in a solipsistic world. It is actually hinted at in physical law. To be able to derive the Lorentz transformation we have to assume that there are two \emph{different} observers or subjects in uniform motion in relation to each other who are able to observe the \emph{same} events. An event is always associated with an object, so that the derivation presupposes a relation $r_{l}^{kk'}$ between two subjects $k$ and $k'$ according to Fig. \ref{Figure3}. The fact that physical law is invariant under Lorentz transformations therefore indicates that the world is constructed in order to contain several subjects whose perspectives on the world are equally valid. This lack of hierarchy among subjects reflects the lack of hierarchy among objects, in the sense that they are all assumed to adhere to the same physical laws. This is another expression of the symmetry between the subjective and the objective aspect of the world.  

\begin{figure}[tp]
\begin{center}
\includegraphics[width=80mm,clip=true]{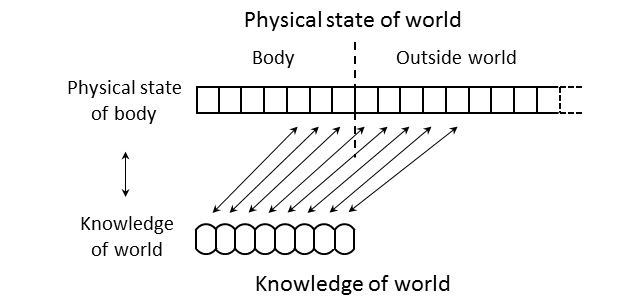}
\end{center}
\caption{Relation of the knowledge of the world in terms of perceived objects, and the physical state of the world. The tilted arrows indicate the one-to-one correspondence between these. The knowledge of the world is encoded in the physical state of the body according to detailed materialism, so that there is also a vertical correspondece. This correspondence is not object-to-object, since one object of perception usually corresponds to many objects in the body. The vision of a black dot involves the eye, the visual nerve and the visual cortex. The unmatched physical objects in the top row illustrates our incomplete knowledge of the body and of the world.}
\label{Figure4}
\end{figure}

A consequence of the ideas expressed above is that both individual and collective potential knowledge are fundamentally incomplete. This conclusion is central to the attempt in this paper to use an epistemic framework to motivate quantum mechanics.

We can arrive at the conclusion in the following way. Assume that the potential knowledge about all objects in the world is complete. According to the assumption of detailed materialism, this knowledge has to be a function of the physical state of the body, which in turn is in one-to-one correspondence to the assumed perfectly known objects of the body. Therefore, the complete potential knowledge becomes a proper subset of itself (Fig. \ref{Figure4}). This can be acounted for if the structure of knowledge is fractal, if it lacks foundation. But if we further assume that the body cannot be divided into more than a finite number of objects that can be described by a finite number of attributes, then we  arrive at a contradiction. In that case individual potential knowledge is always incomplete.

If we further assume that there are objects in the world that are not part of the body of any aware subject, then the same goes for collective potential knowledge. The last assumption corresponds to an assumption that there is no panpsychism.

\begin{figure}[tp]
\begin{center}
\includegraphics[width=80mm,clip=true]{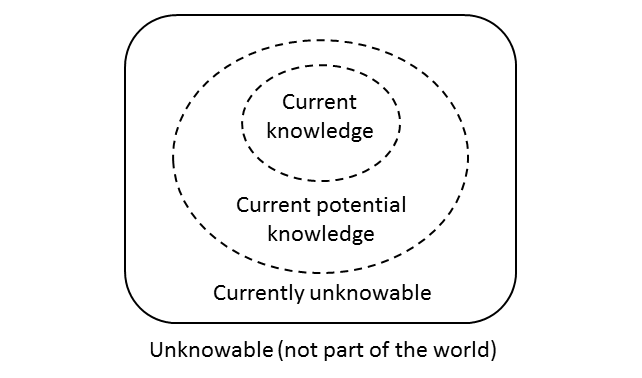}
\end{center}
\caption{An onion of knowledge. The actual or current knowledge corresponds to the properly interpreted current state of awareness. The potential knowledge represents those things that are knowable in principle at a given moment. It is part of potential knowledge that there is something currently unknowable. The solid boundary of the outermost layer of knowability corresponds to boundary of the world. There is such a boundary since the existence of proper and improper interpretation of awareness is assumed (Fig. \ref{Figure2}). In other words, not everything that is conceivable is part of the world.}
\label{Figure5}
\end{figure}

Even if there is always something outside potential knowledge, the contents of this something may change from time to time. That is to say, the currently unknowable may become part of potential knowledge at a later time. The boundary of this currently unknowable may be identified with the boundary of the properly interpreted world (Fig. \ref{Figure5}).

The argument for the incompletness of knowledge given above rests upon the assumption that knowledge is always encoded in bodies of subjects, which together form a proper subset of all the objects in the world. The popular argument for Heisenberg's uncertainty principle rests on the same assumption. In that case we imagine a spatially limited observer reaching out to the objects surrounding her. In so doing she inevitably disturbs them and lose some other knowledge about them. This can be seen as the dynamic version of the static argument given here, where we argue that complete knowledge cannot be encoded in a proper subset of itself.

The picture sketched above means that we abandon the traditional materialistic world view in which the physical objects are primary and the subjective perceptions of these objects are secondary or emergent. The price to pay for this shift of perspective is that we have to identify and give primary scientific weight to a set of interpretational abilities, or forms of perception. This fact seemingly makes the conceptual foundations of physics more complex, but this is not necessarily a bad thing, since these forms of perception make it posssible to give form to physical law, as we will see below. They are closely related to those statements that Kant called \emph{synthetic a priori}, statements which are seen to be true without empirical observations, but which are nevertheless not logical necessities \cite{kant}. These forms of perception rather provide the necessary framework that makes it possible to give meaning to empirical observations. They therefore transcend these observations. The main idea at the heart of this study can be reformulated as the idea that there is a one-to-one correspondence between these forms of perceptions and the form of physical law.

\section{Knowledge space and state space}
\label{knowledgestate}

Knowledge is assumed to consist of knowledge about a set of objects and their attributes. An attribute can be defined as a set of qualities that can be associated with each other and ordered. One quality in such a set can be called an attribute value. Red, green and blue are three different such qualities or attribute values that can be ordered into a spectrum that defines the attribute \emph{colour}. Attributes may be internal or relational. An internal attribute, such as colour, refers to the object itself, wheras a relational attribute, such as distance or angle, relates two or more objects.

An object may \emph{divide}, meaning that potential knowledge arises at some time $n$ that it has two different values of the same attribute, or that there is a relational attribute that refers to the object itself. Using a magnifying glass, we can decide that a grain of sand has internal spatial structure, so that we can associate distances between its different parts. An object that can divide may be called \emph{composite}, whereas as an object that cannot be divided may be called \emph{elementary}. Clearly, it is meaningless to associate a size to an elementary object.

Any two objects can also \emph{merge}, in a process that is the reverse of division. We may say that the knowledge about an object increases when it divides, and that the knowledge the two objects decrease when they merge. The field of vision becomes more blurred.

\subsection{Minimal objects}
\label{minimalobjects}

If we assume that elementary objects are the basic building blocks of the perceived world, then it contains a fixed number of such objects since they cannot divide. Each perceived object can then be divided a finite number of times into a given set of such elementary objects. But we argued above that potential knowledge is always incomplete. One thing we cannot know exactly is how many elementary objects are contained in each perceived object. We cannot verify or refute the hypothesis that the number of the basic building blocks of nature indeed stays fixed. In Section \ref{guiding} we introduce the principle of \emph{explicit epistemic minimalism} which says that the reliance on any such hypothesis that cannot be verified or refuted gives rise to wrong physical predictions, and should therefore be discarded. Nevertheless, it is possible to introduce a weaker form of elementarity that is epistemically acceptable.

This is the notion of \emph{minimal objects}. Let $\mathcal{M}$ be a set of objects defined exclusively by their internal attributes. Then $\mathcal{M}$ is a \emph{minimal set of objects} if and only if it fulfils the following conditions. 1) The number of objects in $\mathcal{M}$ is finite. 2) Division of any object in $\mathcal{M}$ gives rise to objects that are elements in $\mathcal{M}$ themselves. 3) There is no proper subset of $\mathcal{M}$ that fulfils condition 2). We identify the elementary fermions with such a minimal set of objects.

Using these minimal objects we can formulate an `axiom of foundation' for knowledge. It can always be expressed as the knowledge of a number of perceived objects and their internal and relation attributes. The knowledge of each of these objects can in turn always be expressed as the knowledge about a finite, but unspecified, number of minimal objects of which the object is composed, in the sense that we can in principle arrive at these minimal objects after a finite number of divisions of the perceived object. We note that the existence of a bottom level of knowledge of this kind was used in the motivation of the fundamental incompleteness of knowledge given in the preceding section.

\subsection{Quasiobjects}
\label{quasiobjects}

The minimal objects that we may use in a representation of a state of knowledge need not be directly perceived themselves. Their existence and their attributes may be deduced from directly perceived objects using physical law. For example, in a photography of a bubble chamber, given knowledge of the experimental setup, the visible traces can be deduced to be elementary fermions with specific energies and charges which sometimes divide. Such deduced objects are called \emph{quasiobjects}. They need not be microscopic objects, but may be everyday, large objects like the moon. We can deduce that the moon is there and calculate its orbit even if we do not always see it, using our previous observations and physical law, including the fact that objects that can be considered to consist of minimal objects do not suddenly disappear.

Even if quasiobjects are indispensable in order to formulate physical law in a simple and general way, it important to note that from the strict epistemic perspective that we adopt here, they do not add any new knowledge about the physical state, which itself corresponds to the properly interpreted potential perceptions at a given time. All we can know about the quasiobjects is deduced from this potential knowledge. This knowledge is primary and the quasiobjects are secondary abstractions. This perspective is opposite to the conventional scientific viewpoint from which the quasiobjects are seen as primary - they are really out there - and the perceptions that give rise to knowledge are secondary, capturing only a small subset of the information stored in the state of all the quasiobjects.

\begin{figure}[tp]
\begin{center}
\includegraphics[width=80mm,clip=true]{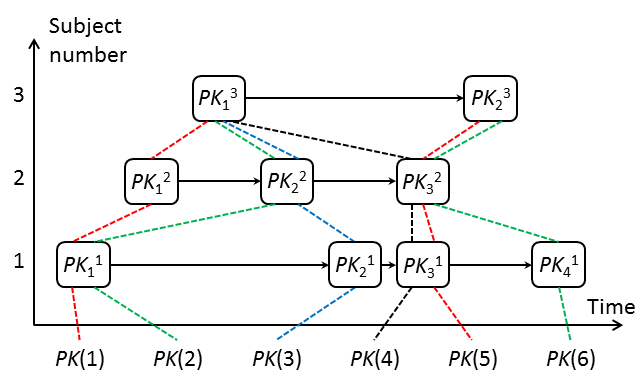}
\end{center}
\caption{The potential knowledge $PK(n)$ is updated each time any of the individual states of knowledge $PK_{m_{k}}^{k}$ are updated. For example, $PK(2)$ is the union of $PK_{1}^{1}$, $PK_{2}^{2}$, and $PK_{1}^{3}$. This state is updated to $PK(3)$ when $PK_{1}^{1}\rightarrow PK_{2}^{1}$. If two individual updates have space-like separation, they can sometimes be considered to occur simultaneously, as the updates $PK_{2}^{1}\rightarrow PK_{3}^{1}$ and $PK_{2}^{2}\rightarrow PK_{3}^{2}$. Overlaps between individual states of potential knowledge typically occur (Fig. \ref{Figure3}), but this is not shown here for clarity. }
\label{Figure6}
\end{figure}

\subsection{Potential knowledge and knowledge space}
\label{knowledge space}

After setting the stage, let us be more formal. Denote by $PK^{k}$ the state of potential knowledge of subject $k$. When $k$ potentially perceives a change corresponding to an event $e^{k}$, her individual sequential time $m_{k}$ is updated according to $m_{k}\rightarrow m_{k}+1$, and so is her potential knowledge, according to $PK^{k}_{m_{k}}\rightarrow PK^{k}_{m_{k}+1}$. Such an update means that $PK^{k}$ has changed, so that we always have  $PK^{k}_{m_{k}+1}\neq PK^{k}_{m_{k}}$. The collective potential knowledge $PK(n)$ is updated according to $PK(n)\rightarrow PK(n+1)$ whenever any individual state $PK^{k}$ is updated (Fig. \ref{Figure6}). We get $PK(n)\neq PK(n+1)$. This update defines the update of sequential time according to $n\rightarrow n+1$.

We have chosen to encode the instants of sequential time as a discrete sequence of intergers. This is appropriate since a subjectively perceived change that defines a temporal update corresponds to a binary distinction between before and after. 

Suppose that subject $k$ may send a message $\mu$ to subject $k'$ telling $k'$ that $e^{k}$ has occurred, and $k'$ may receive this message before or at the same time as she potentially perceives the change corresponding to the event $e^{k'}$ that defines the update of $PK^{k'}$. Then the temporal ordering of the events $e^{k}$ and $e^{k'}$ can be determined epistemically, and they cannot be considered to occur simultaneously. The event defined by the delivery of the message may be called $e_{\mu}^{k}$, and the event defined by the reception of the message may be called $e_{\mu}^{k'}$. The temporal ordering of the four events may be expressed as

\begin{equation}
n[e^{k}]\leq n[e_{\mu}^{k}]< n[e_{\mu}^{k'}]\leq n[e^{k'}].
\label{messengeroredering}
\end{equation}
The first and last temporal relation arise from the assumed inherent ability of any given subject $k$ or $k'$ to order her perceptions temporally. We have $n[e^{k}]=n[e_{\mu}^{k}]$ if and only if $e^{k}=e_{\mu}^{k}$, and $n[e_{\mu}^{k'}]=n[e^{k'}]$ if and only if $e_{\mu}^{k}=e^{k'}$. The middle temporal relation $n[e_{\mu}^{k}]< n[e_{\mu}^{k'}]$ arises from the finite maximum speed of messages defined by the speed of light, together with the assumption that the bodies of two different subjects are spatially separated. 

If such an ordering of the events $e^{k}$ and $e^{k'}$ via a pair of messenger events $(e_{\mu}^{k},e_{\mu}^{k'})$ is not possible, then we may sometimes say that $e^{k}$ and $e^{k'}$ occur \emph{simultaneously}. If so, the updates $PK^{k}_{m_{k}}\rightarrow PK^{k}_{m_{k}+1}$ and $PK^{k'}_{m_{k'}}\rightarrow PK^{k'}_{m_{k'}+1}$ together defines the update $PK(n)\rightarrow PK(n+1)$. In conventional language, this situation may occur if and only if the events $e^{k}$ and $e^{k'}$ have space-like separation at sequential time $n$.

It is important to note that such a pair of events does not necessarily occur simultaneously, in the sense described above. In some situations it becomes self-contradictory to say that they are simultaneous. Consider three events $e^{k}$, $e^{k'}$  and $e^{k''}$. Suppose that $e^{k}$ and $e^{k'}$ have space-like separation, as well as $e^{k}$ and $e^{k''}$. In contrast, suppose that $e^{k'}$ and $e^{k''}$ have time-like separation. If we would say that $e^{k}$ and $e^{k'}$ are simultaneous, as well as $e^{k}$ and $e^{k''}$, then $e^{k'}$ and $e^{k''}$ would also be simultaneous. But they cannot be, because of Eq. [\ref{messengeroredering}].

Even though we cannot use the ability to send a message $\mu$ according to the above discussion in order to decide in each case whether a pair of events perceived by two subjects are simultaneous or not, we assume that the question always has a definite answer. This is necessary in order to construct a universal temporal ordering of all events, as defined by sequential time $n$.

Even if $PK(n)\rightarrow PK(n+1)$ whenever $PK^{k}_{m_{k}}\rightarrow PK^{k}_{m_{k}+1}$, the converse is not true. We may have $PK(n)\rightarrow PK(n+1)$ even if $PK^{k}_{m_{k}}$ stays the same. Of course, this happens when the change that defines $n\rightarrow n+1$ is perceived by another subject $k'$. In this case we may write $PK^{k}(n)=PK^{k}(n+1)$. Otherwise $PK^{k}(n)\neq PK^{k}(n+1)$. These considerations make the expression $PK^{k}(n)$ well-defined. At each sequential time $n$ we may then express the collective potential knowledge $PK(n)$ as

\begin{equation}
PK(n)=\bigcup_{k}PK^{k}(n).
\label{cpk}
\end{equation}

\begin{figure}[tp]
\begin{center}
\includegraphics[width=80mm,clip=true]{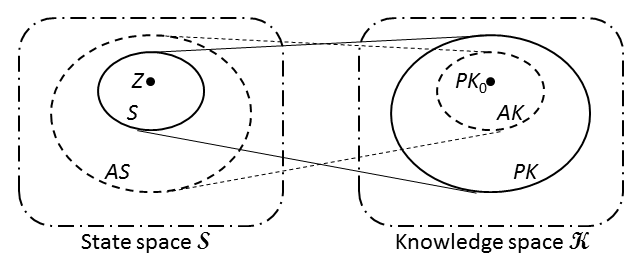}
\end{center}
\caption{The state space $\mathcal{S}$ has exact states $Z$ as elements. Larger states of knowledge $PK$ are mapped to smaller physical states $S$ and vice versa. In particular, the actual or aware knowledge $AK$ is contained in the potential knowledge $PK$, whereas the corresponding `aware physical state' $AS$ contains the physical state $S$. Since knowledge is always incomplete, $S$ never shrinks to a single element $Z$.}
\label{Figure7}
\end{figure}

The states of potential knowledge $PK(n)$ and $PK^{k}(n)$ can be seen as subsets of a knowledge space $\mathcal{K}$, which corresponds to the knowledge contained in a hypothetical state of complete potential knowledge (Fig. \ref{Figure7}). The relation $K\subset K'\subset\mathcal{K}$ means that $K'$ contains the same knowledge as $K$, but also more than that. The additional knowledge in $K'$ may consist of more knowledge about the same objects as those that are part of $K$, or knowledge about other objects. That $K\cap K'\neq\varnothing$ means that the two states contain knowledge about the same objects. Two knowledge states $K\subset\mathcal{K}$ and $K'\subset\mathcal{K}$ never contradict each other, even if $K\cap K'=0$. The non-overlap corresponds to knowledge about different objects rather than different or contradictory knowledge about the same objects.

To even speak about such contradictory knowledge we have to be able to say that two objects that are part of $K$ and $K'$, respectively, are the same. This notion is only possible to define according to Fig \ref{Figure3}; it presupposes two subjects who make different interpretations of what they see at a given time. This possibility is excluded by the assumed distinction between proper and improper interpretations, which transends the individual perceptions, and thus are common to any pair of subjects.

If $K$ and $K'$ would correspond to two states of collective knowledge for which $K\cap K'=0$, it does not even make sense to say that the states are contradictory. An object that is part of $K$ then cannot be associated with another object that is part of $K'$. As discussed above, that notion requires two subjects that observe the same object. Such a pair of different perspectives on the same object are not available at the collective level.

Since we have concluded that knowledge is always incomplete, a set $K\subseteq\mathcal{K}$ may correspond to a state of knowledge only if $K\subset\mathcal{K}$. We should not refer explicitly to $\mathcal{K}$ since it is not related to any physical state. The only thing we know about $\mathcal{K}\backslash PK(n)$ is that $\mathcal{K}\backslash PK(n)\neq 0$, and that it contains things that we do not know anything about at time $n$.

Knowledge may be completed in different ways. If we see a tree at a long distance, it may turn out to be an oak, a beech or a chestnut tree on a closer look. The individual leaves may turn out to be arranged in countless different ways. We let $\mathcal{K}$ represent the union of all such completions $\mathcal{K}_{PK}$ of the potential knowledge $PK$ at hand.

There is only one basic element of knowledge $PK_{0}$, in contrast to the many knowledge completions. This element corresponds to the rudimentary state of knowledge `There is something', corresponding to the dim, undifferentiated first light of awareness.

\subsection{The physical state and state space}
\label{statespace}

We now turn to the state space $\mathcal{S}$, which is kind of an inverse to $\mathcal{K}$ (Fig. \ref{Figure7}). The elements of $\mathcal{S}$ are knowledge completions $\mathcal{K}_{PK}$. Each such completion corresponds to an \emph{exact state} $Z$. The state space $\mathcal{S}$ is the set of all possible states of complete knowledge

\begin{equation}
\mathcal{S}=\{Z\}=\{\mathcal{K}_{PK_{0}}\}.
\end{equation}

Just as we should not refer explicitly to the unphysical knowledge completions, we should not refer explicitly to exact states $Z$, but only to physical states $S$. The physical state that corresponds to the state of potential knowledge $PK$ is the set of all knowledge completions consistent with $PK$, that is

\begin{equation}
S=\{\mathcal{K}_{PK}\}.
\end{equation}
We may say that $S$ is the set of all exact states $Z$ that cannot be ruled out given the potential knowledge $PK$. The incompleteness of knowledge implies that $S$ always contains more than one element $Z$. The inverse relation between $\mathcal{K}$ and $\mathcal{S}$ can be expressed as

\begin{equation}
PK'\subset PK \Leftrightarrow S'\supset S,
\label{inversespaces}
\end{equation}
where $PK\leftrightarrow S$ and $PK'\leftrightarrow S'$. When knowledge shrinks, the physical state expands, and vice versa. The actual or aware knowledge $AK$ is assumed to fulfil $AK\subseteq PK$, according to the discussion in Section \ref{basic}. This means that the corresponding `aware physical state' $AS$ fulfils $S\subseteq AS$, according to Eq. [\ref{inversespaces}] (Fig. \ref{Figure7}). In terms of the physical state, Eq. [\ref{cpk}] is re-expressed as

\begin{equation}
S(n)=\bigcap_{k}S^{k}(n).
\label{cs}
\end{equation}
The requirement that the knowledge of different subjects cannot contradict each other can be expressed as the fact that the physical state $S(n)$ as defined in Eq. [\ref{cs}] is always a non-empty set, that is

\begin{equation}
\bigcap_{k}S^{k}(n)\neq\varnothing
\label{cs2}
\end{equation}
for all $n$. In other words, there are exact states $Z$ that conform with the knowledge of all subjects $k$.

\begin{figure}[tp]
\begin{center}
\includegraphics[width=80mm,clip=true]{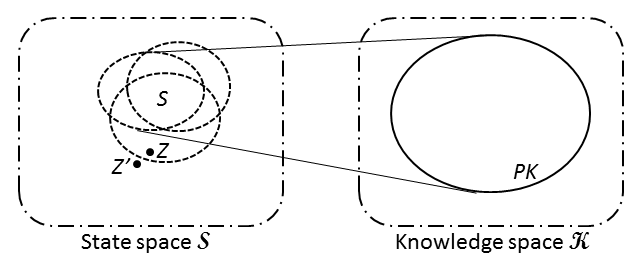}
\end{center}
\caption{The boundary $\partial PK$ of a state of potential knowledge $PK$ is precisely defined, but the boundary $\partial S$ of a physical state $S$ is not. A precise definition would require a precise discrimination between very similar exact states $Z$ and $Z'$, which would require complete knowledge.}
\label{Figure8}
\end{figure}

There is an asymmetry between the mirror images of knowledge represented in knowledge space $\mathcal{K}$ and state space $S$, respectively. The state of potential knowledge is supposed to be well-defined in principle: either we potentially know something or not. Therefore the boundary $\partial PK$ of $PK$ in $\mathcal{K}$ must be considered precisely given. The same is not true for the boundary $\partial S$ of $S$ (Fig. \ref{Figure8}). Suppose that one exact state $Z$ cannot be ruled out given the potential knowledge $PK$, so that $Z\in S$, but a very similar exact state $Z'$ can, so that $Z'\not\in S$. To be able to discriminate between such neighboring exact states on each side of the boundary $\partial S$ we must have acess to complete knowledge, which we have not. Therefore the $\partial S$ must be considered fuzzy. There are several boundaries $\partial S$ that cannot be excluded given $PK$. Furthermore, we cannot tell exactly which boundaries cannot be exluded. The set $\{\partial S\}$ has a fuzzy boundary for the same reason as $S$ has one. The argument can be repeated, of course, to conclude that boundary of the set of possible boundaries of $\{\partial S\}$ is also fuzzy, and so on.

Denote by $I$ a general item of knowledge. The logical relations between two such items $I_{1}$ and $I_{2}$ can then be expressed in state space and knowledge space as shown in Table \ref{itemrelations}.

Since all knowledge is assumed to be possible to express in terms of objects and their attributes, the incompleteness of knowledge must correspond to uncertainty about the number of objects or uncertainty about their attribute values. Such an uncertainty can always be expressed as the fact that there are several numbers of objects that cannot be excluded by the knowledge at hand, or that there is more than one combination of attribute values that cannot be excluded.

\begin{table}
\caption{Set-theoretical symbols in state space and knowledge space relating items of knowledge $I_{1}$ and $I_{2}$, where $\tilde{K}_{2}$ represents the item of knowledge `not $I_{2}$' in knowledge space.}
\label{itemrelations}       
\begin{tabular}{lll}
\hline\noalign{\smallskip}
Logical relation & State space $\mathcal{S}$ & Knowledge space $\mathcal{K}$ \\
\noalign{\smallskip}\hline\noalign{\smallskip}
		$I_{1}$\,AND\,$I_{2}$ & $S_{1}\cap S_{2}$ & $K_{1}\cup K_{2}$ \\
		$I_{1}$\,OR \,$I_{2}$ & $S_{1}\cup S_{2}$ & $K_{1}\cap K_{2}$ \\
		$I_{1}$\,NOT\,$I_{2}$ & $S_{1}\setminus S_{2}$ & $K_{1}\cup\tilde{K}_{2}$ \\
\noalign{\smallskip}\hline
\end{tabular}
\end{table}

For the sake of illustration, consider the case where the uncertainty resides in the attribute values of a given object. Figure \ref{Figure9} shows arrays of those values $\upsilon_{1j}$ and $\upsilon_{i2}$ allowed by physical law of two attributes 1 and 2, respectively. Complete knowledge of these attributes means that all pairs of values $(\upsilon_{1j},\upsilon_{i2})$ except one can be excluded. To have no knowledge means that no such pair of values can be excluded. Partial or incomplete knowledge may be of two different kinds. That knowledge is \emph{defocused} simply means that some set of values $\{\upsilon_{1j}\}$ of a attribute 1 and some set of values $\{\upsilon_{i2}\}$ of attribute 2 can be excluded. That knowledge is \emph{conditional} means that it can be expressed as a condition that relates the values of attributes 1 and 2, for instance `If the value of attribute 1 is $\upsilon_{1j}$, then the values of attribute 2 is $\upsilon_{i2}$. Knowledge can be both defocused and conditional, corresponding to a set of statements fo the form `If the value of attribute 1 is $\upsilon_{1j}$, then the values of attribute 2 cannot belong to the set $\{\upsilon_{i2}\}$'.

\subsection{The structure of state space}
\label{structure}

\begin{figure}[tp]
\begin{center}
\includegraphics[width=80mm,clip=true]{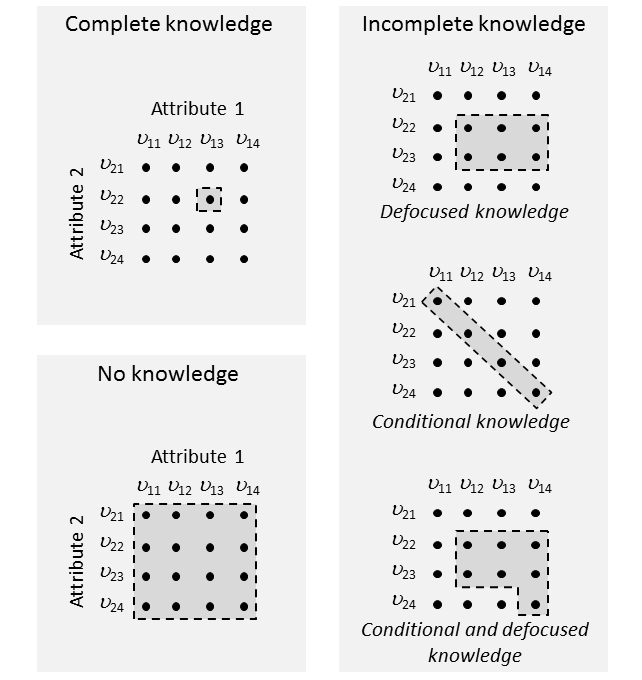}
\end{center}
\caption{Different types of knowledge about a pair of atttributes. Such knowledge always corresponds to the ability to exclude a set $\{(\upsilon_{1j},\upsilon_{i2})\}$ of pairs of attribute values allowed by physical law. Such a pair corresponds to a point in the present graphic representation. The encircled regions that correspond to pairs of values that cannot be excluded can be seen as projections of the physical state $S$ on the subspace of state space $\mathcal{S}$ spanned by the attributes 1 and 2.}
\label{Figure9}
\end{figure}

\begin{figure}[tp]
\begin{center}
\includegraphics[width=80mm,clip=true]{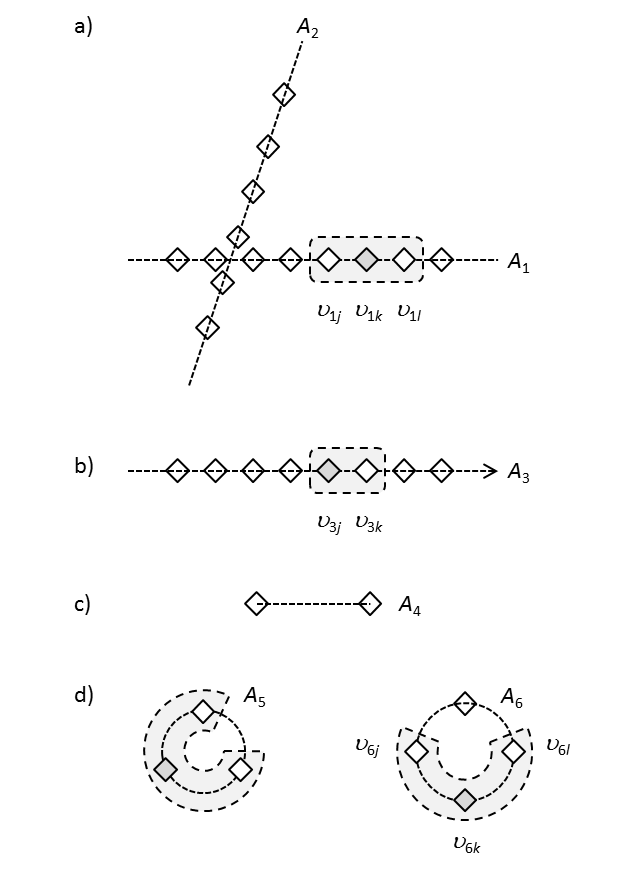}
\end{center}
\caption{The attributes $A_{i}$ can be regarded as axes that span state space. a) The \emph{ordering} of the attribute values along a given axis is determined from the concept of \emph{betweenness}. We assume that it is always possible to decide whether a value $\upsilon_{1k}$ is placed between another pair of values $(\upsilon_{1j},\upsilon_{1l})$ or not. b) The values of sequential time has an additional structure, namely \emph{direction}. For such an attribute $A_{3}$, it is always possible to decide which value $\upsilon_{3k}$ \emph{succeeds} another value $\upsilon_{3j}$. c) For attributes with only two possible values, the betweenness quality cannot be defined. d) Circular attributes are defined by the property that for any pair $(\upsilon_{6j},\upsilon_{6l})$, all other values $\upsilon_{6k}$ are placed between the members of this pair.}
\label{Figure10}
\end{figure}

The encircled regions in Fig. \ref{Figure9} correspond to projections of the physical state $S$ onto a subspace of state space $\mathcal{S}$. Let us discuss in more detail the structure of $\mathcal{S}$. Say that the object $O_{l}$ in the state of potential knowledge $PK$ is described by the set of attributes $\{A_{il}\}_{i}$. Then the set $\{A_{il}\}$ of all attributes of all such objects can be regarded as axes that span state space, according to the assumption that all knowledge can be expressed in terms of objects and their attributes (Fig. \ref{Figure10}).

Let $A_{i}$ be an attribute of a given object. The set of values $\{\upsilon_{ij}\}_{j}$ of $A_{i}$ are assumed to be qualities of perception that can be associated with each other and subjectively ordered. We may say that the set $\{\upsilon_{ij}\}_{j}$ with its ordered elements defines $A_{i}$. To define the ability to order the elements $\{\upsilon_{ij}\}_{j}$ we can use the concept of betweenness.

\begin{defi}[\textbf{Betweenness}]
Betweenness is a relation between three different values $\upsilon_{ij}$, $\upsilon_{ik}$ and $\upsilon_{il}$ of the same attribute $A_{i}$. If $\upsilon_{ik}$ is between the members of the pair $(\upsilon_{ij},\upsilon_{il})$ we write $\upsilon_{ij}\succ \upsilon_{ik}\prec \upsilon_{il}$. Otherwise we write $\upsilon_{ij}\not\succ \upsilon_{ik}\not\prec \upsilon_{il}$.

When the values are permuted in these formal expressions, the following rules hold. We have $\upsilon_{ij}\succ \upsilon_{ik}\prec \upsilon_{il}$ if and only if $\upsilon_{il}\succ \upsilon_{ik}\prec \upsilon_{ij}$. If $\upsilon_{ij} \not\succ \upsilon_{ik}\not\prec \upsilon_{il}$, then $\upsilon_{ik}\succ \upsilon_{ij}\prec \upsilon_{il}$ or $\upsilon_{ij}\succ \upsilon_{il}\prec \upsilon_{ik}$.

Suppose that there are more than three different values of $A_{i}$, and that the betweenness relation is defined for each triplet picked from the quadruplet $\{\upsilon_{ij},\upsilon_{ik},\upsilon_{il},\upsilon_{im}\}$. Then the following transitivity rules hold. If $\upsilon_{ij}\succ \upsilon_{ik}\prec \upsilon_{il}$ and $\upsilon_{ik}\succ \upsilon_{il}\prec \upsilon_{im}$, then $\upsilon_{ij}\succ \upsilon_{ik}\prec \upsilon_{im}$ and $\upsilon_{ij}\succ \upsilon_{il}\prec \upsilon_{im}$. If $\upsilon_{ij}\succ \upsilon_{ik}\prec \upsilon_{il}$ and $\upsilon_{ik}\succ \upsilon_{im}\prec \upsilon_{il}$, then $\upsilon_{ij}\succ \upsilon_{ik}\prec \upsilon_{im}$.
\label{betweenness}
\end{defi}

\begin{defi}[\textbf{Ordered attribute}]
Let $\Upsilon_{i}=\{\upsilon_{ij}\}$ be the set of values of attribute $A_{i}$ allowed by physical law. Suppose that $\upsilon_{ij}\in\Upsilon_{i}$, $\upsilon_{ik}\in\Upsilon_{i}$ and $\upsilon_{il}\in\Upsilon_{i}$, and that these values are all different. The attribute $A_{i}$ is ordered if and only if the betweenness relation is defined for each such triplet $(\upsilon_{ij},\upsilon_{ik},\upsilon_{il})$.
\label{orderedvalues}
\end{defi}

The concept of ordering says nothing about direction. We have not added any arrows to the dashed attribute axes in Fig. \ref{Figure10}(a). If we have a preconceived notion about direction, we can say that Definition \ref{orderedvalues} is symmetric with respect to a direction reversal, according to the rule saying that $\upsilon_{ij}\succ \upsilon_{ik}\prec \upsilon_{il}$ if and only if $\upsilon_{il}\succ \upsilon_{ik}\prec \upsilon_{ij}$. However, the necessity to introduce this symmetry is an artefact of the directedness of the formal notation, which introduces a redundancy in the representation of the betweenness concept. We have to introduce the notion of direction separately via the concept of succession.

\begin{defi}[\textbf{Succession}]
Succession is a relation between two different values $\upsilon_{ij}$ and $\upsilon_{ik}$ of the same attribute $A_{i}$. If $\upsilon_{ik}$ is a successor to $\upsilon_{ij}$ we write $\upsilon_{ij}<\upsilon_{ik}$. Otherwise we write $\upsilon_{ij}>\upsilon_{ik}$.

The rule for value permutation in this formal expression is $\upsilon_{ij}<\upsilon_{ik}$ if and only if $\upsilon_{ik}>\upsilon_{ij}$.

Suppose that there are more than two different values of $A_{i}$, and that the succession relation is defined for each pair picked from the triplet $\{\upsilon_{ij},\upsilon_{ik},\upsilon_{il}\}$. Then the following transitivity rules hold. If $\upsilon_{ij}<\upsilon_{ik}$ and $\upsilon_{ik}<\upsilon_{il}$, then $\upsilon_{ij}<\upsilon_{il}$.
\label{succession}
\end{defi}

\begin{defi}[\textbf{Directed attribute}]
Let $\Upsilon_{i}=\{\upsilon_{ij}\}$ be the set of values of attribute $A_{i}$ allowed by physical law. Suppose that $\upsilon_{ij}\in\Upsilon_{i}$, $\upsilon_{ik}\in\Upsilon_{i}$, and that these values are different. The attribute $A_{i}$ is directed if and only if the succession relation is defined for each such pair $(\upsilon_{ij},\upsilon_{ik})$.
\label{directededvalues}
\end{defi}

The primary example of a directed attribute is sequential time $n$. We have assumed the interpretational ability to distinguish objects in the state of potential knowledge $PK(n)$ that belong to the present from those that belong to the past. This distinction breaks the directional symmetry of time from the epistemic perspective. Accordingly, we can attach an arrow to sequential time, as in Fig. \ref{Figure10}(b).

Sequential time is a discrete attribute where $n+1$ is the neighbouring time following just after $n$. But we do not specify whether the two successive values in Defintion \ref{succession} are neighbours or not, meaning that we do not address the question whether we can squeeze another value between them, according to Definition \ref{betweenness}. This makes the above approach different from that used in the Peano axioms for the natural numbers, where the successor $S(x)$ of $x$ is the neighbour to $x$ in the discrete sequence of natural numbers. The definitions of betweenness and succession suggested here do not address the question whether the set of values $\{\upsilon_{ij}\}$ is discrete or continuous.

There are attributes for which just two values are allowed by physical law, such as the projection of the spin of an electron along a given axis [Fig. \ref{Figure10}(c)]. Another example is the binary \emph{presentness attribute}, which tells whether a perceived object belongs to the present or to the past. In such a case we cannot use the concept of betweenness to say that the attribute is ordered.

Some attributes such as quark colour may be called circular. The values of such attributes lack direction and are naturally represented by equidistant complex numbers of modulus one. They can be defined by the quality that for any pair of values $(\upsilon_{ij},\upsilon_{il})$, all other values $\upsilon_{ik}$ are placed between the members of this pair, so that we may write $\upsilon_{ij}\succ \upsilon_{ik}\prec \upsilon_{il}$. Clearly, an attribute must have at least three possible values to be called circular.

In general, we may say that an attribute is either ordered, or it has just two possible values. A directed attribute is always ordered, but the converse is not true.

\begin{figure}[tp]
\begin{center}
\includegraphics[width=80mm,clip=true]{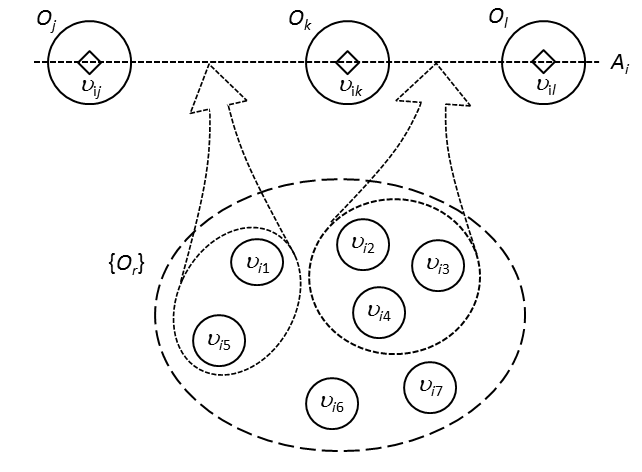}
\end{center}
\caption{A pool $\{O_{r}\}$ of reference objects can be used to measure distances between values of a given attribute $A_{i}$. In this example, we get $d_{jl}=5$, $d_{jk}=2$ and $d_{kl}=3$. We decide that $d_{jk}<d_{kl}$. However, the outcome of this comparison depends on the pool chosen, so that this method cannot be used straightaway to compare distances objectively.}
\label{Figure11}
\end{figure}

We have suggested a couple of qualities of attributes and their values that give structure to state space $\mathcal{S}$. These qualities transcend the individual attributes and the perception of their values. All such transcendent qualities should be used as structural building blocks of $\mathcal{S}$, but nothing else. The actual percpetions are encoded in the physical state $S\subset\mathcal{S}$ rather than $\mathcal{S}$ itself. In the language of Section \ref{basic}, the state space corresponds to the forms of perception, whereas the state corresponds to the perceptions themselves.

This distinction is crucial when it comes to the concept of \emph{distance}. If the values of an attribute $A_{i}$ are knowably discrete, then the distance $d_{jl}$ between two values $\upsilon_{ij}$ and $\upsilon_{ll}$ is naturally defined as the number of other values $\upsilon_{ik}$ that can be fitted between them, the number of such values that fulfil the relation $\upsilon_{ij}\succ \upsilon_{ik}\prec \upsilon_{il}$. The distance is defined by qualitites of state space $\mathcal{S}$, and is therefore an inherent quality of $\mathcal{S}$.

This is not so when we cannot exclude the possibility that the values of $A_{i}$ are continuous. Then we cannot exclude that infinitely many different values $\upsilon_{ik}$ can be fitted between any pair of values $\upsilon_{ij}$ and $\upsilon_{il}$. We get no information about $d_{jl}$. Instead, to obtain a value of $d_{jl}$ we have to consider two objects $O_{j}$ and $O_{l}$ with known values $\upsilon_{ij}$ and $\upsilon_{il}$, and introduce a predefined pool $\{O_{r}\}$ of reference objects $O_{r}$ with known values $\upsilon_{ir}$ of attribute $A_{i}$ (Fig. \ref{Figure11}). With these reference objects at hand, we create a `measurement state' $S$, in which the maximum number $M_{jl}$ of members of $\{O_{r}\}$ which fulfil $\upsilon_{ij}\succ \upsilon_{ir}\prec \upsilon_{il}$ are fitted between the two objects $O_{j}$ and $O_{l}$. Then we identify $d_{jl}=M_{jl}$. The pool $\{O_{r}\}$ defines the unit in which $d_{jl}$ is determined. The use of actual objects in this procedure means that $d_{jl}$ is part of $S$ rather than being a quality of $\mathcal{S}$. Without going into details, we note that an analogous procedure can be devised even if the values $\upsilon_{ij}$, $\upsilon_{il}$, and $\{\upsilon_{rk}\}$ are not precisely known.

This procedure misses an essential aspect of the concept of distance - the subjective ability to decide which of two distances is the larger one. We can decide whether two people are roughly equally tall, and if not, who is taller. If two men and one woman talk, it is most often evident that the pitch of the two male voices are more similar than the pitches of the female voice and one of the male voices. In short, we have an subjective sense of scale. This sense of scale cannot be reconstructed from the building blocks shown in Fig. \ref{Figure11}. When two distances $d_{jk}$ and $d_{jl}$ are compared, the outcome depends on the choice of pool $\{O_{r}\}$ and is therefore arbitrary.

We have to introduce the ability to judge whether $d_{jk}<d_{kl}$, $d_{jk}>d_{kl}$ or $d_{jk}=d_{kl}$ as a primary interpretational ability, as a form of perception that cannot be explained in terms of other forms of perception. The relative size of two distances referring to the same attribute should therefore be seen as part of the structure of state space $\mathcal{S}$. With this ability at hand we can choose pools of reference objects such that their values $\upsilon_{ir}$ are equidistant, like the markings on a ruler. This unit distance $\Delta_{r}$ captures the essential quality of the pool $\{O_{r}\}$ of reference objects, making it unneccessary to refer to the pool itself as a unit of measurment. Note, however, that the numerical distances $d$ obtained in this manner are defined only as parts of physical states $S$. It is only their relative size that reflects the structure of state space $\mathcal{S}$ itself.

Even if we consider the judgement of relative size to be a primary form of perception, it cannot be considered to be part of the \emph{collective} potential knowledge $PK$ for all attributes $A_{i}$ having a possibly continuous set of values. Special relativity teaches us that two subjects can judge spatial and temporal distances differently. In particular, they can come to different conclusions about the relative size of two spatial or temporal intervals. For these relative attributes, the validity of this judgement is therefore restricted to the \emph{individual} potential knowledge $PK^{k}$.

The spatial and temporal attributes $r$ and $t$ need more structure than \emph{betweenness} and \emph{relative size} to be properly characterized. To be able to say in a clearcut way that a spatial position $r_{k}$ is placed \emph{between} two other such positions $r_{j}$ and $r_{l}$, we need to assume that the three positions are placed along a straight line. We therefore have to introduce \emph{straightness} as a primary form of spatial perception. We also have to introduce \emph{orthogonality} on the same footing.

Just as relative size, the corresponding judgements are not universally or collectively valid, but are restricted to the individual potential knowledge $PK^{k}$. The absence of a universal cartesian coordinate system reflects the diffeomorphism invariance of general relativity. It should be noted here that in the present epistemic approach to physics, we distinguish between sequential time $n$ and relational time $t$, where the former attribute is used to evolve the physical state, and the latter is used to measure temporal distances.

We will not refer $t$ in the general formalism developed in this paper. The relation between the two aspects of time is discussed at length in Ref. \cite{epistemic}. We will also avoid explicit reference to $r$, aiming here at a general discussion about attributes $A_{i}$ and the observation of their values. Note that orthogonality as a form of perception applies only to physical space. In general, there is no inherent orthogonality relation defined between two attributes $A_{i}$ and $A_{i'}$ that help to span $\mathcal{S}$, as indicated by the tilted pair of axes in Fig. \ref{Figure10}(a). The attribute values cannot be used to define a general inner product, and there is no general metric for $\mathcal{S}$.

\subsection{State space dimension}
\label{statespacedim}

We may ask about the dimension $D[\mathcal{S}]$ of $\mathcal{S}$. Each attribute of each object adds one to $D[\mathcal{S}]$ and one axis to $\mathcal{S}$. We might try to specify $D[\mathcal{S}]$ as the total number of minimal objects in the world times the number of attributes possessed by each of these. But the total number of minimal ojects is an unknowable, possibly infinite number. The number of minimal objects contained in each perceived object cannot be determined either. Their number is not fixed, just as the number of elementary particles is not fixed.

\begin{figure}[tp]
\begin{center}
\includegraphics[width=80mm,clip=true]{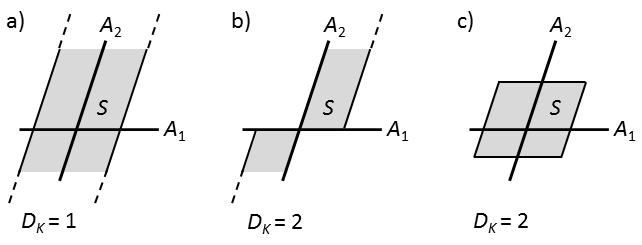}
\end{center}
\caption{The knowledge dimension $D_{K}$ in a two-dimensional state space, spanned by attributes $A_{1}$ and $A_{2}$. There is knowledge associated with $A_{1}$ in all three cases, meaning that some of its values can be excluded, so that $D_{K}\geq 1$. a) There is no knowledge associated with $A_{2}$, so that $D_{K}=1$. b) There is conditional knowledge associated with $A_{2}$, so that $D_{K}=2$. c) There is knowledge associated with $A_{2}$, so that $D_{K}=2$.}
\label{Figure12}
\end{figure}

Therefore we introduce the knowledge dimension $D_{K}[\mathcal{S}](n)$. Put simply, it equals the number of distinctions in the state of potential knowledge $PK(n)$. Only distinctions pertaining to directly perceived objects count. Quasiobjects like minimal objects are disregarded since they do not add anything to $PK$; our knowledge about them is a function of what we perceive directly, according to the discussion in Section \ref{quasiobjects}.

We may also say that $D_{K}[\mathcal{S}](n)$ is the number of potentially perceived attributes at time $n$. If two different values of the same attribute are perceived, then they count as two different attributes, since they are associated with two different objects. If no values of a given attribute $A_{i}$ of a given object can be excluded, then this attribute is not perceived at all from an epistemic perspective, and does not contribute to $D_{K}[\mathcal{S}](n)$ (Fig. \ref{Figure12}).

Even so, we may have conditional knowledge pertaining to $A_{i}$. Say that we know that the value of $A_{1}$ is $\upsilon_{11}$ or $\upsilon_{12}$, but that we know nothing about the value of $A_{2}$. Nevertheless, we know that $A_{1}=\upsilon_{11}\Rightarrow A_{2}=\upsilon_{21}$. Already such indirect knowledge about $A_{2}$ lifts it from the darkness of complete ignorance. We may therefore regard a condition involving $A_{i}$ as a distinction associated with $A_{i}$, since a condition picks an implication $A\Rightarrow B$ and drops the alternative $A\not\Rightarrow B$ [Fig. \ref{Figure12}(b)]. 

\begin{defi}[\textbf{Independent attribute}]
An attribute of an object is independent if and only if physical law allows it to take several values even if all other attributes of the given object are completely known.
\label{indattributes}
\end{defi}

\begin{defi}[\textbf{Knowledge dimension}]
Let $N_{O}(n)$ be the number of objects $O_{l}$ contained in $S(n)$, and let $N_{Al}(n)$ be the number of independent and distinct attributes that can be associated with $O_{l}$ at sequential time $n$. An attribute $A_{i}$ of $O_{l}$ is distinct if and only if at least one of the following two condistions is fulfilled: 1) $A_{i}$ is directly observed by some subject, or 2) there is conditional potential knowledge that involves $A_{i}$. Then the knowledge dimension of $\mathcal{S}$ at time $n$ is given by $D_{K}[\mathcal{S}](n)=\prod_{l=1}^{N_{O}(n)}N_{Al}(n)$.
\label{knowdimnum}
\end{defi}

There is no reason \emph{a priori} to assume that potential knowledge is conserved. It may grow or shrink. In particular, the number of objects contained in $PK$ may change, both because the number of perceiving subjects may change, and since the perception of a given subject may become more refined or more blurred. Therefore the knowledge dimension $D_{K}[\mathcal{S}]$ must be seen as a dynamical quantity, as a function of $n$.

The known distinct attributes that determines $D_{K}[\mathcal{S}](n)$ according to Definition \ref{knowdimnum} span a space that may be called the knowledge state space $\mathcal{S}_{K}$. We may see $\mathcal{S}_{K}$ as a subspace of an underlying state space $\mathcal{S}$ which is such that \emph{any} conceivable knowledge state space fulfils

\begin{equation}
\mathcal{S}_{K}\subseteq\mathcal{S}.
\end{equation}
We clearly have

\begin{equation}
D_{K}[\mathcal{S}](n)\leq D[\mathcal{S}]
\end{equation}
at each sequential time $n$. Since there is \emph{a priori} no upper limit on the number of subjects that may contribute to $PK$, and no upper limit of the size of the body that encodes the knowledge of each subject, there is no upper limit on the number of known distinct attributes. Therefore we must set $D[\mathcal{S}]=\infty$.

We will argue in Section \ref{evolution} that when it comes to representations of physical law, the only important part of $\mathcal{S}$ is spanned by the $D_{K}[\mathcal{S}]$ attributes that we know anything about. This conclusion conforms with the strict epistemic perspective advocated in this study.

\subsection{A measure on state space}
\label{volmeasure}

We have argued that there is no inherent metric in state space. It is nevertheless possible to define a measure. It will be essential in our discussions about probability.

\begin{defi}[\textbf{Attribute value space} $\mathcal{S}(A,\upsilon)$]
Let $A$ be an independent attribute according to Definition \ref{indattributes}. $\mathcal{S}(A,\upsilon)\subseteq \mathcal{S}$ is the set of exact states $Z$ for which there is at least one object for which $A$ is defined, and for which the value of $A$ is $\upsilon$.
\label{valuespacedef}
\end{defi}

\begin{defi}[\textbf{State space volume}]
The measure $V[S]\geq 0$ is defined for any state $S\in\mathcal{S}$ belonging to a $\sigma$-algebra $\Sigma_{V}$ over $\mathcal{S}$, and is such that $V[\mathcal{S}(A,\upsilon)]=V[\mathcal{S}(A,\upsilon')]$ for any independent attribute $A$, and any pair of values $(\upsilon,\upsilon')$ of $A$ allowed by physical law. Like for any measure, $V[\Sigma_{1}\cup \Sigma_{2}]=V[\Sigma_{1}]+V[\Sigma_{2}]$ for any two disjoint sets $\Sigma_{1}\in\Sigma_{V}$ and $\Sigma_{2}\in\Sigma_{V}$. For any exact state $Z$ we have $V[Z]=1$.
\label{voldef}
\end{defi}

The condition $V[S(A,\upsilon)]=V[S(A,\upsilon')]$ can be interpreted as a statement that for each exact state $Z$ for which the value of $A$ is $\upsilon$ there is exactly one exact state $Z'$ for which the value is $\upsilon'$. We thus compare state space volumes in the same way as we compare the sizes of two sets $\Sigma_{1}$ and $\Sigma_{2}$ by putting elements of $\Sigma_{1}$ into one-to-one correspondence with elements of $\Sigma_{2}$. Nevertheless, we avoid reference to the individual elements $Z$ of the space $\mathcal{S}$ in this condition. We do so because they lack epistemic meaning if considered one by one.

Nevertheless, we let $V[Z]=1$ to ensure that a physical state $S$ consistent with at least one exact state gets positive volume. This assignment is allowed theoretically since the exact state is a well-defined concept. However, it cannot be used to calculate the volume of actal physical states since exact states cannot be knowably counted one by one. Instead we have to compare the volumes of different states. Such a relative volume specification will be sufficient in our treatment of probability, and can also be used to make the concept of entropy epistemically well-defined \cite{epistemic}. 

Note that Definition \ref{voldef} does not refer to the structure of the set of possible values $\upsilon$, as to whether they are continuous or discrete. We avoid such references since we want to keep the conceptual framework as general as possible.

\subsection{Object states and object state space}
\label{objectspace}

Typically, we want to model and predict the evolution of a particular object $O_{l}$ rather than the whole world. The object $O_{l}$ of interest may be composite, of course, meaning that it is or can be divided. We can define the complement $\Omega_{O}$ to any object $O$, where $\Omega_{O}$ corresponds to `the rest of the world'. The entire world may be denoted $\Omega$.

If the world $\Omega$ as a whole is regarded to be an object, then it can only contain a finite number of minimal objects, according to the `axiom of foundation' discussed in Section \ref{minimalobjects}. This is too restrictive. We want to allow the world to be infinite in size, so that we can come across a (countably) infinite number of objects if we travel along a straight line forever. To make this possible we have to make a distinction between the division of an object and the appearance of new objects among those already perceived. Even if the depth of knowledge is finite, we do not want to put boundaries to the scope of knowledge by assumption. Thus neither $\Omega$ nor $\Omega_{O}$ are considered to be objects.

There is another sense in which the world $\Omega$ cannot be regarded as an object. By definition, the entire universe lacks a complement. In contrast, all objects have a complement which contains at least one other object. This is so since each perceived object corresponds to another object in the body of the perceiving subject according to the assumption of detailed materialism (Section \ref{basic}).

Even if the world cannot be seen as an object, our actual \emph{potential knowledge} $PK$ of the world \emph{may} correspond to a (composite) object. This knowledge corresponds to an object if the number of aware subjects is finite. This is so since each of the finite bodies of these subjects can only perceive a finite number of objects.

We define the \emph{object state} $S_{O}$ as the physical state that would result if the knowledge about all the other objects was erased. More precisely, $S_{O}$ is the union of all exact states $Z$ in state space that do not contradict the fact that $O$ exists, or the potential knowledge of its internal attributes. Correspondingly, we may define the \emph{environment state} $S_{\Omega_{O}}$ as the union of all exact states $Z$ in state space that do not contradict the existence of any of the perceived objects in the complement $\Omega_{O}$ to $O$, or the potential knowledge of the attributes internal to this complement. Generally, $S\subseteq S_{O}\cap S_{\Omega_{O}}$.

\begin{figure}[tp]
\begin{center}
\includegraphics[width=80mm,clip=true]{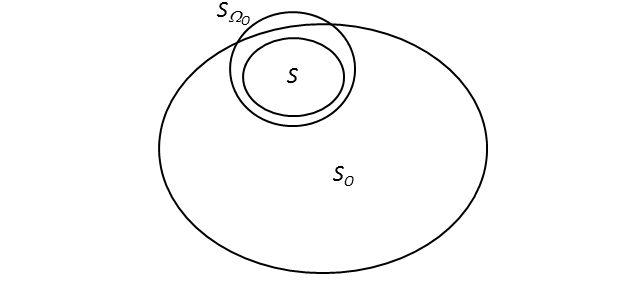}
\end{center}
\caption{$S_{O}$ is the state corresponding to knowledge of object $O$. $S_{\Omega_{O}}$ is the state corresponding to all other objects in a state $S$ of composite knowledge. $S_{O}$ is shown large since it corresponds to `small' knowledge of a single object (c.f. Fig. \ref{Figure7}). The state $S$ is a proper subset of the intersection between $S_{O}$ and the environment $S_{\Omega_{O}}$ whenever $O$ is related to its environment.}
\label{Figure13}
\end{figure}

If there would be no potential knowledge $PK_{R}$ about the relational attributes that relate $O$ to its environment $\Omega_{O}$, and if there would be no conditional knowledge $PK_{C}$ that relates $O$ and $\Omega_{O}$, then we would have $PK=PK_{O}\cup PK_{\Omega_{O}}$. This is the same as to say $S=S_{O}\cap S_{\Omega_{O}}$. However, whenever $S_{O}$ is defined, there is also some knowledge about the relation between $O$ and its environment. We must therefore always write $PK=PK_{O}\cup PK_{\Omega_{O}}\cup PK_{R}$. This means that $PK$ is larger than $PK_{O}\cup PK_{\Omega_{O}}$, so that

\begin{equation}
S\subset S_{O}\cap S_{\Omega_{O}},
\label{relatedobject}
\end{equation}
as illustrated in Fig. \ref{Figure13}.

Considering the set $\{O_{l}\}$ of all objects that are part of $PK(n)$ we may also write

\begin{equation}
S(n)\subset\bigcap_{l}S_{Ol}(n),
\label{cso}
\end{equation}
where $S_{Ol}$ is the state of object $O_{l}$. This expression should be compared to Eq. [\ref{cs}], in which we consider a set of subjects rather than objects.

In the scientific modelling of the behavior of an object we often assume that it is isolated. In our terminology this approximation corresponds to the assumption that $S=S_{O}\cap S_{\Omega_{O}}$. We know, of course, that this is never quite true, as expressed in Eqs. [\ref{relatedobject}] and [\ref{cso}].

Apart from knowledge $PK_{R}$ about relational attributes that connect object $O$ to its environment $\Omega_{O}$, there may also be conditional knowledge that relates object $O$ and $\Omega_{O}$. In that case we should write  

\begin{equation}
PK=PK_{O}\cup PK_{\Omega_{O}}\cup PK_{R}\cup PK_{C}.
\label{generalpk}
\end{equation}

The object state $S_{O}$ is a subset of state space $\mathcal{S}$ och all possible exact states $Z$ of the world. We may, however, also consider the space $\mathcal{S}_{O}$ of all possible exact states $Z_{O}$ of an \emph{object} rather than of the entire \emph{world}. This makes a qualitative difference, since we cannot consider the world to be an object, as discussed above.

We may embed the object state in this object state space $\mathcal{S}_{O}$ rather than in $\mathcal{S}$. Let us call the corresponding object state $S_{OO}$. It is the set of exact object states $Z_{O}\in \mathcal{S}_{O}$ that are not excluded by the potential knowledge of the attributes of object $O$. Since $S_{OO}$ is a subset of another space, we have to give it a different name than $S_{O}$. If we describe $S_{O}$ as all exact states of the object $O$ and the surrounding world $\Omega_{O}$ that is consistent with the knowledge about $O$, we may describe $S_{OO}$ as all exact states of $O$ that is consistent with the knowledge about $O$, ignoring the rest of the world.

\begin{figure}[tp]
\begin{center}
\includegraphics[width=80mm,clip=true]{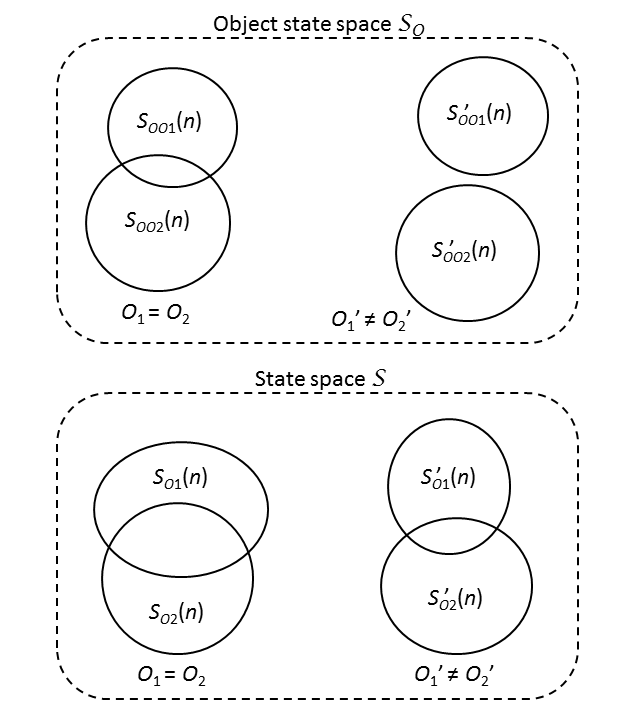}
\end{center}
\caption{The object state space $\mathcal{S}_{\mathcal{O}}$ and the state space $\mathcal{S}$. If two object states $S_{OO1}$ and $S_{OO2}$ overlap in $\mathcal{S}_{\mathcal{O}}$, then they represent the same object. Otherwise they represent different objects. In contrast, states of objects $S_{O}$ perceived at the same time always overlap, even if they are known to be distinct.}
\label{Figure14}
\end{figure}

What difference does this distinction make? Consider two knowably different objects $O_{1}$ and $O_{2}$ that are observed at the same time $n$. Clearly there is a physical state $S(n)$ that is consistent with the simultaneous existence of both these objects, and therefore there are exact states $Z$ that are consistent with these objects. In other words, $S_{O1}\cap S_{O2}\neq\varnothing$. On the other hand, there is no exact object state $Z_{O}$ that is consistent with both these objects. If there were, we would be unable to distinguish them. We would not be able to give them the different names $O_{1}$ and $O_{2}$. This means that $S_{OO1}\cap S_{OO2}=\varnothing$. These relations are illustrated in Fig. \ref{Figure14}. The represention of object states in object state space will be useful when we discuss the evolution of individual objects, and is crucial in the definition of \emph{identifiable objects} (Section \ref{objectevolution}).

In terms of object states $S_{OO}$ embedded in object state space $\mathcal{S}_{O}$, the definitions [\ref{valuespacedef}] and [\ref{voldef}] should be reformulated as follows. 

\begin{defi}[\textbf{Object attribute value space} $\mathcal{S}_{O}(A,\upsilon)$]
Let $A$ be an independent attribute according to Definition \ref{indattributes}. $\mathcal{S}_{O}(A,\upsilon)\subseteq \mathcal{S}_{O}$ is the set of exact object states $Z_{O}$ for which $A$ is defined for object $O$, and for which the value of $A$ in this object is $\upsilon$.
\label{ovaluespacedef}
\end{defi}

\begin{defi}[\textbf{Object state space volume}]
The measure $V[S_{OO}]\geq 0$ is defined for any state $S_{OO}\in\mathcal{S}_{O}$ belonging to a $\sigma$-algebra $\Sigma_{VO}$ over $\mathcal{S}_{O}$, and is such that $V[\mathcal{S}_{OO}(A,\upsilon)]=V[\mathcal{S}_{OO}(A,\upsilon')]$ for any independent attribute $A$, and any pair of values $(\upsilon,\upsilon')$ of $A$ in object $O$ allowed by physical law. Like for any measure, $V[\Sigma_{1}\cup \Sigma_{2}]=V[\Sigma_{1}]+V[\Sigma_{2}]$ for any two disjoint sets $\Sigma_{1}\in\Sigma_{VO}$ and $\Sigma_{2}\in\Sigma_{VO}$. For any exact object state $Z_{O}$ we have $V[Z_{O}]=1$.
\label{ovoldef}
\end{defi}

\section{Guiding principles for physical law}
\label{guiding}

We have argued in general terms that a strict epistemic perspective can be employed to construct a scientific world view. But such a perspective is scientifically useful only if it provides a unified way to understand physical law as we already know it, and also makes it feasible to extend our understanding of it. To this end we need to formulate epistemic principles that limit the possibilities, that exclude some conceivable physical laws.

\subsection{Implicit epistemic minimalism}

History of physics teaches that it is a dead end to introduce objects that cannot be perceived, neither directly nor indirectly, such as the aether, or to make use of attributes that cannot be operationally defined or measured, such as absolute positions or velocities. The same goes for operations with no perceivable effect, such as the interchange of two identical particles. Physical law is epistemically picky. This principle may be called \emph{epistemic minimalism}. There are two levels of this principle: implicit and explicit epistemic minimalism.

\emph{Implicit epistemic minimalism} is the principle that proper physical models can always be expressed without the introduction of entities or distinctions that cannot be subjectively perceived as such, or deduced from such perceptions. This is true, in particular, when it comes to the introduction of objects, attributes and attribute values.

The primary example is Galilean invariance. There is no need to use the attribute \emph{absolute speed} to formulate physical law: it is invariant under the transformation $x\rightarrow x+vt$ for any constant $v$. Even if physical law does not need the idea of absolute speed, it is nevertheless compatible with Galilean invariance. Therefore Newton could uphold the idea of absolute space. This is the reason this level of epistemic minimalism is called \emph{implicit}. To arrive at special relativity we need something more.

\subsection{Explicit epistemic minimalism}
\label{explicitminimalism}

The principle of \emph{explicit epistemic minimalism} means that the introduction in a physical model of entities or distinctions that cannot be subjectively perceived, or deduced from such perceptions, leads to conflict with physical law. In other words, a model which relies on such entities or distinctions gives rise to wrong predictions. This is true, in particular, when it comes to the introduction of objects, attributes and attribute values.

This principle constrains the form of physical law further. It means that Nature explicitly answers "yes" or "no" if we ask her whether a proposed entity or distinction has epistemic meaning.

To obey explicit epistemic minimalism, physical law must be inconsistent with the notion of absolute speed, since this concept is epistemically empty. The speed of an object can only be operationally defined in relation to another object. The notion of absolute speed implies the addition law for velocities, and whenever the addition law holds it is possible to uphold the notion of absolute speed. A bit more generally, we may say that the notion of absolute space and time can be kept alive if and only if the addition law for velocities always holds.

The addition law may expressed as follows. Consider any three objects $O_{1}$, $O_{2}$ and $O_{3}$. Let $u_{12}$ be the relative velocity of $O_{1}$ and $O_{2}$, as judged in the rest frame of $O_{1}$. In the same way, let $u_{23}$ be the relative velocity of $O_{2}$ and $O_{3}$, as judged in the rest frame of $O_{2}$. Then the relative velocity of $O_{1}$ and $O_{3}$ is

\begin{equation}
u_{13}=u_{12}+u_{23},
\label{additionlaw}
\end{equation}
as judged in the rest frame of $O_{1}$.

To rule out the notion of absolute speed, it is thus necessary and sufficient that physical law sometimes break Eq. (\ref{additionlaw}). One way to do this is to introduce a maximum speed $c$ that no object ever exceeds. To give such a concept epistemic meaning, all subjects $k$ must agree that a given entity, under given circumstances that all agree upon, always travels at speed $c$ in their own reference frame. We may let $O_{1}$ be the body of a subject $1$, and $2$ be the body of a subject $2$, whose reference frames should be equally valid according to the discussion in Section \ref{basic}. Of course, this ansatz leads to the Lorentz transformation, which breaks Eq. [\ref{additionlaw}].

In the limit $c\rightarrow \infty$ we get $\sqrt{1-(v/c)^{2}}\rightarrow 1$, and we get back the Galileo transformation, for which Eq. (\ref{additionlaw}) holds. Thus, the introduction of a maximum speed is the only possible way to break Eq. (\ref{additionlaw}). In other words, the finite speed of light and special relativity can be seen as an expression of explicit epistemic minimalism.

\begin{figure}[tp]
\begin{center}
\includegraphics[width=80mm,clip=true]{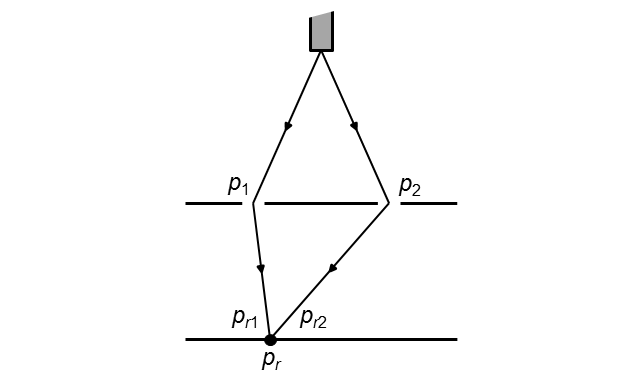}
\end{center}
\caption{Probabilities that can be used to illuminate the difference between implicit and explicit epistemic minimalism in the double-slit experiment. See text for further explanation.}
\label{Figure15}
\end{figure}

The statistics of identical particles is another example of explicit epistemic minimalism. Since it is epistemically meaningless to treat permutations of identical particles as different states, the principle implies that it must lead to wrong statistics if such permutations are included as distinct states in the calcultation of statistical weights. And, of course, it does, since it gives rise to Maxwell-Boltzmann statistics, which in many experimental situations is physically very different from the correct Bose-Einstein and Fermi-Dirac statistics of bosons and fermions, respectively. In contrast, if the minimalism would have been implicit, it would have made no physical difference whether permutations of identical particles were included or not.

The same principle can be used to motivate the Pauli exclusion principle. It does not make epistemic sense to say that two objects are found in the same state. From the epistemic perspective we have to be able to tell the objects apart to be allowed to say that we are indeed dealing with two objects. Their attributes must differ in some respect, be it their their spatio-temporal positions, their momentum or energy, or their internal attributes. The fact that electrons and other building blocks of matter follow Fermi-Dirac rather than Bose-Einstein statistics can therefore be seen as an expression of explicit epistemic minimalism. One may ask why we cannot argue in the same way that elementary bosons cannot share the same state either. The reason is that they should not be seen as objects in the epistemic sense of the word that we use here. These matters are discussed at length in Ref. \cite{epistemic}.

As a fourth example of explicit epistemic minimalism, we may take the fact the orbital angular momentum of a possibly rotating object has to be set to zero if we have no potential knowledge at all where in its orbit the object is positioned at a given time, that is, if the probability distribution of its position is spherically symmetric. Allowing for non-zero angular momentum in such a case gives rise to erroneous physical predictions.

In section \ref{basic} we have already used the same principle to argue that it is impossible to invent an adequate physical model with a fixed number of elementary objects that do not divide or merge.

In the present paper, explicit epistemic minimalism is used mainly in order to analyse the double-slit experiment from an epistemic point of view. Consider Fig. \ref{Figure15}, and assume that it is forever outside potential knowledge which slit the particle actually passes. However, there is (potential) knowledge that it passes slit 1 with probability $p_{1}$ and slit 2 with probability $p_{2}$. (Strictly speaking, according to the discussion in Section \ref{probability}, these probabilities are not defined in such a case, but this is not essential for the present argument.) Implicit epistemic minimalism would mean that it does not matter for the evolution of the system whether we assume that it actually takes one of the paths, even if we can never know which. The only option is then to combine probabilities as if events 1 and 2 are mutually exclusive. That is, for any pair of probabilities $p_{k}$ and $p_{rk}$ we must have

\begin{equation}
p_{r}=p_{1}p_{r1}+p_{2}p_{r2},
\label{normalprob}
\end{equation}
where $p_{r}$ is the probability that the particle finally hits the point $r$ on the detector screen, $p_{k}$ is the probability that it passes slit $k$, and $p_{rk}$ is the probability that it hits $r$ given that it has passed slit $k$.

In contrast, explicit epistemic minimalism means that physical law must contradict the possibility that there is (unknowable) path information. The only way to get the message through is to let

\begin{equation}
p_{r}\neq p_{1}p_{r1}+p_{2}p_{r2}
\label{quantprob}
\end{equation}
for some choice of probabilities. In Section \ref{bornsrule} we discuss how this condition leads to Born's rule.

\subsection{Epistemic completeness}
\label{epcomplete}

Epistemic minimalism is all about the fact that physical law seems to make use of \emph{no more} than what can be perceived and distinguished in principle. Turning the perspective around, we may argue that physical law should make use of \emph{everything} than can be perceived and distinguished from something else. Loosely speaking, if two things can be distinguished, there will be a corresponding distinction in physical law.

This idea can be promoted to the principle of \emph{Epistemic completeness}. All subjectively perceived distinctions, or distinctions deduced from such perceptions, correspond to distinctions in proper models of physical law and of the physical state. This is true, in particular, when it comes to objects, attributes and attribute values.

Since we can deduce from the incompleteness of knowledge that there is a distinction between potential knowledge and the currently unknowable (Fig. \ref{Figure5}), a corresponding distinction should be made in physical law. This is accomplished by the distinction between Eqs. [\ref{normalprob}] and [\ref{quantprob}], which therefore can be seen as a consequence of epistemic completeness as well as a consequence of explicit epistemic minimalism. 

Epistemic completeness can also be applied in order to treat time in a proper scientific way. Since there are primary subjective distinctions between the past, the present and the future, these distinctions should be present also in proper representations of time in physical models. We have tried to do this by introducing sequential time $n$ as an inherently directed attribute according to Definition \ref{directededvalues}.

The attribute $n$ is distinguished from the numerical measure $t$ of the temporal distance between two events. Like all other distances between attribute values, $t$ is not considered to be a primary attribute in state space $\mathcal{S}$, but is defined in certain physical states $S$ in which there are reference objects according to Fig. \ref{Figure11} which defines the distance. Thus knowledge about the distance $t$ between two prior events becomes an object in the state of potential knowledge $PK$ at the present time $n$. This distinction between $n$, which formalise the perceived flow of time and our ablity to order events temporally, and $t$, which formalise our ability to make temporal measurements, can also be seen as an expression of epistemic completeness.  

\subsection{Epistemic closure}
\label{closure}

Comparing the principles of explicit epistemic minimalism and epistemic completeness, the former basically says `everything physical is epistemic', while the latter says `everything epistemic is physical'. Any entity or distinction introduced in a physical model should correspond to a knowable entity or distinction, and any knowable entity or distinction chould have a counterpart in a physical model. We get a one-to-one correpondence between knowable entities and distinctions, and entities and distinctions in proper physical models. We may call this correspondence \emph{epistemic closure}. It reflects the world view expressed in Fig. \ref{Figure1}, where the subjective and the objective aspects of the world are seen as inseparable and equally fundamental.

\subsection{Epistemic consistency}
\label{consistency}
An axiomatic mathematical system is consistent if and only if the given set of axioms and the given set of deduction rules cannot be used to conclude that one and the same theorem $T$ is both true and false. In other words, it is never proper to arrive at both $T$ and $\neg T$. In an analogous way, we may say that a physial theory is consistent if and only if the state of world at a given time $n$ never becomes properly described by two different physical states $S$ and $S'$ if we follow the given physical law.

Such a contradiction can occur only if the physical theory contains different perspectives on the same situation, so that these perspectives may be consistent or not. Otherwise the consistency of the theory is self-evident and does not constitute a condition that constrains physical law. This is so in realistic and deterministic physical theories like Newtonian mechanics. The physical state at a later time $t'$ is a function of the state at a previous time $t$.
It is meaningless to require consistency, since no conflict of perspectives is possible.

In the present epistemic approach to physics the situation is different. We have different subjects who observe the same objects according to Fig. \ref{Figure3}, and the requirement that their knowledge about these objects is consistent is expressed by Eq. [\ref{cs2}].

There is another way in which we can have different perspective on the same situation, on the same set of objects or events. We may observe them as they happen, or we may remember them afterwards. According to the discussion in Section \ref{structure}, each object in the state of potential knowledge $PK(n)$ is equipped with the binary presentness attribute. We may therefore divide $PK(n)$ into two parts, one containing all objects that belong to the present time $n$, and the other containing all objects which corresponds to memories of the past. We may call the first part $PKN(n)$ and the other part $M(PK(n-1))$, writing

\begin{equation}
PK(n)=PKN(n)\cup M(PK(n-1)),
\end{equation}
where $M(\ldots)$ denotes the memory of the state of knowledge within brackets. If we ignore the presentness attribute of all objects in $M(PK(n-1))$ and $PK(n-1)$, we must require that $M(PK(n-1))\subseteq PK(n-1)$. In terms of the physical state, we may write 

\begin{equation}
S(n)=SN(n)\cap M(S(n-1))
\end{equation}
in an analogous notation. We then have the following criterion for epistemic consistency:

\begin{equation}
M(S_{-Pr}(n-1))\supseteq S_{-Pr}(n-1),
\label{epcon2}
\end{equation}
where the subscript $-Pr$ indicates that the value of the presentness attribute is ignored, or, in other words, that we project the physical state from the entire state space $\mathcal{S}$ to the subspace spanned by all other attributes than the presentness attribute. Equation [\ref{epcon2}] represents a criterion that constrains physical law in the sense that it provides a condition that $S(n)$ must fulfil given $S(n-1)$.

Apart from the different perspectives on the same situation provided by different subjects, and those provided by the memory of the situation as compared to the first hand experience of it, there is a third way to get such different perspectives: we may experience or remember the situation at the same time as we deduce knowledge about it from \emph{other} experiences. Memories and deductions have the thing in common that they refer to something else, upon which they provide their own perspective. In more abstract terms, memories and deductions are objects that point to other objects. In this language, epistemic consistency means that the source objects and the target objects must conform.

Let us call the deduced object $\tilde{O}$, and let $O$ be the directly perceived object to which $\tilde{O}$ refer. Then $\tilde{O}$ should never be described by properly deduced attribute values that are knowably different from those of $O$. Employing the object state space $\mathcal{S}_{O}$ introduced in Section \ref{objectspace} and the object state $S_{OO}$ referred to in Fig. \ref{Figure14}, we get the condition

\begin{equation}
(S_{OO})_{-Pr}\cap (S_{\tilde{O}\tilde{O}})_{-Pr}\neq\varnothing.
\label{epcon3}
\end{equation} 
Just as in the concistency criterion [\ref{epcon2}], we ignore the presentness attribute, since it does not matter whether $\tilde{O}$ or $O$, or both, are memories.

\begin{figure}[tp]
\begin{center}
\includegraphics[width=80mm,clip=true]{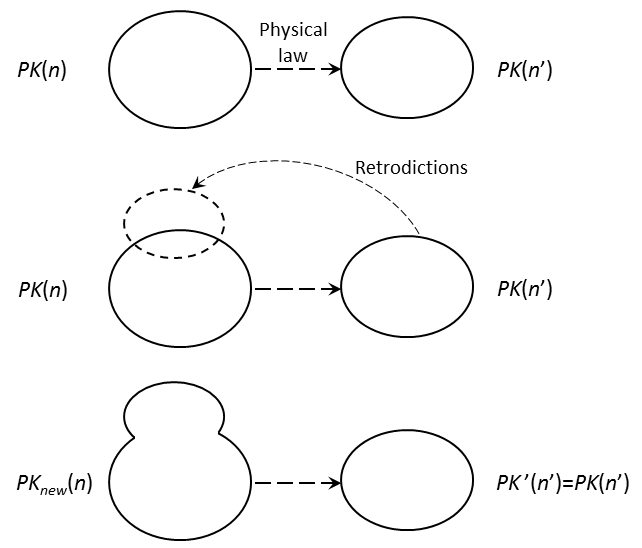}
\end{center}
\caption{Illustration of the consistency criterion expressed in Eq. [\ref{epcon4}]. Anything that can be deduced about time $n$ from any possible future state of potential knowledge $PK(n')$ must be insignificant enough, so that it does not cause a future state $PK'(n')$ that is different in any way from the state $PK(n')$ we actually see. Such a forbidden difference might be an outright contradiction, but it might also amount to more (or less) focused knowledge, or more or less conditional knowledge.}
\label{Figure16}
\end{figure}

There is one way to use epistemic consistency to constrain physical law where the temporal relations are crucial, though, where the deduction is a retrodiction, where it is pointing towards an object belonging to the past. In the present epistemic approach, physical law acts via an evolution rule that constrains the potential knowledge $PK(n')$ at time $n'$ given the potential knowledge $PK(n)$ at a previous time $n<n'$. These matters are discussed in the following section. Suppose that the knowledge $PK(n')$ that appears at time $n'$ allows a retrodiction about the physical state at the previous time $n$. Together with the remembered knowledge about time $n$, this retrodicted or deduced knowledge concerning time $n$ may become greater than the actual state of potential knowledge $PK(n)$ was at the time. This extended knowledge concerning time $n$ may give rise to another state $PK'(n')$ of potential knowledge at the subsequent time $n'$, since the physical evolution law then acts on a different initial state. If that were the case, then there would be two inconsistent corresponding physical states $S(n')$ and $S'(n')$ that properly described the world at time $n'$. This is forbidden, which means that the new knowledge that may appear at time $n'$ is constrained so that it does only allow retrodictions about time $n$ which evolves into a state of knowledge $PK'(n')=PK(n')$. The idea is illustrated in Fig. \ref{Figure16}.

To make the argument more precise and formal, let $\tilde{PK}_{new}(n;n')$ be the new potential knowledge about time $n$ that physical law make it possible to deduce at a later time $n'>n$. That the deduced knowledge is new means that $\tilde{PK}_{new}(n;n')\cap PK(n)=\varnothing$. Let $PK_{new}(n)=PK(n)\cup \tilde{PK}_{new}(n;n')$. Then, for any time $n''>n$, we have

\begin{equation}
PK_{new}(n'')=PK(n'')
\label{epcon4}
\end{equation}
for any potential knowledge $PK_{new}(n'')$ that follows from $PK_{new}(n)$ via physical law.

This more subtle kind of consistency is relevant in discussions of the double-slit experiment (Fig. \ref{Figure15}). Suppose that we see an interference pattern in such an experiment at time $n$. If it were possible to gain knowledge at the later time $n'$ that makes it possible to retrodict which slit the particle actually passed at time $n$, then the future state that would follow if this knowledge were there alreday at time $n$ would contradict the state that we actually perceive and remember now, since then physical law dictates that we get no interference pattern. To avoid such a contradiction we have to impose the quantum mechanical rule that tells us that interference patterns only appear when it is impossible in principle to gain path information at a later time, be it via regained memories, later deduction from memories, or deduction from knowledge acquired later.

This rule can thus be seen as a constraint on physical law imposed by the requirement of epistemic consistency. Even though it may seem that this rule transcends time, anticipating at time $n$ in a non-deterministic world what may or may not happen at time $n'$, this is not really the case, since the impossibility in principle to gain path information at such a later time is a function of the physical state at time $n$. The appearance of an interference pattern at time $n$ in such a situation is an expression of this fact, via the principle of explicit epistemic minimalism (Section \ref{explicitminimalism}).

\section{The evolution operator}
\label{evolution}

Any physical law allows us to define an evolution operator $u$ that acts on any physical state $S$ and gives another state $uS$ such that the next physical state is a subset of $uS$. In other words, given the present physical state, any physical law limits the possibilities for the future. The sharpest formulation of physical law limits these possibilities as much as possible without giving rise to predictions which are sometimes not fulfilled. This idea is captured in the definition of $u_{1}$.

\begin{defi}[\textbf{The evolution operator} $u_{1}$]
The evolution operator $u_{1}$ is defined by the condition that $u_{1}S(n)$ is the smallest possible set $C\subseteq\mathcal{S}$ such physical law dictates that $S(n+1)\subseteq C$.
\label{evolutionu1}
\end{defi}

\begin{figure}[tp]
\begin{center}
\includegraphics[width=80mm,clip=true]{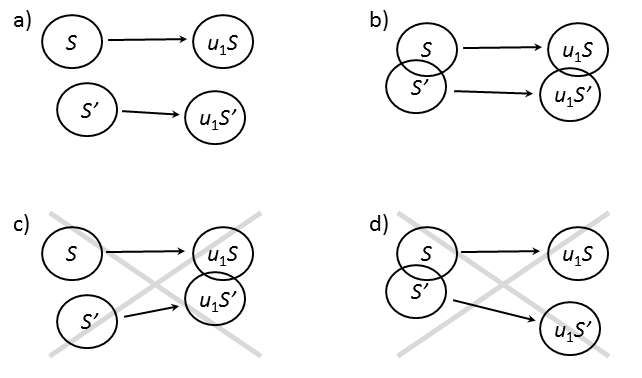}
\end{center}
\caption{Physical law as expressed by $u_{1}$ is analogous to an invertible function. Evolution of two states $S$ and $S'$ according to cases a) and b) are allowed by Assumption \ref{uniqueu1}, whereas evolution according to c) and d) are forbidden.}
\label{Figure17}
\end{figure}

The generality of this definition means that it does not tell us anything about the form of physical law. To give some basic structure to $u_{1}$, we assume the following (Fig. \ref{Figure17}).

\begin{assu}[\textbf{The operator} $u_{1}$ \textbf{is unique and invertible}]
Two states $S$ and $S'$ overlap if and only if $u_{1}S$ and $u_{1}S'$ overlap: $S\cap S'\neq\varnothing\Leftrightarrow u_{1}S\cap u_{1}S'\neq\varnothing$.
\label{uniqueu1}
\end{assu}

That $S\cap S'\neq\varnothing\Rightarrow u_{1}S\cap u_{1}S'\neq\varnothing$ can be regarded as an assumption that the evolution $u_{1}$ is a function in a subjective sense: two states that are subjectively the same cannot evolve into states that are subjectively different. That $S\cap S'=\varnothing \Rightarrow u_{1}S\cap u_{1}S'=\varnothing$ can be seen as an assumption that $u_{1}$ is subjectively invertible: two subjectively different states cannot evolve into states which are subjectively indistinguishable.

\begin{figure}[tp]
\begin{center}
\includegraphics[width=80mm,clip=true]{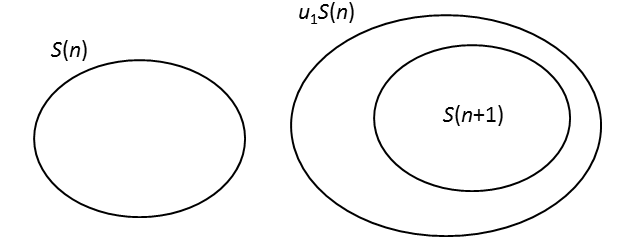}
\end{center}
\caption{Since $S(n+1)\subseteq u_{1}S(n)$ cannot overlap $S(n)$, by definition of successive states, we can define the evolution operator $u_{1}$ such that $u_{1}S(n)$ and $S(n)$ do not overlap either.}
\label{Figure18}
\end{figure}

Time changes if and only if something changes. The sequence of time instants $\{n,n+1,\ldots\}$ is well-defined if and only if we can tell the physical states in the corresponding sequence $\{S(n),S(n+1),\ldots\}$ apart. Each states in this sequence must be knowably different from all the others. It is therefore not possible for one exact state $Z$ to be consistent with both $S(n)$ and $S(n+1)$. That is, we require

\begin{equation}
S(n+1)\cap S(n)=\varnothing.
\end{equation}
It follows that

\begin{equation}
S(n)\cap u_{1}S(n)=\varnothing,
\label{evolutionchange}
\end{equation}
as illustrated in Fig \ref{Figure18}.

One might argue that $S(n+1)$ may correspond to a state in which knowledge \emph{increases} or \emph{decreases} rather than \emph{changes}, as compared to $S(n)$, so that we might have $S(n+1)\subset S(n)$ or $S(n+1)\supset S(n)$, contradicting Eq. [\ref{evolutionchange}]. However, increased or decreased knowledge about the observed objects corresponds to a changed perception of the observer, and must therefore correspond to a changed state of the objects in the body of the same observer, according to the assumption of detailed materialism (Section \ref{basic}). If you put your glasses on to increase your visual knowledge, the state of the neurons in the visual cortex changes. Therefore Eq. [\ref{evolutionchange}] still holds.

The evolution operator $u_{1}$ is a mapping from the power set $\mathcal{P}(\mathcal{S})$ of state space $\mathcal{S}$ to itself. Explicit epistemic minimalism (Section \ref{explicitminimalism}) requires that the evolution should just be defined for elements of $\mathcal{P}(\mathcal{S})$ that may correspond to states of potential knowledge $PK$. Since we argue in Section \ref{basic} that potential knowledge is always incomplete, exact states $Z$ are therefore not in the domain $\mathcal{D}_{u}$ of $u_{1}$.

Since we cannot make $u_{1}S(n)$ any smaller according to Definition \ref{evolutionu1}, we conclude that we may have $S(n+1)=u_{1}S(n)$. And since it must be possible to apply $u_{1}$ to the physical state at all times, we conlude that $u_{1}S(n)\in\mathcal{D}_{u}$. The range of $u_{1}$ becomes the same as its domain.

\begin{equation}
u_{1}: \mathcal{D}_{u}\rightarrow \mathcal{D}_{u}, \,\,\, \mathcal{D}_{u}\subset\mathcal{P}(\mathcal{S}).
\label{transitiveu}
\end{equation}

If $\Sigma\subset\mathcal{S}$ and $\Sigma'\subset\mathcal{S}$ are two sets that may correspond to physical states, then so is $\Sigma\cup \Sigma'$. This set corresponds to the state of potential knowledge $PK \cap PK'$ according to Table \ref{itemrelations}, which in turn correponds to the statement `things are according to $PK$ or to $PK'$', which is always a possible state of knowledge. This closure ot the domain of $u_{1}$ under unions of its elements reminds us of $\sigma$-algebras, and one may try to describe $\mathcal{D}_{u}$ as such an algebra.

According to the definition of a $\sigma$-algebra, that would mean that we have to allow the evolution operator to be applied to the entire state space $\mathcal{S}$, and also to the empty set $\varnothing$. However, it is not possible to fulfil the condition $u_{1}\mathcal{S}\cap\mathcal{S}=\varnothing$ that is implied by Eq. [\ref{evolutionchange}]. In fact, we must have $u_{1}\mathcal{S}\subseteq\mathcal{S}$ according to the definition of the state space. A physical state that equals the entire state space corresponds to a state of no knowledge at all, to a state of non-existence. The evolution operator is not a creation operator; it cannot create something from nothing, but rather evolves the state of a world that already exists. We conclude, therefore, that $\mathcal{D}_{u}$ is not a $\sigma$-algebra. This conclusion is reinforced by the observation that the expression $u_{1}\varnothing$ lacks meaning.

Nevertheless, $\mathcal{D}_{u}$ can be compared to the largest $\sigma$-algebra $\Sigma_{V}$ of measurable subsets of $\mathcal{S}$ (Definition \ref{voldef}). To be able to define probability properly in all relevant circumstances (Section \ref{probability}), we want all physical states $S$ and all evolved states $u_{1}S$ to be measurable. Therefore we assume that $\mathcal{D}_{u}\subseteq\Sigma_{V}$. We have $\mathcal{S}\not\in\mathcal{D}_{u}$ and $\varnothing\not\in\mathcal{D}_{u}$ according to the above discussion. Also $Z\not\in\mathcal{D}_{u}$ for any exact state $Z$. Therefore we conclude that

\begin{equation}
\mathcal{D}_{u}\subset\Sigma_{V}.
\end{equation}

If potential knowledge changes size, it changes content. This observation was used above to validate Eq. [\ref{evolutionchange}]. In general we may say that if $S$ is a possible physical state which can evolve to $u_{1}S$ and corresponds to the state of knowledge $PK$, then $S'$ with $S'\subset S$ or $S'\supset S$ is \emph{not} a conceivable physical state if it corresponds to a different state of potential knowledge $PK'$. It is therefore meaningless to apply $u_{1}$ to such a set $S'$. This means that

\begin{equation}
S\in\mathcal{D}_{u}\Rightarrow S'\not\in\mathcal{D}_{u}
\label{nosubsetev}
\end{equation}
whenever $S'\subset S\subset\mathcal{S}$ or $S\subset S'\subset\mathcal{S}$, and $PK\neq PK'$. The last qualification is added since a given state of potential knowledge $PK$ may correspond to many slightly different physical states $S,S',\ldots$ since we cannot determine the boundary $\partial S$ exactly, as discussed in relation to Fig. \ref{Figure8}. Some of these candidate physical states may be subsets of each other, and we want to be able to apply $u_{1}$ to all of them, of course.

We argued in Section \ref{statespacedim} that the only relevant part of state space $\mathcal{S}$ is the subspace $\mathcal{S}_{K}$ spanned by the known distinct attributes. The number of these attributes defines the knowledge dimension $D_{K}[\mathcal{S}](n)$ (Definition \ref{knowdimnum}). Let $\Pi_{K}u_{1}S(n)$ be the  projection of the state $S(n)$ onto $\mathcal{S}_{K}$. To be epistemically sound, the evolution operator $u_{1}$ should be such that it is sufficient to specify its effect on $\Pi_{K}S$ in order to specify the evolution of the physical state $S$. This condition can be expressed as the commutation relation

\begin{equation}
u_{1}\Pi_{K}S(n)=\Pi_{K}u_{1}S(n),
\end{equation}
which should hold for all $n$.

It is almost self-evident that this condition is fulfilled, when you think about it for a minute. We have identified the physical state which we are going to evolve with a state of knowledge. Thus the evolution $u_{1}$ must depend on this knowledge, and on nothing else. We know nothing about the attributes in the part $\mathcal{S}\setminus \mathcal{S}_{K}$ of state space. Therefore $u_{1}$ cannot depend on these parts. It must depend on something known, it cannot depend on nothing.

To be a bit more formal, consider Fig. \ref{Figure12}. In panel a) there is no knowledge associated with attribute $A_{2}$. We can specify the state $S$ completely by saying $\upsilon_{1}^{\min}<\upsilon_{1}<\upsilon_{1}^{\max}$, where $\upsilon_{1}$ is the value of attribute $A_{1}$. The values of $A_{2}$ does not appear in this specification, and therefore $u_{1}$ cannot depend upon these.

Even though this may seem self-evident, it has peculiar consequences. It means that whenever new attributes become known, their appearance in the `field of vision' is not dictated by the deterministic part of physical law, as given by $u_{1}$. For example, if we suddenly become aware of attribute $A_{2}$, learning that $\upsilon_{2}^{\min}<\upsilon_{2}<\upsilon_{2}^{\max}$, this knowledge appears `out of the blue'. It can be seen as the effect of a \emph{state reduction}, defined by the condition $S(n+1)\subset u_{1}S(n)$.

Let us next discuss such state reductions, and their relation to determinism. 

\begin{defi}[\textbf{Determinism}]
Physical law is deterministic if and only if $S(n+1)=u_{1}S(n)$ for all states $S(n)\in\mathcal{D}_{u}$. The individual state $S(n)$ evolves deterministically if and only if $S(n+m)=u_{m}S(n)$ for all $m\geq 1$, where $u_{m}\equiv (u_{1})^{m}$.
\label{determinism2}
\end{defi}

Assume that $S(n)$ evolves deterministically even though it is not exact. Since we have argued that potential knowledge is incomplete at all times, the evolved state $S(n+1)=u_{1}S(n)$ is not exact either. This means that there are several values of some attribute $A_{i}$ that we know at time $n$ that we cannot exclude at time $n+1$. The same goes for subsequent times $n+2, n+3, \ldots$. But if there is no possibility to gain more knowledge about $A_{i}$ in the future than can be anticipated at time $n$, to exclude some more values, then it is meaningless to say that there are several values that cannot be excluded in the first place. The sequence of states $\{S(n),S(n+1),\ldots\}$ becomes analogous to an atom moving in state space $\mathcal{S}$, which can never be split to see what is inside, even though we assume that it is composite, that it contains several exact states $Z$. Explicit epistemic minimalism (Section \ref{explicitminimalism}) does not allow such a picture, since it makes use of distinctions that can never be probed perceptionwise. Therefore there cannot be any inexact state $S(n)$ that evolves deterministically. Since all true states are inexact due to the incompleteness of knowledge we conclude the following.

\begin{state}[\textbf{Physical law is not deterministic}]
There are times $n$ such that a state reduction occurs at time $n+1$, meaning that $S(n+1)\subset u_{1}S(n)$.
\label{reductionsoccur}
\end{state}

We note that a state reduction is never a deterministic event, according to Definition \ref{evolutionu1}. If the state $S(n+1)$ that emerges from the state reduction $u_{1}S(n)\rightarrow S(n+1)\subset u_{1}S(n)$ were possible to determine at time $n$, then $u_{1}$ should be re-defined to $u_{1}'$ for which $S(n+1)=u_{1}'S(n)$, so that no state reduction would occur.

Suppose next - contrafactually - that potential knowledge is complete, so that we can write $S(n)=Z$ for some exact state $Z$. Suppose further that $S(n)$ does not evolve deterministically even though it is exact. That would correspond to the existence of a stochastic term in the operator $u_{1}$. In other words, such an lack of determinism would not be the result of incomple knowledge, but of an added noise that has no basis in the state of the world, as described by a set of objects, and their internal and relational attributes.

If we assume that there is no such noise in physical law, then we arrive at the conclusion that a state $S(n)$ evolves deterministically if and only if it is exact, so that we may write $S(n)=Z$ for some exact state $Z$. This means that in the absence of noise, complete knowledge and deterministic physical law goes hand in hand. In that case we could equally well \emph{define} an exact state $Z$ as a state that evolves deterministically.

\subsection{Identifiable objects and their evolution}
\label{objectevolution}

To be able to speak about objects whose evolution we follow through time, we have to be able to decide whether two perceptions of an object at different times corresponds to one and the same object. The intuition is the following: if an object in a state of knowledge $PK(n)$ cannot be distinguished from an object in the next state $PK(n+1)$, then these two objects must be seen as the same object. It is impossible to tell them apart, and it is epistemically unsound to give them different names. This can be seen as a consequence of explicit epistemic minimalism (Section \ref{explicitminimalism}). An object $O$ which is possible to track in this way from time $n$ to time $n+1$ may be called \emph{identifiable} at time $n$. It fulfils the criterion

\begin{equation}
S_{OO}(n)\cap S_{OO}(n+1)\neq\varnothing,
\end{equation}
where $S_{OO}\subset\mathcal{S}_{O}$ is the object state embedded in object state space, as discussed in Section \ref{objectspace}.

Note that this definition does not take into account the relational attributes that relate $O$ to its environment $\Omega_{O}$ (Section \ref{objectspace}). The object $O$ can therefore be suddenly moved spatially and still preserve its identity. This means that all minimal objects of a given species are seen as one and the same object, in a certain sense. This view conforms with the conclusion in statistical mechanics that that the exchange of two identical particles does not create a new physical state. However, it is at odds with the fact that electrons can follow distinguishable trajectories through space.

We may therefore introduce the spatio-temporal attributes of an object $O$ as `pseudo-internal' attributes, and include them in the list which specify an expanded internal object state $\hat{S}_{OO}$. This can be achieved if we fix a set of identifiable reference objects $\{O_{r}\}$ and regard them as a rudimentary spatial coordinate system for each other object $O$. (We may alternatively consider the reference objects to be parts of an extended object $\hat{O}$ that is always composite: $\hat{O}=\{O,\{O_{r}\}\}$.) Then we can express the corresponding identifiability condition $\hat{S}_{OO}(n)\cap \hat{S}_{OO}(n+1)\neq\varnothing$. Using this condition, an electron whose existence is deduced in an experiment at time $n$ on Earth should not be identified as being the same as an electron whose existence is deduced at time $n+1$ by a physicist somewhere in the Andromeda Galaxy.

\begin{figure}[tp]
\begin{center}
\includegraphics[width=80mm,clip=true]{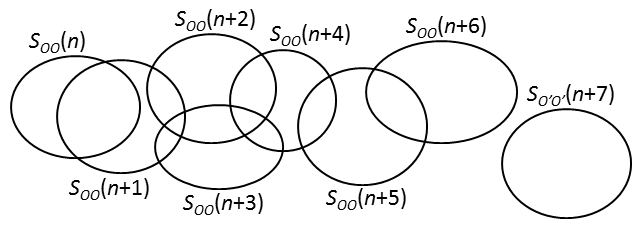}
\end{center}
\caption{The evolution of the state $S_{OO}$ of an object $O$ that is identifiable in the time interval $[n,n+6]$. The object $O'$ at time $n+7$ cannot be identified with object $O$ at time $n+6$, and is therefore given a separate name.}
\label{Figure19}
\end{figure}

It may be possible to track an object during extended periods of time. If object $O$ is identifiable at all times $n, n+1,\ldots,n+m$, then it is said to be identifiable in the time interval $[n,n+m]$. This definition is illustrated in Fig. \ref{Figure19}. Even if the object is identifiable at every time instant in the above sequence, it may be possible to distinguish the object at some time $n+\mu$ from the object at time $n$, that is, we may have $S_{OO}(n)\cap S_{OO}(n+\mu)=\varnothing$, where $1<\mu\leq m$. Due to its identifiability at each instant, the object nevertheless preserves its identity.

If an object $O$ is identifiable at time $n$, there must be other objects in $PK(n+1)$ that \emph{can} be distinguished from objects in $PK(n)$, since sequential time $n$ is updated only if a subjective distinction can be made between now and then. These other objects can collectively be seen as a clock that ticks each time sequential time is updated.

Conversely, if an object changes and time is updated according to $n\rightarrow n+1$, there must be other objects that subjectively stay the same, that are identifiable at time $n$. Otherwise there is nothing that makes it epistemically meaningful to say that the states $S(n)$ and $S(n+1)$ refer to the same world. More formally, a world specified by a sequence of states $S(n)$ preserves its identity during time interval $[n,n+m]$ if and only if there is an identifiable object at all times $n, n+1,\ldots,n+m-1$.

A leave may fall from the tree, but the stem of the tree does not move. When the leave falls, the immobility of the stem helps to preserve the identity of the tree, and of the entire world. When the leave has fallen, the tree may be cut down, and the leave resting on the ground preserves the identity of the world. Even if everything around us seem to change at once, some internal objects may stay the same, such as mood and memories. In this way, the evolution of a given, identifiable world may be compared to walking: when one foot is lifted, the other is resting on the ground.

If object $O$ is identifiable at time $n$, the following expression is defined:

\begin{equation}
S_{O}(n+1)\subseteq u_{1}[S(n)]S_{O}(n).
\label{evolveobject}
\end{equation}
We have to let the evolution operator depend on the state $S(n)$ of the entire world, since object $O$ is always related to the world to which it belongs. Of course, we may define the corresponding expression

\begin{equation}
S_{OO}(n+1)\subseteq u_{O1}[S(n)]S_{OO}(n)
\label{evolveoobject}
\end{equation}
if we work in object state space $\mathcal{S}_{O}$ rather than $\mathcal{S}$.

The above discussion concerns directly perceived objects $O$ rather than deduced quasiobjects $\tilde{O}$. We argued in Section \ref{basic} that objects $O$ are always sufficient to specify the physical state $S$, and to determine its evolution $u_{1}S$. We added, however, that quasiobjects $\tilde{O}$ such as minimal objects are nevertheless useful when we want to express the evolution operator $u_{1}$ in a way that is generally valid, that does not depend on the specific state $S$ or object state $S_{O}$ to which it is applied. The same reasoning applies to the operator $u_{O1}$ as applied to the object state $S_{OO}$ in object state space $\mathcal{S}_{O}$ (Eq. [\ref{evolveoobject}]).

Since we have concluded that all objects cannot be identifiable at all times, it is necessary to introduce a notion of identifiability and evolution of quasiobjects if we want to uphold the idea that no perceived object pops in our out of existence, and that objects preserve their identity even if we do not look at them each moment $n$ of time. That the evolution operator $u_{1}$ is such that this idea can be upheld is supported by primitive experience: we do not need to stare constantly at a flower in the kitchen window to say that it is the same flower we see each morning as we drink our coffee. If it is not there one morning, we conclude that someone must have removed it rather than that it simply vanished.

It is a basic assumption in this study that all properly interpreted objects can be modelled as being composed of minimal objects $\tilde{O}_{M}$ (Section \ref{minimalobjects}). These minimal objects are quasiobjects which are assumed to be possible to track through time in the sense that the evolved minimal-object state $u_{O1}\hat{S}_{\tilde{O}_{M}\tilde{O}_{M}}(n)$ is defined at each time $n$ for each minimal object $\tilde{O}_{M}$ that appears in the model of the physical state at that time. This expression is assumed to be such that

\begin{equation}
u_{O1}[S(n)]\hat{S}_{\tilde{O}_{M}\tilde{O}_{M}}(n)\cap \hat{S}_{\tilde{O}_{M}\tilde{O}_{M}}(n)\neq\varnothing.
\label{minimaloverlap}
\end{equation}

Minimal objects are then assumed to be identifiable in the sense that there is always a minimal object $\tilde{O}_{M}'$ in the model of the state at time $n+1$ whose object state $\hat{S}_{\tilde{O}_{M}'\tilde{O}_{M}'}(n+1)$ fulfils

\begin{equation}
\hat{S}_{\tilde{O}_{M}'\tilde{O}_{M}'}(n+1)\subseteq u_{O1}[S(n)]\hat{S}_{\tilde{O}_{M}\tilde{O}_{M}}(n).
\label{minimalchain}
\end{equation}

This means that we may identify the two minimal objects with each other: $\tilde{O}_{M}'=\tilde{O}_{M}$. Further, there is no minimal object $\tilde{O}_{M}'$ in the model of the state at time $n+1$ that cannot be identified in this way with a minimal object $\tilde{O}_{M}$ belonging to the model at time $n$. 

There is a complication, though. Minimal objects may divide or merge in particle reactions. There has to be a way to identify a set $\{\tilde{O}_{M}'\}$ of outgoing minimal objects from such a reaction at time $n+1$ with a set of incoming minimal $\{\tilde{O}_{M}\}$ objects at time $n$. This is clearly a more tricky matter than to identify a single minimal object with the evolved state of another one according to Eq. [\ref{minimalchain}]. We will not pursue the matter here, but refer the reader to the discussion in Ref. \cite{epistemic}.

Since the only attribute values that can differ between two minimal objects of the same species $M$ (like an electron) are the spatio-temporal ones, we treat them as pseudo-internal attributes according to the above discussion, and put hats above the object states in Eqs. [\ref{minimalchain}] and [\ref{minimaloverlap}]. Otherwise two subsequent states of a minimal object of a given species $M$ always fulfil  $S_{\tilde{O}_{M}\tilde{O}_{M}}(n+1)=S_{\tilde{O}_{M}\tilde{O}_{M}}(n)$, so that the overlap expressed in Eq. [\ref{minimaloverlap}] becomes trivial.

If all minimal objects fulfil the conditions discussed above they may be called \emph{quasi-identifiable}. In effect, each minimal object can then be treated as if it follows a trajectory like that in Fig. \ref{Figure19} with the spatio-temporal interpretation that the object moves to the right, if we allow that the worm splits into several branches at an object division, and we do \emph{not} allow that the chain breaks the way it does at time $n+7$ in that figure.

\begin{figure}[tp]
\begin{center}
\includegraphics[width=80mm,clip=true]{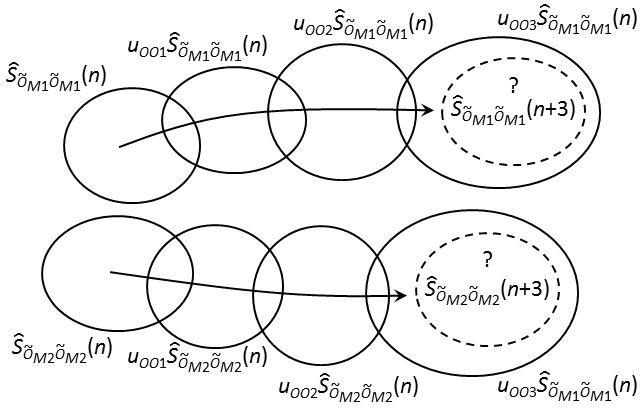}
\end{center}
\caption{An object $O$ is assumed to be composed of two minimal objects $\tilde{O}_{M1}$ and $\tilde{O}_{M2}$. Each of them is identifiable as its state evolves in the object state space $\mathcal{S}_{O}$ according to the evolution rule $u_{O1}$. If $O$ undergoes a perceivable change as $n+2\rightarrow n+3$, then at least one of the minimal objects must also undergo a sudden change at the same moment. However, it is impossible to say which if they are both deduced quasiobjects which are not directly perceived.}
\label{Figure20}
\end{figure}

We can, in principle, follow the trajectory of individual elementary fermions, for example in bubble chambers. Therefore it is epistemically sound to speak about such trajectories and to base physical models on these, even though we very seldom see such trajectories in practice, and we therefore cannot keep track of individual minimal objects that make up a perceived object $O$. In the language used in this study, such a model is epistemically sound since it is possible by assumption to divide each object a finite number of times until we reach the level of minimal objects (Section \ref{minimalobjects}).

We may ask ourselves how perceivable changes may occur at all if all objects can be described as being composed of quasi-identifiable minimal objects. If no such changes occur, it is impossible to define any temporal update $n\rightarrow n+1$ since there is no object that can act as a clock. In other words, how is it possible that a perceived object suddenly changes, like the one in Fig. \ref{Figure19} at time $n+7$? If all minimal objects were directly perceived this would indeed be impossible. All we would see would be minimal objects with trajectories that floated seeminlgy continuously through space. But an object may change suddenly or discretely if it is assumed that the minimal objects of which it is composed are quasiobjects. Then these mimimal objects may follow seemingly continous, but invisible trajectories that interpolate between the perceived distinct states of the object now and then. In this way we may make a proper model of the flower in the kitchen window as being the same as the one we saw yesterday, even if it has dropped a leaf during the night.

We may say that an object which can be modelled as being composed of quasi-identifiable minimal objects is itself quasi-identifiable. It is this weaker notion of identifiability that we most often use in our everyday life. The evolution of a quasi-identifiable object can be defined according to Eq. [\ref{evolveobject}] just like for an identifiable object.

We need to make the discussion a little bit more precise. Suppose that object $O$ knowably changes according to $S_{OO}(n+2)\rightarrow S_{OO}(n+3)$ with $S_{OO}(n+2)\cap S_{OO}(n+3)=\varnothing$, defining the temporal update $n+2\rightarrow n+3$ (Fig. \ref{Figure20}). Even if $O$ is composed of a set of minimal objects $\{\tilde{O}_{M}\}$, at least one of these minimal objects must also change suddenly according to $\hat{S}_{\tilde{O}_{M}\tilde{O}_{M}}(n+2)\cap \hat{S}_{\tilde{O}_{M}\tilde{O}_{M}}(n+3)=\varnothing$ in order to get $S_{OO}(n+2)\cap S_{OO}(n+3)=\varnothing$. Such a sudden change is guaranteed if a state reduction occurs at time $n+3$ so that $u_{O2}\hat{S}_{\tilde{O}_{M}\tilde{O}_{M}}(n)\cap \hat{S}_{\tilde{O}_{M}\tilde{O}_{M}}(n+3)=\varnothing$, where $u_{O2}\equiv u_{O1}u_{O1}$. If all minimal objects in $\{\tilde{O}_{M}\}$ were directly perceived, the suddenly changing ones would not be identifiable, and it would be impossible to uphold the picture that they always follow continuous trajectories. However, if the minimal objects are quasiobjects, it becomes impossible to decide by direct perception \emph{which} of them changes abruptly. All we can say is that at least one of them must do so. Thus we cannot \emph{exclude} the picture that if we choose any one individual minimal object, it is indeed identifiable, following a continuous path.

\section{The measurement process}
\label{measurements}

What we have done until now is to introduce a general epistemic formalism that does not say anything specific about physical law. We have also discussed some epistemic guiding principles that limits the form of physical law to some extent (Section \ref{guiding}). In what follows we will use the formalism to define certain experimental contexts in which the guiding principles can be employed to motivate quantum mechanics as a convenient mathematical representation of the rules that govern observations in such contexts. 

\subsection{Properties}

Given any inexact physical state $S$ we can imagine a set of alternative states that result if additional knowledge is gained. For instance, if we see a raptor in the sky but cannot decide which kind, an exhaustive set of alternatives consists of all species of raptors that live in our country. Any such set of alternatives divides $S$ into distinct subsets $S_{j}$ such that

\begin{equation}\begin{array}{c}
S=\bigcup_{j}S_{j},\\
\forall j\neq j':\;\;S_{j}\cap S_{j'}=\varnothing.
\label{division}
\end{array}\end{equation}

If the list of alternatives is not exhaustive to begin with, it can trivially be completed by a last alternative `not any of the above'. For instance, the bird may be `a golden eagle, a white-tailed eagle, or some other raptor'. Formally, an incomplete set $\{S_{j}\}$ is completed by adding $S\setminus \bigcup_{j}S_{j}$ as the last alternative.

\begin{figure}[tp]
\begin{center}
\includegraphics[width=80mm,clip=true]{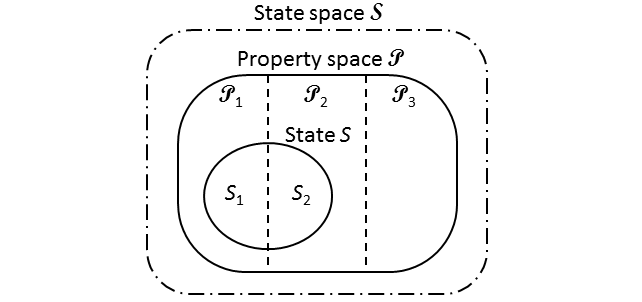}
\end{center}
\caption{The property space $\mathbf{\mathcal{P}}$ of a property $P$ is the union of all states for which there are objects such that $P$ is defined. $\mathbf{\mathcal{P}}=\bigcup_{k}\mathbf{\mathcal{P}}_{j}$, where $\mathbf{\mathcal{P}}_{k}$ is the union of states for which $P$ has value $p_{j}$. An analogous picture could be drawn in object state space $\mathcal{S}_{O}$ with object states $S_{OO}$, showing the object property space $\mathcal{P}_{O}$ with object property value spaces $\mathcal{P}_{Oj}$. The latter spaces are always disjoint, in contrast to the spaces $\mathcal{P}_{Oj}$.}
\label{Figure21}
\end{figure}

Consider the property $P$ that defines the set of alternatives $\{S_{j}\}$. This property defines an abstract property space $\mathcal{P}\subseteq\mathcal{S}$ (Fig. \ref{Figure21}). It is the union of all exact states $Z$ for which there are objects for which the property $P$ can be defined. Each possible value $p_{j}$ of $P$ defines a set $\mathcal{P}_{j}$ as the union of those $Z\in \mathcal{P}$ for which there are objects for which the value of $P$ is $p_{j}$. We get $\mathcal{P}=\bigcup_{j}\mathcal{P}_{j}$.

A property $P$ is a statement about attributes of objects. It may concern one attribute of one object, or several attributes of several objects. In general,

\begin{equation}
p=f(\{\upsilon_{il}\}),
\label{pstatement}
\end{equation}
where $\upsilon_{il}$ is the value of attribute $A_{i}$ of object $O_{l}$. Clearly, any attribute is a property, but the opposite is not necessarily true. By definition, the values $\upsilon_{i}$ of a given attribute $A_{i}$ are always possible to order (Definition \ref{orderedvalues}). In contrast, the values $p_{j}$ of a property $P$ cannot always be ordered. Ordering is impossible when $p_{j}$ is a function of attribute values $\upsilon_{il}$ and $\upsilon_{i'l}$ belonging to different attributes $A_{i}$ and $A_{i'}$. For example, the color of the tail feathers of a bird can be ordered according to the spectrum. However, letting $P$ represent bird species, the different species $p_{j}$ cannot be ordered, since the classification depends on other attributes than feather colors.

If $P$ is the species of the raptor, then $\mathcal{P}$ is the union of all exact states $Z$ for which there is a raptor. We have $S\subset\mathcal{P}$ for any physical state $S$ with raptors, since in any situation we know more than just `there is a raptor'. We know the landscape in which we see the bird, we have self-awareness, and so on.

Since a group of objects can always be formally described as a single, composite object, we can always say that $P$ applies to a given object $O$. We can therefore always define an object property space $\mathcal{P}_{O}$ and embed it in the object state space $\mathcal{S}_{O}$ rather than in $\mathcal{S}$ (Fig. \ref{Figure21}). In this case we should write $S_{OO}\subseteq\mathcal{P}_{O}$, since the only thing we may know about the object $O$ in the sky is that it is a raptor, so that $S_{OO}=\mathcal{P}_{O}$. The object property value space $\mathcal{P}_{O}$ is easier to work with than $\mathcal{P}$, since we have

\begin{equation}
\forall j\neq j':\;\;\mathcal{P}_{Oj}\cap \mathcal{P}_{Oj'}=\varnothing.
\label{disjointvalues}
\end{equation}
In contrast, the property value spaces $\mathcal{P}_{j}$ are not necessarily disjoint. There may be one exact state $Z$ in which there is one object with value $p_{j}$ of property $P$, and another object with value $p_{j'}$. Then $Z\in\mathcal{P}_{j}$ and $Z\in\mathcal{P}_{j'}$.

\subsection{Alternatives}
\label{alternatives}

To make the concept of alternatives meaningful in a formalism aimed at expressing physical law, the alternatives should have the potential to be realized. To be able to talk about such realizable alternatives in the first place, it should be possible to follow them through time until one of them eventually comes true. In our vocabulary this means that it should be possible to apply the evolution operator to them, and that they should be identifiable.   

Suppose that $S$ is a given physical state. Then there is no proper subset of $S$ which is a conceivable physical state, as discussed in Section \ref{evolution}. Such a subset would correspond to enlarged potential knowledge. But if potential knowledge changes size, it changes content. This means that it is epistemically meaningless to consider the evolution of alternatives $S_{j}\subset S$, according to Eq. [\ref{nosubsetev}].

Instead, we should consider alternatives associated to the specific object $O$ to which the alternatives apply. In the example above the observed raptor is the relevant object. This object $O$ should be identifiable or quasi-identifiable, as discussed in Section \ref{objectevolution}. Then its evolution is defined according to Eq. [\ref{evolveobject}]. We may then define the object alternative $S_{Oj}$ as

\begin{equation}
S_{Oj}(n)=S_{OO}(n)\cap\mathcal{P}_{Oj},
\label{altdef}
\end{equation}
and its evolution as

\begin{equation}
u_{O1}S_{Oj}(n)\equiv [u_{O1}S_{OO}(n)]\cap\mathcal{P}_{Oj}.
\label{altevdef}
\end{equation}
We suppress the dependence $u_{O1}=u_{O1}[S(n)]$ to make the expressions less cluttered, but have to remember that this dependence is crucial in some considerations below. The two definitions [\ref{altdef}] and [\ref{altevdef}] mean that for all $j\neq j'$, and for all states $S(n)$ of the entire world,

\begin{equation}\begin{array}{c}
S_{Oj} \cap S_{Oj'}=\varnothing\\
u_{O1}S_{Oj} \cap u_{O1}S_{Oj'}=\varnothing.
\end{array}
\end{equation}

Definition [\ref{altevdef}] is reasonable since the property value states $\mathcal{P}_{Oj}$ are abstract subsets of $\mathcal{S}_{O}$ that do not refer to any particular object state $S_{OO}$ (Fig. \ref{Figure21}). Property $P$ can be defined for $O$ whenever $O$ is the relevant object to which the alternatives apply, and stays relevant as time passes. This means that $S_{OO}\subseteq\mathcal{P}_{O}$ for all such times $n,n+1,\ldots$ of interest. We may therefore write $S_{OO}(n)=\bigcup_{j}[S_{OO}(n)\cap\mathcal{P}_{Oj}]$ and $u_{O1}S_{OO}(n)=\bigcup_{j}[u_{O1}S_{OO}(n)\cap\mathcal{P}_{Oj}]$. Definitions [\ref{altdef}] and [\ref{altevdef}] then imply

\begin{equation}
u_{O1}S_{OO}(n)=u_{O1}\bigcup_{j}S_{Oj}(n)=\bigcup_{j}u_{O1}S_{Oj}(n)
\label{linearev}
\end{equation}
for all states $S(n)$. These relations express that the object evolution operator $u_{O1}$ is linear in a set-theoretical sense.

\begin{table}
\caption{Three knowability levels of alternatives}
\label{levels}
\begin{tabular}{ll}
\hline\noalign{\smallskip}
		1 & \emph{It will never become known which alternative is true}. No pro-\\
		  & perty value $p_{j}$ that corresponds to an alternative $S_{Oj}$ defined\\
			& at time $n$ and referring to object $O$ will ever be observed.\\
\noalign{\smallskip}
		2 & \emph{It may become known which alternative is true}. There is a time\\
		  & $n_{1}>n$ such that it is possible that such a property value $p_{j}$ is\\
		  & observed at some time $n'\geq n_{1}$, so that $S_{OO}(n')=S_{Oj}(n')$, but\\
		  & it is not dictated by physical law that this will happen. We let \\
			& $\hat{n}$ be the smallest possible such time $n_{1}$.\\
\noalign{\smallskip}
		3 & \emph{It will become known which alternative is true}. There is also a\\
		  & time $n_{2}>n$ such that physical law dictates that one of the pro-\\
			& perty value $p_{j}$ will be observed at some time $n''\leq n_{2}$, so that\\
			& $S_{OO}(n'')=S_{Oj}(n'')$. We let $\check{n}$ be the smallest possible such\\
			& time $n_{2}$.\\
\noalign{\smallskip}\hline
\end{tabular}
\end{table}

From a strict epistemic perspective, it is only meaningful to speak about alternatives at some time $n$ if they are imagined at that time by some subject $k$, who also observes $O$, to which the alternatives refer. The alternatives can therefore be seen as objects $\{O_{j}\}$ which are present in the state of potential knowledge $PK(n)$, but which are separate from $O$. Rather, they belong to its complement $\Omega_{O}$, which includes the body of $k$, and they contribute to $PK_{\Omega_{O}}$ (Eq. [\ref{generalpk}]). If we imagine a situation in which subject $k$ continuously observes $O$ during a period of time to check whether an alternative comes true, we must assume that the imagined alternatives $\{O_{j}\}$ are identifiable objects during this period of time, just as the object $O$ must be an identifiable or quasi-identifiable object.

It is clear from these considerations that the fate of the alternatives depends crucially on other things than the state $S_{OO}(n)$ of the object itself. The alternatives are external to object $O$, but are related to it by means of reference. The object $O$ is, of course, also physically related to and affected by all other parts of its environment $\Omega_{O}$, just like any other object. As discussed above, these dependences are summarized in the expression $u_{O1}=u_{O1}[S(n)]$.

Depending on $S(n)$, all possible fates of a set of alternatives defined at time $n$ can be put into one of the three classes listed in Table \ref{levels}. They define the three levels of knowability of the alternatives. Comparing the definitions of knowability levels 2 and 3 in Table \ref{levels}, we conclude that $\check{n}\geq\hat{n}$.

Up until now we have only considered alternatives which in principle can be realized at once, meaning that $S_{OO}(n)\cap\mathcal{P}_{Oj}\neq\varnothing$ for each relevant $j$. This is sufficient when we consider examples such as the raptor in the sky. The bird can certainly be assumed to belong to a certain species at each time we observe it, even if we cannot decide which at the present moment $n$.

However, in other cases this is not true. Playing roulette at the casino, the alternatives $p_{1}$ and $p_{2}$ that the ball end up at a red or a black number can only be realized after a certain time, when it has lost enough momentum as it circulates in the wheel. Before that $S_{OO}(n)\cap\mathcal{P}_{Oj}\neq\varnothing$ even though the alternatives are defined in the sense that we know that one of them will eventually come true (knowability level 3). The same situation occurs in many scientific setups. In the double-slit experiment, alternatives $p_{1}$ and $p_{2}$ that the particle passes one slit or the other can only come true after the particle has had time to travel from the particle gun to the screen with the slits. We clearly have to broaden and sharpen the definition of alternatives to make the concept generally applicable.

\begin{figure}[tp]
\begin{center}
\includegraphics[width=80mm,clip=true]{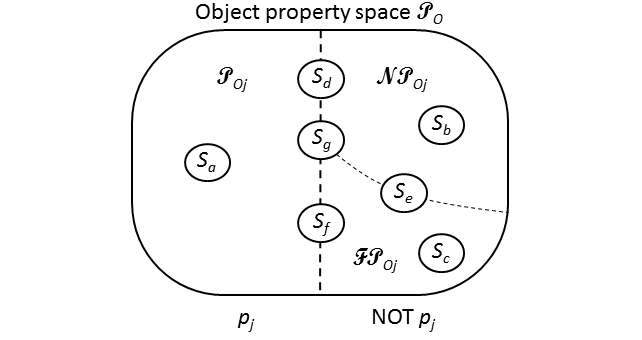}
\end{center}
\caption{Present and possible future properties. Each property value $p_{j}$ spans the object property space $\mathcal{P}_{O}$ in itself if we replace all other values by the complement `not $p_{j}$'. The region $\mathcal{FP}_{Oj}$ is the union of those object states $S_{OO}$ which do not have property value $p_{j}$ at present, but will have it at some future time. The region $\mathcal{NP}_{Oj}$ is the union of those states for which the property value is not, and will never be $p_{j}$. The object state $S_{OO}$ may relate to the three regions in seven different ways.}
\label{Figure22}
\end{figure}

We may divide object property space $\mathcal{P}_{O}$ into two parts: $\mathcal{P}_{Oj}$ and $\mathcal{P}_{Oj}^{c}$, where $\mathcal{P}_{Oj}^{c}$ is the region in which the value of property $P$ is not $p_{j}$ (Fig. \ref{Figure22}). For object states $S_{OO}$ that are embedded in one of these regions, we know for sure whether the property value at present is $p_{j}$ or not. For states that overlap both regions, we do not know.

We may define the region $\mathcal{FP}_{Oj}$ as the union $\bigcup_{j}\Sigma_{j}$ of those sets $\Sigma_{j}\subset\mathcal{P}_{Oj}^{c}$ for which $\Sigma_{j}\in\mathcal{D}_{u}$, and for which there is a positive integer $m$ such that $u_{Om}\Sigma_{j}\subset\mathcal{P}_{Oj}$, where $u_{Om}\equiv u_{O1}$. That is, $\mathcal{FP}_{Oj}$ is the region of object property space consisting of states that we know will have property value $p_{j}$ at some future time, but which do not have it now. Further, we may define $\mathcal{NP}_{Oj}=\mathcal{P}_{Oj}^{c}/\mathcal{FP}_{Oj}$, meaning that it is the union of those object states for which we know that the property value is not $p_{1}$ now, and that it will never be.

An object state $S_{OO}$ may overlap these three regions in various ways, as expressed in Fig. \ref{Figure22}. The state $S_{b}$ will, by definition, be a subset of $\mathcal{NP}_{Oj}$ forever. The evolved states $u_{Om}S_{d}$, $u_{Om}S_{e}$ and $u_{Om}S_{g}$ will forever (for any $m\geq 1$) at least partially belong to $\mathcal{NP}_{Oj}$. However, physical law may or may not allow that there is a time $n+m$ such that $S_{OO}(n+m)\subset\mathcal{P}_{Oj}$ given that $S_{OO}(n)$ belongs to one of the classes $S_{d}$, $S_{e}$ or $S_{g}$. If it is allowed, then property $p_{j}$ may be called \emph{realizable}.

Let

\begin{equation}
\vec{\mathcal{P}}_{Oj}=\mathcal{P}_{Oj}\cup\mathcal{FP}_{Oj}
\label{pfp}
\end{equation}
be the region in which object states have property value $p_{j}$ now, or will have it at some future time. We may call these regions \emph{future property value spaces}. Even if two property value spaces $\mathcal{P}_{Oj}$ and $\mathcal{P}_{Oj'}$ never overlap, two future property value spaces $\vec{\mathcal{P}}_{Oj}$ and $\vec{\mathcal{P}}_{Oj'}$ may or may not overlap. If they do not overlap, meaning that $\vec{\mathcal{P}}_{Oj}\cap \vec{\mathcal{P}}_{Oj'}=\varnothing$, the property values may be called called \emph{mutually exclusive}. By this we mean not only that the two property values cannot occur at the same time - they cannot occur in succession either.

To exemplify, the property values $p_{1}$ and $p_{2}$ that a given particle in a given double slit experiment passes slit 1 and 2, respectively, are mutually exclusive. The setup is such that if the particle passes one of the slits, it cannot pass the other at a later time. In contrast, if $p_{1}$ corresponds to the fact that the distance between two objects is $x_{1}$ and $p_{2}$ corresponds to the fact the distance is $x_{2}$, the property values can occur one after the other if the objects are moving, even if they, of course, cannot occur simultaneously.

Making use of the future property value spaces $\vec{\mathcal{P}}_{Oj}$, we define the corresponding future alternatives $\vec{S}_{Oj}$ as follows.

\begin{defi}[\textbf{Future alternative}]
Let $\vec{S}_{Oj}=S_{OO}(n)\cap\vec{\mathcal{P}}_{Oj}$ and $u_{O1}\vec{S}_{Oj}\equiv [u_{O1}S_{OO}(n)]\cap\vec{\mathcal{P}}_{Oj}$. Then $\vec{S}_{Oj}$ is a future alternative if and only if $\vec{S}_{Oj}\neq\varnothing$ and $\vec{S}_{Oj}\subset S_{OO}(n)$. Further, physical law must allow an object state $S_{OO}$ such that $S_{OO}\subseteq \vec{S}_{Oj}$.
\label{futurealt}
\end{defi}

The last sentence in the definition means that a future alternative $\vec{S}_{Oj}$ is realizable in the sense that there is a state $S_{OO}$ of the relevant object $O$ in which the corresponding value $p_{j}$ of its property $P$ is actually observed. Note, however, that we do not require that $p_{j}$ can be observed at some time in the future given the \emph{particular} object state $S_{OO}(n)$. We allow $\vec{S}_{Oj}$ to have any of the three knowability levels 1, 2 and 3 listed in Table \ref{levels}.

We may want to consider a complete set of future alternatives such that it is guaranteed that one of the alternatives eventually come true, that one of the property values is eventually revealed. To do so, we require that the future alternatives $\vec{S}_{Oj}$ do not overlap. Otherwise the quality of completeness would be hard to use effectively in the experimental contexts that we are going to discuss below.

If the corresponding property values $p_{j}$ are mutually exclusive, so that $\vec{\mathcal{P}}_{Oj}\cap \vec{\mathcal{P}}_{Oj'}=\varnothing$ for all $j\neq j'$, the future alternatives $\vec{S}_{j}$ are automatically disjoint. We may, however, define disjoint future alternatives $\vec{S}_{Oj}$ even if this is not the case. Consider again the property \emph{distance}, and call it $P$. Two different distances may be found in succession, so that the property values are not mutually exclusive. But in a physical setup prepared to measure the distance between the objects at a given time, the possible outcomes nevertheless define disjoint future alternatives, defined as `the outcome of the measurement that we have prepared'. These alternatives correspond to a slightly different property $P'$, defined as `the first distance to be measured after the present time $n$ between a given pair of identifiable objects'.

\begin{defi}[\textbf{Complete set of future alternatives}]
A set of future alternatives $\{\vec{S}_{Oj}\}$ is complete if and only if $S_{OO}(n)=\bigcup_{j}\vec{S}_{Oj}$ and $\vec{S}_{Oj}\cap \vec{S}_{Oj'}=\varnothing$ for all $j\neq j'$.
\label{setfuturealt}
\end{defi}

From this definition, and from the condition $\vec{S}_{Oj}\subset S_{OO}(n)$ in Definition \ref{futurealt}, we see that a complete set of future alternatives always contains more than one element, meaning that the value of $P$ is unknown to begin with, at time $n$. To make the concept of alternatives meaningful, several such alternatives should have a chance to come true. On the other hand, a complete set of future alternatives cannot contain more than a finite number of elements. Even if the corresponding property values $p_{j}$ might be continuous, the resolution of any actual observation is finite.

This may seem self-evient, but let us discuss the statement in little bit more detail. Each outcome in a complete set of alterntives is subjectively distinguishable from all the others by definition. A distiction always have two ends; we distinguish this from that. Consequently, we can label each alternative with a unique symbol, for example an integer. We see therefore that the number of alternatives is always countable. If they are countably infinite, then it must be possible to make an infinite number of distinctions at the time $n$ at which the alternatives is defined, and to keep track of all these distinctions until one alternative is realized. To encode an infinite number of distinctions requires an infinite number of objects. These objects must all be part of the observed object $O$ to which all alternatives apply. However, the assumed existence of minimal objects makes it impossible to divide an object an infinite number of times (Section \ref{minimalobjects}). We conclude therefore that the number of alternatives has to be finite.

\begin{state}[\textbf{The number of alternatives in a complete set}]
A complete set of future alternatives $\{\vec{S}_{Oj}\}$ always contains more than one element, but it never contains more than a finite number of elements.
\label{finitenoalt}
\end{state}

\begin{figure}[tp]
\begin{center}
\includegraphics[width=80mm,clip=true]{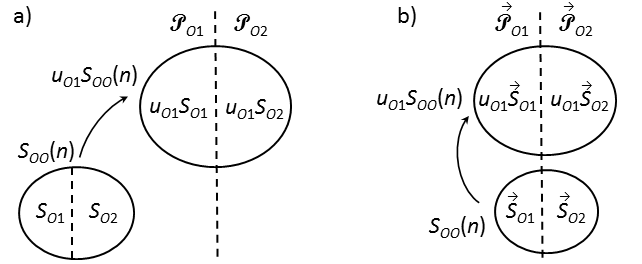}
\end{center}
\caption{The evolution of an object state $S_{OO}$ embedded in object state space $\mathcal{S}_{O}$ when it is divided into two parts by a) the object property values spaces $\mathcal{P}_{O1}$ and $\mathcal{P}_{O2}$, and b) the future property values spaces $\vec{\mathcal{P}}_{O1}$ and $\vec{\mathcal{P}}_{O2}$. In the first case a) the alternatives $S_{O1}$ and $S_{O2}$ cannot be realized at time $n$, but only at the subsequent time $n+1$ when the object state has `floated' across $\mathcal{S}_{O}$ so that it becomes divided by the boundary between $\mathcal{P}_{O1}$ and $\mathcal{P}_{O2}$. b) The corresponding future alternatives $\vec{S}_{O1}$ and $\vec{S}_{O2}$ cannot be realized at time $n$ either, but by the definition of $\vec{\mathcal{P}}_{O1}$ and $\vec{\mathcal{P}}_{O2}$, the object state is `locked onto' the boundary of these spaces. In this representation, $S_{OO}$ cannot `float' from value to value, like the position of a ball rolling on the ground.}
\label{Figure23}
\end{figure}

A future alternative may be said to be invariant in time. We have `integrated away' its time dependence, so to say. We have $\vec{S}_{Oj}\subseteq\vec{\mathcal{P}}_{Oj}$. By definition of the region $\vec{\mathcal{P}}_{Oj}$ we also have $u_{1}\vec{S}_{Oj}\subseteq\vec{\mathcal{P}}_{Oj}$. If no state reduction occurs at time $n+1$, meaning that $S_{OO}(n+1)=u_{1}S_{OO}(n)$, then the set $\{u_{1}\vec{S}_{Oj}\}$ is also a complete set of future alternatives. When we consider a complete set of future alternatives, we may therefore say that the object state is `locked onto' the boundaries between the disjoint regions $\vec{\mathcal{P}}_{Oj}$, as illustrated in Fig. \ref{Figure23}(b).

These boundaries represent a kind of coordinate system in $\mathcal{S}_{O}$. If we use the coordinate system defined by the set of ordinary property value spaces $\{\mathcal{P}_{Oj}\}$, then the object state $S_{OO}$ may instead `float around', as indicated in Fig. \ref{Figure23}(a). The object state may intersect a given region $\mathcal{P}_{Oj}$ at some times, but not at others. If it intersect $\mathcal{P}_{Oj}$ at time $n+1$ but not at the preceding time $n$, we may nevertheless define the corresponding alternative $S_{Oj}(n)$ at time $n$ as $S_{Oj}(n)\equiv (u_{O1})^{-1}\{[u_{1}S_{OO}(n)]\cap\mathcal{P}_{j}\}$. The inverse $(u_{O1})^{-1}$ of the evolution operator is assumed to exist according to the discussion in relation to Fig. \ref{Figure17}.

If we let $\{\vec{S}_{Oj}(n)\}$ be a complete set of future alternaties, we may show that

\begin{equation}
u_{O1}S_{OO}(n)=u_{O1}\bigcup_{j}\vec{S}_{Oj}(n)=\bigcup_{j}u_{O1}\vec{S}_{Oj}(n)
\label{linearfev}
\end{equation}
for all states $S(n)$ in the same way as we demonstrated Eq. [\ref{linearev}]. This means that the object evolution operator $u_{O1}[S(n)]$ is linear in a set-theoretical sense regardless whether we decompose the object state $S_{OO}$ according to the `coordinate system' defined by $\{\mathcal{P}_{Oj}\}$ or that defined by $\{\vec{\mathcal{P}}_{Oj}\}$.

A complete set of future alternative belongs to knowability level 2 or 3 according to Table \ref{levels}. Depending on the physical state $S(n)$ of the world at the time $n$ at which the alternatives are defined, one of them \emph{may} come true, or \emph{must} come true. Two future alternatives $\vec{S}_{Oj}$ and $\vec{S}_{Oj'}$ in such a set are subjectively distinguishable by definition. More than that, once one alternative is realized at some time $n+m$, all future states of the world $S(n')$ with $n'\geq n+m$ becomes subjectively distinguishable from those that would follow if another alternative was realized. We may say that the choice between alternatives is always definitive.

\begin{figure}[tp]
\begin{center}
\includegraphics[width=80mm,clip=true]{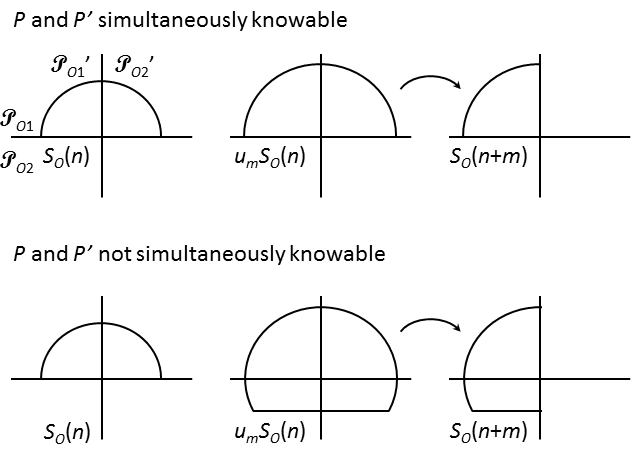}
\end{center}
\caption{The object state $S_{OO}\subset\mathcal{S}_{O}$ just after the value of property $P$ is decided to have value $p_{1}$. The object state space $\mathcal{S}_{O}$ is divided into four quadrants by the pairs of object property value spaces $(\mathcal{P}_{O1},\mathcal{P}_{O2})$ and $(\mathcal{P}_{O1}',\mathcal{P}_{O2}')$. a) If $P$ and $P'$ are simultaneously knowable, the state may be contained in a single quadrant just after the value of $P'$ has been subsequently decided. b) This may never happen if the properties are not simultaneously knowable.} 
\label{Figure24}
\end{figure}

This follows from the assumed invertibility of the evolution operator (Fig. \ref{Figure17}). The realization of one future alternative $\vec{S}_{Oj}$ defines the temporal update $n+m-1\rightarrow n+m$, since it corresponds to a subjective change of perception as the value $p_{j}$ of property $P$ is revealed, and any such change defines a temporal update. Thus, the revelation of value $p_{j}$ also defines the update $S(n+m-1)\rightarrow S(n+m)$ of the physical state, whereas the alternative revelation of value $p_{j}'$ would have defined the alternative update $S(n+m-1)\rightarrow S'(n+m)$. Since the values $p_{j}$ and $p_{j}'$ of the given identifibale object $O$ are subjectively distinguishable by definition, we have $S(n+m)\cap S'(n+m)=\varnothing$. The invertibility of the evolution operator then implies that $u_{M} S(n+m)\cap u_{M} S'(n+m)=\varnothing$ for all $M\geq 1$, where $u_{M}\equiv(u_{1})^{M}$. This in turn implies that $S(n+m+M)\cap S'(n+m+M)=\varnothing$ for all $M\geq 0$.

\subsection{Simultaneous knowability}
\label{simultaneous}

We argued in Section \ref{basic} that potential knowledge is fundamentally incomplete. This means that all properties cannot be known at the same time. This incompletness is guaranteed only if there are pairs of properties $P$ and $P'$ such that precise knowledge about the value of one of these properties makes it impossible to know the precise value of the other at the same time, as illustrated in Fig. \ref{Figure24}. Therefore we argue that this is indeed the case. This fact will be important to keep in mind when we try use our conceptual framework to describe experimental contexts in which several properties are observed in succession.

\subsection{The experimental context}
\label{expcontext}

We are most often interested in a particular aspect of a future alternative $\vec{S}_{Oj}$, an aspect that can be coded as a value $p_{j}$ of a property $P$. Other information that is part of the potential knowledge that corresponds to $\vec{S}_{Oj}$ is considered irrelevant.

For example, in a scientific experiment the object $O$ to which the alternatives $\vec{S}_{Oj}$ consists of the experimental apparatus together with the specimen $OS$ to be examined. If the specimen is a single electron, the property of interest may be its spin direction or its position when it hits a detector. As we perceive the outcome of the experiment, it is irrelevant if we, at the same moment, perceive a new scratch on some metal part of the detector.

\begin{figure}[tp]
\begin{center}
\includegraphics[width=80mm,clip=true]{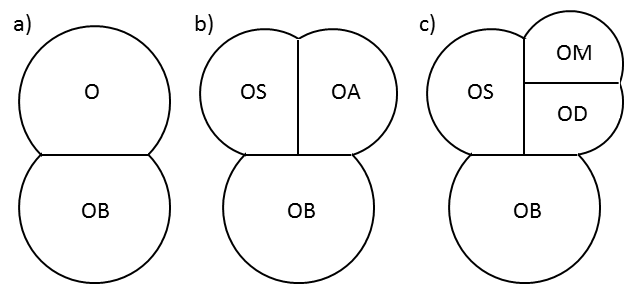}
\end{center}
\caption{Objects that have to be or may be parts of an observational setup. a) The observed object $O$ and the body of an observer $OB$ are necessary parts. b) The observed object may sometimes be divided into a specimen $OS$ and an apparatus $OA$ with which we decide a property of the specimen. We may then say that the observation is the outcome of an experiment. c) When the specimen is a deduced quasiobject, the apparatus can be divided into a machine $OM$ and a detector $OD$, where the subjective change in the state of the detector defines the outcome of the experiment.} 
\label{Figure25}
\end{figure}

Let us schematically discuss the role of the specimen in the observational setup (Fig. \ref{Figure25}). Such a setup necessarily consists of at least two objects: the object $O$ that we observe, and the body $OB$ of an observer. In a controlled, scientific setting, $O$ is divided into at least two parts: the specimen $OS$ and the apparatus $OA$, with which we study the specimen, decide some of its properties. Our aim is to formulate quantum mechanics as a convenient language to analye the behavior of specimens.

The division of $O$ into a specimen $OS$ and an apparatus $OA$ opens up the possibility that $OS$ is a quasiobject like an electron (Section \ref{quasiobjects}). However, $OA$ cannot be a quasiobject, since something has to be actually observed in the experiment. If $OS$ is a quasiobject the setup is such that there is conditional knowledge (Section \ref{statespace}) that relates the states of $OS$ and $OA$, so that new knowledge about the state of $OS$ is gained by deduction (using physical law) when new knowledge of $OA$ is gained by observation.

This description applies not only when the specimen $OS$ is a microscopic quasiobject like an electron. Consider, for example, a paleontologist who finds a fossil of a dinousaur. The fossil can be seen as the apparatus $OA$, and the dinosaur is the specimen $OS$, which must be seen a quasiobject whose properties are deduced from $OA$ via conditional knowledge known \emph{a priori} by the expert.

The dinosaur belongs to the past, meaning that the knowledge gained in the observational setup refers to a past time. The same situation occurs when an astronomer observes a distant galaxy in a telescope. The perceived luminous blob is a directly perceived object that belong to the present part $PKN(n)$ of potential knowledge according to the discussion in Section \ref{consistency}, but the galaxy iself is interpreted to be located in a distant part of space. Therefore the properties of the galaxy that we infer refer to a distant past, and the galaxy itself must be considered to be a quasiobject from the perspective of the astronomer. When genuinely new knowledge about the past is gained in this way, it must be assumed that this knowledge is epistemically consistent in the sense expressed in Fig. \ref{Figure16}. That the gained knowledge is genuinely new simply means that it corresponds to a state reduction $S_{OO}(n+m-1)\rightarrow S_{OO}(n+m)\subset u_{O1}S_{OO}(n+m-1)$, so that other alternatives could have been realized. 

\begin{defi}[\textbf{The specimen} $OS$ \textbf{and its state} $S_{OS}$]
Assume that $O$ is a composite object, and let $S_{OS}$ be the state of an object $OS$ that is part of $O$, and whose possible property values are used to define a complete set of future alternatives for $O$. Then $OS$ is a specimen with specimen state $S_{OS}\subset \mathcal{S}_{O}$.
\label{specstatedef}
\end{defi}

From this definition, we immediately conclude that

\begin{equation}
S_{OS}\supset S_{OO}.
\end{equation}

When the specimen $OS$ is a quasiobject, it is possible to divide the apparatus $OA$ into one detector $OD$ and one machine $OM$ (Fig. \ref{Figure25}). The term `machine' may not be very illuminating, and we define it negatively as those parts of the apparatus $OA$ that is not a detector.

Let $OD$ be the detector for property $P$, which is observed at time $n+m$. Suppose that an experiment starts at time $n$, and that $OS$ is a quasiobject in the time interval $[n,n+m]$. Then $OD$ are those objects that are part of the apparatus $OA$, and are such that $S_{OD}(n)\cap S_{OD}(n')\neq\varnothing$ whenever $n\leq n'<n+m$ and $S_{OD}(n+m-1)\cap S_{OD}(n+m)=\varnothing$, where $S_{OD}\subset\mathcal{S}_{O}$ is the object state of $OD$. The perceived change of the detector state $S_{OD}\subset\mathcal{S}_{O}$ thus defines the observation of $P$, and also defines the temporal update $n+m-1\rightarrow n+m$. The state of the machine $S_{OM}$, on the other hand, may undergo perceived changes during the course of the experiment, but may not change subjectively at time $n+m$.

\begin{defi}[\textbf{Property value state} $S_{Pj}$ \textbf{of a specimen}]
Consider a set of properties $\{P_{OS}\}$ that specify the nature of the specimen $OS$, with fixed, limited value ranges $\{\Upsilon_{POS}\}$ that are considered known \emph{a priori}. Consider also another property $P$ that can be defined for $OS$ so that $S_{OS}\subset\mathcal{S}_{O}$, but whose value may vary. Let $S_ {Pj}\subset\mathcal{S}_{O}$ be the state of $OS$ that corresponds to the knowledge that the value of $P$ is $p_{j}$, in addition to knowledge about $\{\Upsilon_{POS}\}$.
\label{propertyvaluestate}
\end{defi}

\begin{figure}[tp]
\begin{center}
\includegraphics[width=80mm,clip=true]{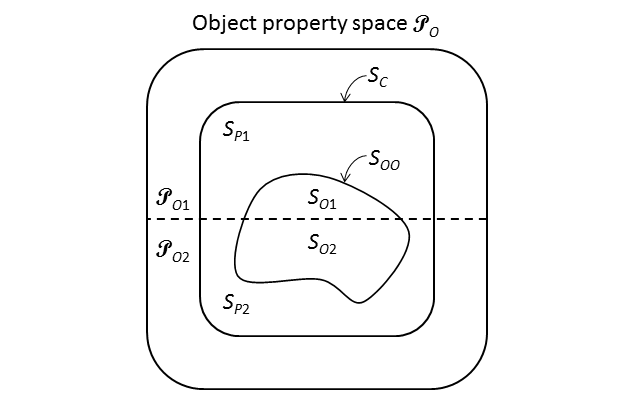}
\end{center}
\caption{The property value states $S_{Pj}$ and the contextual state $S_{C}(n)$ of the specimen $OS$ we investigate. If we forget all knowledge about the composite object $O$ except the nature of a specimen $OS$ that is part of it, and also know the present value $p_{j}$ of a property $P$ that applies to $OS$, we get the state $S_{Pj}$. If there is no knowledge of the value of $P$ (and no knowledge of any other property $P'$), then $S_{C}=\bigcup_{j}S_{Pj}$. If forget the very nature of $OS$ we get a state that equals the entire object property space $\mathcal{P}_{O}$. Assuming that $P$ has two possible values we have $V[S_{P1}]=V[S_{P2}]$, whereas we may have $V[S_{O1}]\neq V[S_{O2}]$ for the corresponding alternatives $S_{Oj}$ that apply to the entire object $O$ (Fig. \ref{Figure25}).} 
\label{Figure26}
\end{figure}

The fact that $S_{Pj}$ corresponds to knowledge about nothing more than that the value of $P$ is $p_{j}$ means that $S_{Oj}\subset S_{Pj}$
according to the definition of the object alternative given in Eq. [\ref{altdef}], as illustrated in Fig. \ref{Figure26}. The union $\bigcup_{j}S_{Pj}$ is a state that corresponds to knowledge about the fixed ranges $\{\Upsilon_{POS}\}$ of the values of $\{P_{OS}\}$, that is, to knowledge of the nature of the specimen. It consists of all exact object states $Z_{O}$ that are not excluded by the existence of a specimen of the given nature.

\begin{defi}[\textbf{An experimental context} $C$]
The context $C$ is the potential knowledge contained at time $n$ in the state $S_{OO}(n)$ of the observed object $O$ (Fig. \ref{Figure25}), together with potential knowledge at the same time $n$ about a sequence of complete sets of future alternatives $\{\vec{S}_{Oj}\}, \{\vec{S}_{Oj}'\},\ldots,\{\vec{S}_{Oj}^{(F)}\}$ that correspond to values of properties $P, P',\ldots, P^{(F)}$, that are attained in sequence. These properties are defined for a specimen $OS$ that is a part of $O$, but they do not belong to $\{P_{OS}\}$. Further, at time $n$ the knowability level associated with the values of each property should be 1 or 3, and the knowability level associated with $P^{(F)}$ should be 3 (Table \ref{levels}).
\label{observationalcontext}
\end{defi}

We may say that the context $C$ is \emph{initiated} at time $n$. This is the time of no return; after that the property values will be attained in sequence, whether we like it or not. There may, however, be intermediate unobservable properties in the sequence, like $P$ in Fig. \ref{Figure28}(c). The important thing is that there are no propertes in the sequence that may or may not be observed; the observational context should correspond to a well-defined experiment. 

\begin{defi}[\textbf{Contextual state} $S_{C}$ \textbf{of a specimen}]
Consider a context $C$ in which $P, P',\ldots, P^{(F)}$ are observed in sequence at times $n+m,n+m',\ldots,n+m^{(F)}$. Then $S_{C}(n')\subset\mathcal{S}_{O}$ is defined for $n\leq n'\leq n+m^{(F)}$ and corresponds to the potential knowledge of these properties at time $n'$, in addition to knowledge about the values of $\{P_{OS}\}$.
\label{contextualstate}
\end{defi}

\begin{figure}[tp]
\begin{center}
\includegraphics[width=80mm,clip=true]{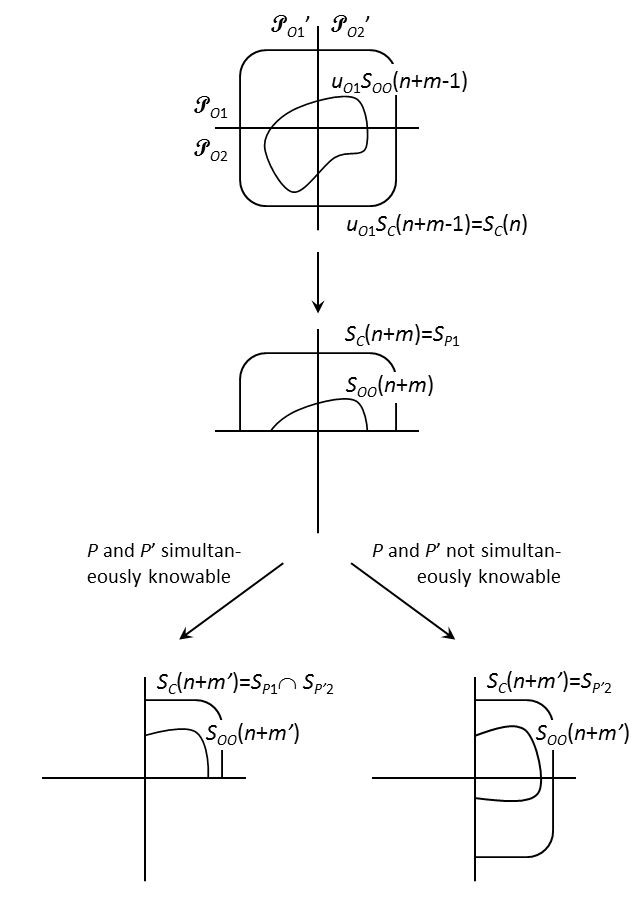}
\end{center}
\caption{Sequences of state reductions of the contextual state $S_{C}$ when properties $P$ and $P'$ with two possible values each are observed at times $n+m$ and $n+m'$, respectively. The final state $S_{C}(n+m')$ depends on whether $P$ and $P'$ are simultaneously knowable or not. Compare Figs. \ref{Figure24} and \ref{Figure26}.} 
\label{Figure27}
\end{figure}

At time $n$, before the first property $P$ is observed, we have

\begin{equation}
S_{C}(n)=\bigcup_{j}S_{Pj}=\bigcup_{j}S_{P'j}=\ldots=\bigcup_{j}S_{P^{(F)}j},
\end{equation}
as illustrated in Fig. \ref{Figure26}. When the value of $P$ is observed to be $p_{j}$ at time $n+m$, the contextual state reduces to

\begin{equation}
S_{C}(n+m)=S_{Pj}.
\end{equation}
If $P$ and $P'$ are simultaneously knowable (Fig. \ref{Figure24}), the contextual state reduces further to

\begin{equation}
S_{C}(n+m')=S_{Pj}\cap S_{P'j'}
\end{equation}
when the value of $P'$ is observed to be $p_{j'}'$ at time $n+m'>n+m$. In contrast, if $P$ and $P'$ are not simultaneously knowable, then we may loose all knowledge of the value of $P$ at time $n+m'$, so that

\begin{equation}
S_{C}(n+m')=S_{P'j'}.
\end{equation}
We may, for example, know and remember that the specimen is an electron, and let $P$ be the spin in the $z$-direction and $P'$ be the spin in another direction that will be measured subsequently. These state reductions are illustrated in Fig. \ref{Figure27}. In general, we see that

\begin{equation}
S_{C}\supseteq S_{OS}\supset S_{OO}.
\end{equation}

If $P$ and $P'$ indeed correspond to the spin of an electron along two different directions, there are only two possible values of each property that are allowed by physical law. These properties also correspond to an independent attribute of an individual object according to Definition \ref{indattributes}. Such a property may be called \emph{fundamental}. A complete set $\{\vec{S}_{Oj}\}$ of future alternatives will be called \emph{fundamental} if and only if it is defined by the values $p_{j}$ of some property $P$, such that there is one alternative $\vec{S}_{j}$ for each property value allowed by physical law. We will talk about a \emph{fundamental context} when the experimental context $C$ is such that all involved properties are fundamental, and all sets of future alternatives are also fundamental.

Depending on the observed properties, it may or may not be possible to create an experimental context $C$ such that the set of future alternatives becomes fundamental. It is impossible if the values of the corresponding property $P$ are continuous, since the set of possible outcomes $\{p_{j}\}$ of an actual observation of $P$ must always form a discrete set. This is so because each outcome must be subjectively distinguishable from all the alternatives that were not realized.

\section{Probability}
\label{probability}

Probability is considered to be a measure on a future alternative $\vec{S}_{Oj}$ (Definition \ref{futurealt}). To be able to assign a probability, the alternative should have knowability level 3 according to Table \ref{levels}. There has to be a predefined moment of decision $\check{n}$, as defined in that table, after which we know whether the corresponding event has occurred or not. Further, probabilities should only be assigned to a complete set $\{\vec{S}_{Oj}\}$ of such alternatives (Definition \ref{setfuturealt}). This requirement corresponds to the condition that assigned probabilities should always add up to one.

These conditions are quite conventional, even though they are expressed in the somewhat unconventional vocabulary used in this paper. However, the strict epistemic approach employed here forces us to add the condition that there has to be actual potential knowledge about the probilities at the time $n$ at which the complete set $\{\vec{S}_{Oj}(n)\}$ of future alternatives is defined, before any of the corresponding events occur. The probability $q_{j}$ that the future alternative $\vec{S}_{Oj}$ occurs may be seen as a property of the corresponding `alternative-object' in the mind of the observer $OB$ that points to the observed object $O$ (Fig. \ref{Figure25}). The probability $q_{j}$ may be said to exist whenever the potential knowledge at time $n$ makes it possible to exclude some of its possible values in the range $[0,1]$.

This necessary \emph{a priori} knowledge about the probabilities may be obtained in different ways. It may arise via repeated experiments before time $n$ in which the observed object $O$ is prepared in nearly identical initial states a large numer of times, and is isolated sufficiently from its environment $\Omega_{O}$ so that the probabilities can be estimated in the frequentist sense. It may also arise via symmetry considerations, like for a dice.

The only entities there are, in the present approach to physics, on which the probability $q_{j}$ to see the propery value $p_{j}$ may depend are the future alternative $\vec{S}_{Oj}$ that corresponds to $p_{j}$, the other future alternatives $\{\vec{S}_{Ok}\}$ in the complete set, the state $S_{OO}$ of the observed object to which $q_{j}$ apply, and the entire physical state $S$. We may write

\begin{equation}
q_{j}=f[\vec{S}_{Oj},\{\vec{S}_{Ok}\}, S_{OO}, S],
\end{equation}
where the range of $f$ is the real interval $[0,1]$.

The physical state $S$ enters the picture if the observed object $O$ is not isolated from the environment. As noted above, such isolation is necessary if $q_{j}$ cannot be deduced from the symmetries of $O$. If $q_{j}$ can indeed be deduced beforehand from such symmetries, we should include in the definition of $O$ all aspects of the world that define these symmetries. This means that we can drop the dependence of $q_{j}$ on $S$.

In this way we also exclude hypothetical 'mental influences' on the outcome. Such influences have to be attributed to states of the bodies $OB$ of observers that does not belong to the object $O$ under study and cannot affect it by means of ordinary physical law. The exclusion of mental influences also makes it possible to drop the dependence of $q_{j}$ on the \emph{other} future alternatives $\{\vec{S}_{Ok}\}$ in the complete set. You cannot influence the probability that something will happen by imagining other alternatives.

Complete sets of future alternatives $\{\vec{S}_{Oj}\}$ (Definition \ref{setfuturealt}) correspond to mutually exclusive events. The axioms of probability can therefore be expressed as  

\begin{equation}\begin{array}{rcl}
f[\vec{S}_{Oj},S_{OO}] & \geq & 0\\
f[S_{OO},S_{OO}] & = & 1\\
f[\vec{S}_{Oj}\cup \vec{S}_{Ok},S_{OO}] & = & f[\vec{S}_{Oj},S_{OO}]+f[\vec{S}_{Ok},S_{OO}],
\end{array}
\label{probcond}
\end{equation}
whenever $j\neq k$.

We have to relate the function $f$ to the only measure that we have defined on sets $\Sigma$ in object state space $\mathcal{S}_{O}$, namely $V[\Sigma]$ (Section \ref{volmeasure}). The relations [\ref{probcond}] are fulfilled for all complete sets of future alternatives (Definition \ref{setfuturealt}) for which probabilities can be assigned if and only if we let

\begin{equation}
q_{j}=f[\vec{S}_{Oj},S_{OO}]=V[\vec{S}_{Oj}]/V[S_{OO}]\equiv v[\vec{S}_{Oj},S_{OO}]
\label{probdef}
\end{equation}
whenever $q_{j}$ exists, according to the conditions given at the beginning of this section. The last equality defines the \emph{relative volume} 

\begin{equation}
v[\Sigma_{1},\Sigma_{2}]\equiv V[\Sigma_{1}]/V[\Sigma_{2}],
\label{relvol}
\end{equation}
assuming that the two sets in the argument are measurable, so that $\Sigma_{1}\in\Sigma_{VO}$ and $\Sigma_{2}\in\Sigma_{VO}$.

We have not included any time dependence in Eq. [\ref{probdef}] since it follows from the definition of a complete set of future alternatives that all relative volumes $\{v[\vec{S}_{Oj},S_{OO}]\}$ stay the same until one alternative is realized. Formally,

\begin{equation}
v[u_{O1}\vec{S}_{Oj}(n'),u_{O1}S_{OO}(n')]=v[\vec{S}_{Oj}(n'),S_{OO}(n')]
\label{invariantv}
\end{equation}
whenever $n\leq n'<n+m$, assuming that one alternative comes true at time $n+m$, revealing the value $p_{j}$ of property $P$.

Clearly, $q_{j}$ does not depend on the exact shapes of the boundaries $\partial\vec{S}_{Oj}$ and $\partial {S}_{OO}$, but only on the measure $V$ on these sets. This is necessary, since the exact shape of the boundaries cannot be exactly known, according to the discussion in relation to Fig. \ref{Figure8}. Nevertheless, these boundaries are conceptually well-defined. In a similar way, the relative volume $v[\vec{S}_{Oj}(n'),S_{OO}(n')]$ is always well-defined, even if it happens to be unknown, even if it is not part of $PK(n')$ in the form of a probability.

In the present approach, probability is always associated with a macroscopic physical state, since it is a function of a subjectively perceived future alternative $\vec{S}_{Oj}$ by definition. This means that we cannot assign probabilities to the states of microscopic quasiobjects such as electrons. Consider the statement \emph{The spin in the $x$-direction of an electron is} $+1/2$. If this statement should have any chance to be verified in a knowable sense, the corresponding future alternative $\vec{S}_{Oj}$ must involve a macroscopic detector $OD$ mounted on a macroscopic machine $OM$, and also an observer $OB$, according to Fig. \ref{Figure25}. Therefore, from the present perspective, the probability we assign to the spin direction of an electron is a measure not on the state of the electron itself, but on the state of the electron together with the states of the detector, the machine and the observers.

The fact that it is meaningless to assign probabilites to the states of microscopic quasiobjects like electrons \emph{by themselves} becomes evident from the following consideration. Let $\tilde{O}$ be the quasiobject, and let $S_{\tilde{O}\tilde{O}}$ be its state embedded in object state space $\mathcal{S}_{O}$. Assume that the property $P$ of interest is an independent attribute of $\tilde{O}$ (Definition \ref{indattributes}), and that we are interested in the likelihood of each of the values $p_{j}$ of $P$ that are allowed by physical law. However, from the definition of object state space volume (Definition [\ref{ovoldef}]) we pick the relation $V[\mathcal{S}_{OO}(A,\upsilon)]=V[\mathcal{S}_{OO}(A,\upsilon')]$. It means that for each exact object state $Z_{O}$ with value $p_{j}$ of the fundamental property $P$ there is another exact object state with value $p_{k}$ of $P$. In this sense, each value $p_{j}$ of $P$ must be considered equally likely, and there is no other sense in which we can compare the likelihood of different values if we consider $\tilde{O}$ in isolation. This basic fact can be seen as an expression of the \emph{assumption of a priori equal probabilities} for microscipic states that is used in statistical mechanics. 

However, in the present approach to physics quasiobjects $\tilde{O}$ should never be considered in isolation; they are abstract entities secondary to the directly perceived macroscopic objects $O$ (Section \ref{quasiobjects}). For the sake of discussion, let us nevertheless treat a quasiobject like an electron \emph{as if} it were a specimen $OS$ that could be observed directly. Then we could throw away the apparatus $OA$ with the detector $OD$ and decide its state with our bare eyes. In that case we could write $OS=O$ in Fig. \ref{Figure25}, and we would get $S_{C}=S_{OO}$, as well as $V[\vec{S}_{Oj}]=V[S_{Pj}]$.

Referring to Fig. \ref{Figure26}, we may say that the state $S_{OO}$ would expand until it reached the boundary of $S_{C}$. If we were about to observe a fundamental property $P$ of the specimen with $m$ possible values, we would get $V[S_{Pj}]=V[S_{Pk}]$ for all pairs of indices $(j,k)$, since for each exact object state $Z_{O}$ for which the value of $P$ is $p_{j}$ there is another such state $Z_{O}'$ for which the value is $p_{k}$ and all the other properties of the specimen are kept fixed. This means that $V[\vec{S}_{Oj}]=V[\vec{S}_{Ok}]$ for all pairs $(j,k)$. In terms of the relative volume $v[\vec{S}_{Oj},S_{OO}]$ we would get $v[\vec{S}_{Oj},S_{OO}(n)]=v[S_{Pj},S_{C}(n)]=1/m$ for $1\leq j\leq m$. In other words, all probabilites $q_{j}$ would always be equal to $1/m$ according to Eq. \ref{probdef}.

This consideration again shows that the non-trivial content of the concept of probability cannot reside in a quasiobject like an electron itself, but must be attributed to the macroscopic apparatus $OA$ with which is probed. Put differently, knowledge about probabilities cannot be encoded in the state of the electron itself, but only in the state of the apparatus $OA$ with which we observe it. For example, if we have an experimental context $C$ in which we measure the electronic spin $P$ in a given direction, and after that the spin $P'$ in another direction at an angle $\phi$ relative to the first direction, then the probability to see a given pair of spin values $(p_{j},p_{j'})$ is a function of $\phi$, which in turn is a function of the spatial relation between the two detectors used to measure $P$ and $P'$, respectively. These detectors are parts of the apparatus $OA$, rather than parts of the specimen $OS$. 

To demonstrate that the present notion of probability gives rise to the correct probability in the frequentist sense, we may formulate the equivalent of Borel's law of large numbers. Consider a series of $N$ observations of the same property $P$ of object $O$, where $O$ is prepared in identical states $S_{OO}$ before each observation, and the same complete set of future alternatives $\{\vec{S}_{Oj}\}$ corresponding to the set of property values $\{p_{j}\}$ is defined for each observation of $P$. If the relative volume $v[\vec{S}_{Oj},S_{OO}]$ is known, then the probability to get $p_{j}$ is $q_{j}=v[\vec{S}_{Oj},S_{OO}]$ at each of the $N$ observations.

We can regard this setup as a single experimental context $C$ with an initial state $S_{OON}(n)\subset\mathcal{S}_{O}$ in which $N$ properties $P=P'=\ldots=P^{(N)}$ are observed in sequence, according to Definition \ref{observationalcontext}. For this context we can define the future alternative $\vec{S}_{OK}$ that corresponds to the circumstance that property value $p_{j}$ is observed at least $K-\epsilon N$ times, and no more than $K+\epsilon N$ times. Here we let $\epsilon>0$, and we require that $K-\epsilon N\geq 0$ and $K+\epsilon N\leq N$.

For any such $\epsilon$ we get $v[\vec{S}_{OK},S_{OON}]\rightarrow 1$ as $N\rightarrow\infty$ if and only if $Nq_{j}\in[K-\epsilon N,K+\epsilon N]$. In simple language we would say that if $K$ is the number of times the value $p_{j}$ is observed in $N$ trials, then the probability is one that $K/N\rightarrow q_{j}$ as $N\rightarrow\infty$. Since the relative state space volumes $v[\vec{S}_{Oj},S_{OO}(n)]$ fulfil Kolmogorov's axioms of probability whenever $\{\vec{S}_{Oj}\}$ is a complete set of future alternatives, this statement can be proved in the same way as usual \cite{borel}.

\section{Elements of quantum mechanics}
\label{qmelements}

In this section we will argue that quantum mechanics emerges as the only generally applicable algebraic representation of experimental contexts $C$ that respects the guiding principles for physical law discussed in Section \ref{guiding}.

\subsection{State representations}

A physical state $S$ is a set in state space $\mathcal{S}$ of unattainable states $Z$ of complete knowledge about the world. In order to do physics, we have to represent these states symbolically or algebraically. We let $\bar{S}$ denote such a representation of $S$, which encodes the knowledge contained in $S$.

The representation of a state $S$ may be complete, meaning that all potential knowledge contained in $S$ is represented or encoded in $\bar{S}$. In that case we write

\begin{equation}
\bar{S}\hookrightarrow S.
\label{completerep}
\end{equation}
The representation may also be partial, meaning that only some knowledge of interest is represented. In this case we write

\begin{equation}
\bar{S}\rightharpoonup S.
\label{partialrep}
\end{equation}

Of course, the same state $S$ can be represented in different ways. We may have $\bar{S}\hookrightarrow S$ as well as $\bar{S}'\hookrightarrow S$ even if $\bar{S}\neq \bar{S}'$. A simple example is a state in which there are two objects $O_{1}$ and $O_{2}$ with a known spatial distance $d_{12}$ between them. This state can be partially represented by assigning two positions $r_{1}$ and $r_{2}$ to the two objects. But the choice of positions is arbitrary as long as $|r_{1}-r_{2}|=d_{12}$. A change of the coordinate system does not change the state, just the representation of it. Since physical law by definition acts upon and changes the physical state, it must be insensitive to changes of representation that do not affect the state itself. These matters are further discussed in Ref. \cite{epistemic}.

Analogous considerations apply to object states $S_{O}\subset\mathcal{S}$ and $S_{OO}\subset\mathcal{S}_{O}$. Their representations are denoted $\bar{S}_{O}$ and $\bar{S}_{OO}$, respectively.

Consider a complete set of future alternatives $\{\vec{S}_{Oj}\}$ (Definition \ref{setfuturealt}) with relative volumes $\{v_{j}\}\equiv \{v[\vec{S}_{Oj},S_{OO}]\}$, where these alternatives correspond to the observations of values $\{p_{j}'\}$ of property $P$. For any such state and any such set of alternatives we may write

\begin{equation}
\bar{S}_{OO}\equiv\left[\begin{array}{cccc}
\vec{S}_{O1} & \vec{S}_{O2} & \ldots & \vec{S}_{OM}\\
v_{1} & v_{2} & \ldots & v_{M}
\end{array}\right]\hookrightarrow S_{OO}.
\label{rep}
\end{equation}
This representation is complete since $S_{OO}=\bigcup_{j}\vec{S}_{Oj}$. We may even say that it is over-determined, since the relative volumes $\{v_{j}\}$ are functions of $\{\vec{S}_{Oj}\}$. The alternatives are objects that are external to the observed object to which they refer, as discussed in Section \ref{alternatives}. We may therefore say that $\bar{S}_{OO}$, as defined in Eq. [\ref{rep}] is also a partial representation of the environment $\Omega_{O}$ to $O$, so that we may write $\bar{S}_{OO}\rightharpoonup S$.

If another complete set of future alternatives $\{\vec{S}_{Oj}'\}$ that corresponds to the values $\{p_{j}'\}$ of another property $P'$ of $O$ is defined by some subject, we may, of course, express the alternative representation

\begin{equation}
\bar{S}_{OO}\equiv\left[\begin{array}{cccc}
\vec{S}_{O1}' & \vec{S}_{O2}' & \ldots & \vec{S}_{OM}'\\
v_{1}' & v_{2}' & \ldots & v_{M}'
\end{array}\right]\hookrightarrow S_{OO}.
\label{rep2}
\end{equation}

The set-theoretical linearity of the evolution operator (Eq. [\ref{linearfev}]) and the invariance of the relative volumes under evolution (Eq. [\ref{invariantv}]) allow us to write

\begin{equation}
\bar{u}_{O1}\bar{S}_{OO}\equiv\left[\begin{array}{cccc}
\bar{u}_{O1}\vec{S}_{O1} & u_{O1}\vec{S}_{O2} & \ldots & u_{O1}\vec{S}_{OM}\\
v_{1} & v_{2} & \ldots & v_{M}
\end{array}\right]\hookrightarrow u_{O1}S_{OO}.
\label{uprep}
\end{equation}

Formally, we may rearrange the schema [\ref{rep}]  as follows:

\begin{equation}
\bar{S}_{OO}=v_{1}\bar{S}_{O1}+v_{2}\bar{S}_{O2}+\ldots + v_{M}\bar{S}_{OM}.
\label{arep}
\end{equation}
However, this representation attains algebraic meaning only if we can show that the forms $AB$ and $A+B$ can be manipulated as if they represent multiplication and addition, respectively. Considering Eq. [\ref{uprep}], we note to begin with that this representation allows us to call the evolution operator $u_{O1}$ linear, at least in a formal sense.

\begin{equation}
\bar{u}_{O1}\bar{S}_{OO}=v_{1}\bar{u}_{O1}\bar{S}_{O1}+v_{2}\bar{u}_{O1}\bar{S}_{O2}+\ldots +v_{M}\bar{u}_{O1}\bar{S}_{OM}.
\label{auprep}
\end{equation}

We have put bars over $u_{O1}$ and $\vec{S}_{Oj}$ at the right hand side of Eqs. [\ref{arep}] and [\ref{auprep}], in contrast to the state representations shown in Eqs. [\ref{rep}] and [\ref{uprep}]. We have chosen to do so in to open up the possibility that the future alternatives $\vec{S}_{Oj}$ may themselves be represented in the same way as $S_{OO}$. In other words, we want the representation [\ref{arep}] to allow recursive representations of arbitrary depth. This is relevant in experimental contexts $C$ in which several properties $P, P',\ldots$ are observed in succession. To keep the notation fairly simple, we have removed the arrow on top of the future alternatives to make room for the bar.

Along this line of thought, suppose that there are two complete sets of future alternatives applying to the same initial state $S_{OO}$ of the same object $O$, as expressed in Eqs. [\ref{rep}] and [\ref{rep2}]. We may then write

\begin{equation}
\bar{S}_{Oj}\equiv\left[\begin{array}{cccc}
\vec{S}_{Oj1} & \vec{S}_{Oj2} & \ldots & \vec{S}_{OjM'}\\
v_{j1} & v_{j2} & \ldots & v_{jM'}
\end{array}\right]\hookrightarrow \vec{S}_{Oj},
\label{rep12}
\end{equation}
where

\begin{equation}
\vec{S}_{Ojj'}\equiv \vec{S}_{Oj}\cap \vec{S}_{Oj'}
\label{sojj}
\end{equation}
and $v_{jj'}\equiv v[\vec{S}_{Ojj'},\vec{S}_{Oj}]$. The set  $\{\vec{S}_{Ojj'}\}_{j}$ can be regarded as a complete set of future alternatives for the state $\vec{S}_{Oj}$ in which it has become known that the value of property $P$ is $p_{j}$, and we are about to observe property $P'$, provided $P$ and $P'$ are simultaneously knowable (Fig. \ref{simultaneous}).

In the case $M=M'=2$, the recursive version of representation [\ref{arep}] can be expressed as

\begin{equation}
\bar{S}_{OO}=v_{1}(v_{11}\vec{S}_{O11}+v_{12}\vec{S}_{O12})+v_{2}(v_{21}\vec{S}_{O11}+v_{22}\vec{S}_{O12}).
\label{recrep}
\end{equation}

Now, the two properties $P$ and $P'$ observed in succession may equally well be regarded as a single property with possible vectorial values $\{(p_{j},p_{j'})\}$. Then $\{\vec{S}_{Ojj'}\}_{jj'}$ becomes a complete set of future alternatives for the state $\vec{S}_{OO}$ before any property $P$ or $P'$ is observed. The corresponding relative volumes can be expressed as $v_{j}v_{jj'}=v[\vec{S}_{Ojj'},\vec{S}_{OO}]$. In this case $S_{OO}$ should be represented according to Eq. [\ref{arep}] as

\begin{equation}
\bar{S}_{OO}=v_{1}v_{11}\vec{S}_{O11}+v_{1}v_{12}\vec{S}_{O12}+v_{2}v_{21}\vec{S}_{O11}+v_{2}v_{22}\vec{S}_{O12}.
\label{recrep2}
\end{equation}
Comparing Eqs. [\ref{recrep}] and [\ref{recrep2}] we see that the distributive law holds in the representation [\ref{rep}].

\subsection{Representation of the contextual state}
\label{staterep}

The above discussion about algebraic respresentations uses two arbitrary complete sets of future alternatives as a starting point, as represented in Eqs. [\ref{rep}] and [\ref{rep2}], disregarding their knowability level and the relation between them in an actual experimental context $C$ (Definition \ref{observationalcontext}). We turn to such matters now.

In so doing, we try to represent the contextual state $S_{C}$ (Definition [\ref{contextualstate}]) of the specimen $OS$, rather than the state $S_{OO}$ of the entire experimental setup $O$ (Fig. \ref{Figure25}). This reason is that $S_{C}(n)$ encodes the relevant knowledge about the values of the properties observed within context, whereas $S_{OO}$ contains a lot of additional knowledge about the details about the apparatus $OA$ used to observe $OS$. This means that the symbols $\bar{S}_{Oj}$ in Eq. [\ref{arep}] themselves represent quite complex knowledge that is most often irrelevant to the outcome of the experiment in terms of bare property values $p_{j}$, $p_{j'}$ and so on.

Nevertheless, there is one quantity encoded in the apparatus $OA$ that we indeed want to represent, namely the probability for the different outcomes (Section \ref{probability}). This rules out the na\"{i}ve representation

\begin{equation}
\bar{S}_{C}\equiv\left[\begin{array}{cccc}
S_{P1} & S_{P2} & \ldots & S_{PM}\\
v_{P1} & v_{P2} & \ldots & v_{PM}
\end{array}\right]\hookrightarrow S_{C},
\label{vpjrep}
\end{equation}
where $\{S_{Pj}\}$ are the property value states (Definition \ref{propertyvaluestate}), since we have $v_{P1}=v_{P2}=\ldots=v_{PM}$ for fundamental properties. These numbers tell us nothing about probabilities, as discussed in Section \ref{probability} in relation to Fig. \ref{Figure26}.

We therefore seek another representation of the same form as [\ref{rep}] that is meaningful in all kinds of contexts $C$, regardless whether the observed properties are fundamental or not (Section \ref{expcontext}). In general we may thus write

\begin{equation}
\bar{S}_{C}=
\left[\begin{array}{cccc}
S_{P1} & S_{P2} & \ldots & S_{PM}\\
a_{1} & a_{2} & \ldots & a_{M}
\end{array}\right],
\label{mrep}
\end{equation}
where the numbers $a_{j}$ are related in some as yet undetermined way to the relative volumes $v_{Oj}$ of the future alternatives $\vec{S}_{Oj}$, and thus to the probability to find property value $p_{j}$ in an actual observation within the context $C$.

In other words, the numbers $\{a_{j}\}$ are contextual, referring not primarily to the contextual `naked' state $S_{C}$ of the specimen $OS$ that we investigate, but to the state of the entire experimental setup $O$, of which the specimen is just a small part. That $\{a_{j}\}$ points outwards from $OS$ to the means $OA$ by which we observe it means that the representation $\bar{S}_{C}$ of $S_{C}$ is even more contextual than $S_{C}$ itself.

This means that $\bar{S}_{C}$ is not only a complete representation of $S_{C}$, but an over-determined one, so that we may safely write $\bar{S}_{C}\hookrightarrow S_{C}$ according Eq. [\ref{completerep}]. On the other hand it is just a partial representation of the entire setup $O$ since we disregard every detail of $O$ but the possible property values and the probabilities to see them, which we hope to capture via the numbers $\{a_{j}\}$. Therefore $\bar{S}_{C}\rightharpoonup S_{OO}$ according Eq. [\ref{partialrep}]. This means that the initial states $S_{OO}(n)$ and $S_{O'O'}(n)$ of different contexts $C$ and $C'$ may have the same representation: $\bar{S}_{C}\rightharpoonup S_{OO}$ and $\bar{S}_{C}\rightharpoonup S_{O'O'}$.

We propose the following re-expression of the representation [\ref{mrep}]:

\begin{equation}
\bar{S}_{C}=a_{1}\bar{S}_{P1}+a_{2}\bar{S}_{P2}+\ldots +a_{M}\bar{S}_{PM}.
\label{carep}
\end{equation}
We do so since the analogous representation expressed in Eq. [\ref{arep}] shows some appealing characteristics. It keeps its form when it is used recursively to represent the observation of several properties in a sequence. In so doing, we saw that it is possible to interpret the expression algebraically in the sense that the distribute laws hold for the real numbers $v_{x}$ (Eqs. [\ref{recrep}] and [\ref{recrep2}]). Also, the evolution operator $u_{O1}$ can be interpreted to be linear (Eq. [\ref{auprep}]). We hope that these characteristics can be carried over to the representation [\ref{carep}], and that even more beneficial characteristics can be found.

More precisely, we formulate the four desiderata for the representation [\ref{carep}] listed in Table \ref{desiderata}.

\begin{table}
\caption{Desiderata for the algebraic representation [\ref{carep}] of the contextual state}
\label{desiderata}
\begin{tabular}{ll}
\hline\noalign{\smallskip}
		1 & The numbers $\{a_{j}\}$ can be used to calculate the probability $q_{j}$ to\\
		  & observe the value $p_{j}$ of any property $P$ observed within context\\
			& whenever this probability exists.\\
\noalign{\smallskip}
		2 & We can define a contextual evolution operator $u_{C}$ whose repre-\\
		  & sentation is formally linear, meaning that\\
			& $\bar{u}_{C}\bar{S}_{C}=a_{1}\bar{u}_{C}\bar{S}_{P1}+a_{2}\bar{u}_{C}\bar{S}_{P2}+\ldots +a_{M}\bar{u}_{C}\bar{S}_{PM}$.\\
\noalign{\smallskip}
		3 & The distributive law holds for the numbers $\{a_{j}\}$ and the repre-\\
		  & sentations of the property value states $\{S_{Pj}\}$, meaning that\\
			& $a_{1}(a_{2}+a_{3})\bar{S}_{Pj}=(a_{1}a_{2}+a_{1}a_{3})\bar{S}_{Pj}=a_{1}a_{2}\bar{S}_{Pj}+a_{1}a_{3}\bar{S}_{Pj}$\\
		  & for any triplet $(a_{1},a_{2},a_{3})$ of such numbers that may appear in\\
			& Eq. [\ref{carep}].\\
\noalign{\smallskip}
		4 & The form of the representation is generally valid: it applies to\\
		  & all kinds of experimental contexts $C$, regardless the number of\\
			& properties observed in succession and their knowability levels\\
		  & (Table \ref{levels}), always respecting the principle of epistemic closure\\
			& (Section \ref{closure}).\\
\noalign{\smallskip}\hline
\end{tabular}
\end{table}

The evolution of $S_{C}$ depends on the entire experimental setup $O$ and its evolution. We may define the contextual evolution operator $u_{C}$ such that $\bar{u}_{C}\bar{S}_{C}(n)$ represents the contextual state just before the observation of property $P$ at time $n+m$. That is, the temporal update $n+m-1\rightarrow n+m$ corresponds to a state reduction

\begin{equation}
u_{C}S_{C}(n)\rightarrow S_{C}(n+m)\subset u_{C}S_{C}(n),
\label{evred}
\end{equation}
where $u_{C}=u_{C}[S_{O}(n)]$. Similarly, $\bar{u}_{C}\bar{S}_{C}(n+m)$ represents the contextual state just before the observation of property $P'$ at time $n+m'$, so that the temporal update $n+m'-1\rightarrow n+m'$ corresponds to a state reduction

\begin{equation}
u_{C}S_{C}(n+m)\rightarrow S_{C}(n+m')\subset u_{C}S_{C}(n+m).
\label{evred2}
\end{equation}

\subsection{Born's rule}
\label{bornsrule}

To fulfil the desideratum in Table \ref{desiderata} that the representation [\ref{carep}] makes it possible to calculate the probability for a given alternative, a given set of numbers $\{a_{j}\}$ must correspond to a single set of relative volumes $\{v_{j}\}$, and thus to a given set of probabilities $\{q_{j}\}$. That is, we require

\begin{equation}
\{v_{j}\}=f(\{a_{j}\}).
\end{equation}
Of course, $S_{C}$ can still be seen as a legitimate state of an object (the specimen), and can therefore be represented as in Eq. [\ref{arep}] using relative volumes $v_{j}$ rather than as in [\ref{carep}] using numbers $a_{j}$. This situation occurs when all knowledge of the context dissipates, so that $S_{OO}$ grows to fill the entire state $S_{C}$ (Fig. \ref{Figure26}). To make these two representations consistent we require

\begin{equation}
v_{j}=f(a_{j}).
\label{vfa}
\end{equation}
There is no need to require \emph{a priori} that the function $f$ is invertible, given the purpose of the representation [\ref{carep}] that we seek, which is to keep track of property values and probabilities.

\begin{figure}[tp]
\begin{center}
\includegraphics[width=80mm,clip=true]{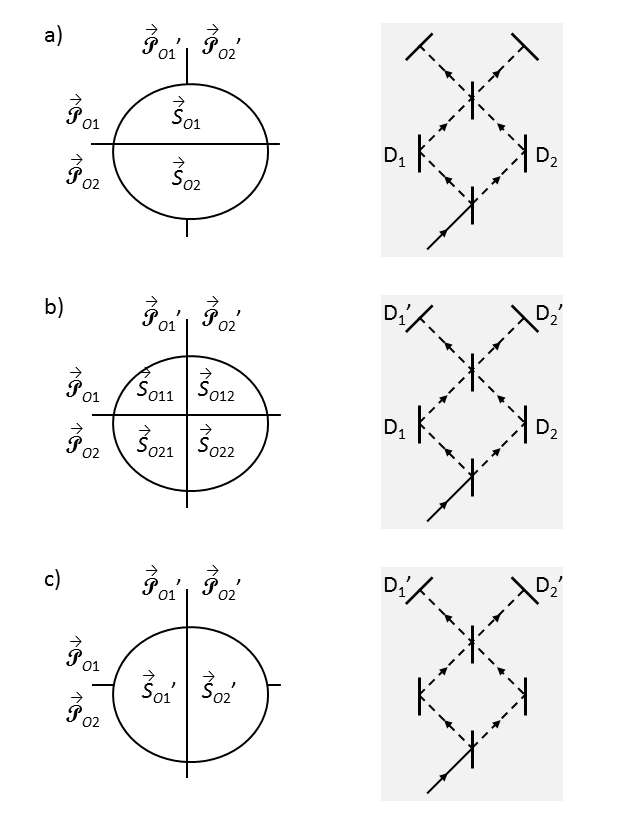}
\end{center}
\caption{Graphical representation of experimental contexts in which binary values of two properties $P$ and $P'$ are attained. Solid lines divide future alternatives corresponding to a property for which the value will be decided (knowability level 3). Interrupted lines divide alternatives corresponding to a property for which the value can never be decided (knowability level 1) a) A single property $P$ at knowability level 3 is defined. b) Two such properties $P$ and $P'$ are defined, where the value of $P$ is decided first, then the value of $P'$. c) A single property $P'$ at knowability level 3 is defined, but its value is revealed after the unknowable value of $P$ is attained. Each of the cases can be implemented by an adjustable Mach-Zehnder interferometer. Property $P$ corresponds to the passage of the left or right mirror, and $P'$ corresponds to the final absorption to the left or right. The prescence of a detector to decide the value of the property is marked by the letter $\mathrm{D}$.}
\label{Figure28}
\end{figure}

It turns out that there is only one function $f$ that makes it possible to fulfil the four desiderata listed in Table \ref{desiderata}, as well as a couple of other conditions to be discussed below, which concern the freedom to choose the details of the experimental setup. To demonstrate this fact it is sufficient to consider a simple family of experimental contexts $C$ in which only two properties $P$ and $P'$ are observed, having two observable values each. After having done that we discuss why this unique choice of $f$ is acceptable also in more general families of contexts (Definition \ref{observationalcontext}).

A concrete example of such a family of contexts is given by an adjustable Mach-Zehnder interferometer, as illustrated in Fig. \ref{Figure28}.

\begin{figure}[tp]
\begin{center}
\includegraphics[width=80mm,clip=true]{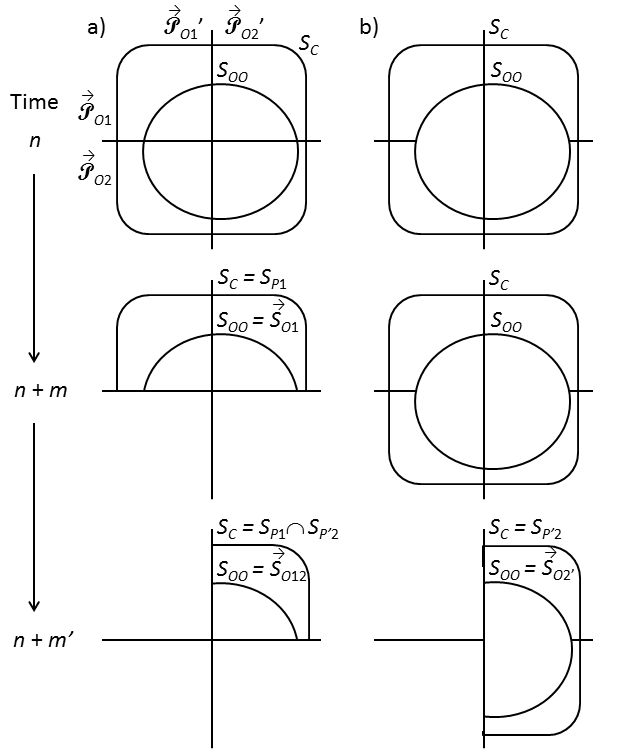}
\end{center}
\caption{The evolution in the setup depicted in Fig. \ref{Figure28} in terms of the contextual state $S_{C}$ and the property value states $S_{Pj}$. Panel a) shows the case in which properties $P$ and $P'$ both have knowability level 3 according to Table \ref{levels}, and the values $p_{1}$ and $p_{1}'$ are realized in succession. Panel b) shows the case in which property $P$ has knowability level 1. Again, property $P'$ has knowability level 3, and the value $p_{1}'$ is realized.} 
\label{Figure29}
\end{figure}

Consider Fig. \ref{Figure29}. As usual, the observational context is assumed to be initiated at time $n$, and properties $P$ and $P'$ attain their values at times $n+m$ and $n+m'$, respectively (Definitions \ref{observationalcontext} and \ref{contextualstate}). We use the vague phrase `attain their values' rather than `are observed', since we allow for the case where $P$ has knowability level 1 (Table \ref{levels}).

Consider first the case where property $P$ has knowability level 3 [Fig. \ref{Figure29}(a)]. At initial time $n$ we have

\begin{equation}
\bar{S}_{C}(n)=a_{1}\bar{S}_{P1}+a_{2}\bar{S}_{P2}.
\label{initialsc}
\end{equation}
Evolving this contextual state we get $\bar{u}_{C}\bar{S}_{C}(n)=a_{1}\bar{u}_{C}\bar{S}_{P1}+a_{2}\bar{u}_{C}\bar{S}_{P2}=a_{1}\bar{S}_{P1}+a_{2}\bar{S}_{P2}$ since the property value states $S_{Pj}$ are not time-dependent, where we have used the linearity of $\bar{u}_{C}$ according to the list of desiderata in Table \ref{desiderata}. A state reduction then takes place according to Eq. [\ref{evred}], and value $p_{1}$ or $p_{2}$ is revealed:

\begin{equation}
\bar{S}_{C}(n+m)=\bar{S}_{Pj}
\label{reducedsc}
\end{equation} 
for $j=1$ or $j=2$. If the probabilities $q_{1}$ and $q_{2}$ of these alternatives are defined, we have $q_{j}=v_{j}=f(a_{j})$. At time $n+m'$ a second state reduction takes place, at which the value of $P'$ is revealed. The state that reduces is 

\begin{equation}
\bar{u}_{C}\bar{S}_{C}(n+m)=a_{j1}\bar{S}_{PP'j1}+a_{j2}\bar{S}_{PP'j2},
\label{finalreduction}
\end{equation}
where $j'=1$ or $j'=2$, and

\begin{equation}
S_{PP'jj'} \equiv S_{Pj}\cap S_{P'j'}.
\label{sppjj}
\end{equation}
We get

\begin{equation}
\bar{S}_{C}(n+m')=\bar{S}_{PP'jj'},
\end{equation}
according to Eq. [\ref{evred2}]. If the probabilities $q_{1}'$ and $q_{2}'$ associated with the alternatives for $P'$ are also defined, we get

\begin{equation}
q_{j'}' = f(a_{1})f(a_{1j'})+f(a_{2})f(a_{2j'})
\label{classicprob}
\end{equation}
according to the classical axioms of probability, where $f(a_{j})=v[\vec{S}_{Oj},S_{OO}(n)]$ and $f(a_{jj'})=v[\vec{S}_{Ojj'},S_{OO}(n+m)]=v[\vec{S}_{Ojj'},\vec{S}_{Oj}]$ [Figs. \ref{Figure28}(b) and \ref{Figure29}(a)].

Consider now the case where property $P$ has knowability level 1 [Fig. \ref{Figure29}(b)]. As before, we may write $\bar{S}_{C}(n)=a_{1}\bar{S}_{P1}+a_{2}\bar{S}_{P2}$ and $\bar{u}_{C}\bar{S}_{C}(n)=a_{1}\bar{S}_{P1}+a_{2}\bar{S}_{P2}$. The question is how to handle what happens at time $n+m$. We know (contextually) that $P$ has attained value $p_{1}$ or $p_{2}$ but it is forever outside potential knowledge which of these values apply. No state reduction occurs, but epistemic completeness (Section \ref{epcomplete}) requires that we somehow account for the knowledge that `one of the two alternatives has occurred' in the representation of the contextual state. Most natural is to do it by writing

\begin{equation}
\bar{S}_{C}(n+m)=a_{1}\bar{S}_{C1}(n+m)+a_{2}\bar{S}_{C2}(n+m)
\label{hypostates}
\end{equation}
where $S_{C1}$ and $S_{C2}$ are two hypothetical contextual states that would have applied if the corresponding alternative had occurred. This means that

\begin{equation}
\bar{S}_{Cj}(n+m)=\bar{S}_{Pj}
\label{scjpj}
\end{equation}
for $j=1$ or $j=2$.

At time $n+m'$ a state reduction does finally take place. The state that reduces is

\begin{equation}
\bar{u}_{C}\bar{S}_{C}(n+m)=a_{1}\bar{u}_{C}\bar{S}_{C1}(n+m) + a_{2}\bar{u}_{C}\bar{S}_{C2}(n+m)
\end{equation}
according to the linearity of $u_{C}$ in the formal algebraic representation. Comparing with Eq.[\ref{scjpj}] we still have

\begin{equation}
\bar{u}_{C}\bar{S}_{Cj}(n+m)=\bar{S}_{Pj}
\label{uscjpj}
\end{equation}
for $j=1$ or $j=2$. We should also write

\begin{equation}
\bar{u}_{C}\bar{S}_{Cj}(n+m)=a_{j1}\bar{S}_{P'1} + a_{j2}\bar{S}_{P'2}.
\label{umrep2}
\end{equation}
This expression is different from Eq. [\ref{finalreduction}], since in the present case we have no knowledge about the value of $P$ after the final state reduction within context. We get

\begin{equation}\begin{array}{rcl}
\bar{u}_{C}\bar{S}_{C}(n+m) & = & a_{1}\bar{u}_{C}\bar{S}_{C1}(n+m)+a_{2}\bar{u}_{C}\bar{S}_{C2}(n+m)\\
& = & a_{1}(a_{11}\bar{S}_{P'1}+a_{12}\bar{S}_{P'2})+a_{2}(a_{21}\bar{S}_{P'1}+a_{22}\bar{S}_{P'2})\\
& = & (a_{1}a_{11}+a_{2}a_{21})\bar{S}_{P'1}+(a_{1}a_{12}+a_{2}a_{22})\bar{S}_{P'2},
\end{array}
\label{algu}
\end{equation} 
where we have made use of the distributive laws expressed in the list of desiderata in Table \ref{desiderata}. These laws give the representation algebraic meaning. Finally,

\begin{equation}
\bar{S}_{C}(n+m')=\bar{S}_{P'j'}
\end{equation} 

for $j'=1$ or $j'=2$. Equation [\ref{algu}] means that the probability that the evolved contextual state $u_{C}S_{C}(n+m)$ will reduce to $S_{P'1}$ is $f(a_{1}a_{11}+a_{2}a_{21})$, and the probability that it will reduce to $S_{P'2}$ is $f(a_{1}a_{12}+a_{2}a_{22})$, provided that these probabilities exist. In short,

\begin{equation}
q_{j'}'=f(a_{1}a_{1j'}+a_{2}a_{2j'}).
\label{quantprob2}
\end{equation}
We argued in Section \ref{explicitminimalism} that we should not treat the case where the value of property $P$ is unknowable as if we can actually know it. According to the principle of explicit epistemic minimalism we get the wrong answers if we do. In the present sitution this means that the probability [\ref{quantprob2}] must be different than in the case [\ref{classicprob}] where the value of $P$ becomes known, and can sometimes be associated with a probability $q_{j}$. That is,

\begin{equation}
f(a_{1}a_{1j'}+a_{2}a_{2j'})\neq f(a_{1})f(a_{1j'})+f(a_{2})f(a_{2j'}). 
\end{equation}
We may put the reason why different equations must hold in the two cases another way, as a consequence of epistemic completeness (Section \ref{epcomplete}). The fundamental epistemic distinction between knowability levels 1 and 3 must correspond to a distinction in physical law. Such a distinction can be expressed only if different equations hold for the probabilities of the values of $P'$. To fulfil this condition we must require that

\begin{equation}
f(a)\neq a.
\label{fanota} 
\end{equation}

Given this fact, let us discuss which other functions $f$ are possible in Eq. [\ref{vfa}]. All relative volumes must add to one: $1=\sum_{j}v_{j}$ and $1=\sum_{j'}v_{jj'}$. We get the conditions

\begin{equation}\begin{array}{lll}
1 & = & f(a_{1})+f(a_{2})\\
1 & = & f(a_{11})+f(a_{12})\\
1 & = & f(a_{21})+f(a_{22})\\
1 & = & f(a_{1}a_{11}+a_{2}a_{21})+f(a_{1}a_{12}+a_{2}a_{22}).
\end{array}
\label{acond}
\end{equation}

In this connection, we may ask whether the expressions $f(a_{jj'})$ in the above system of equations are well-defined. In the case where both $P$ and $P'$ have knowability level 3 they are defined in terms of a realizable future alternative for of a given object state, as shown in relation to Eq. [\ref{classicprob}]. But here $P$ has knowability level 1, so that there are no future alternatives $\vec{S}_{jj'}$ (Fig. \ref{Figure28}). Instead we must define $f(a_{jj'})=v[\Sigma_{jj'},\vec{S}_{j'}]$, where $\Sigma_{jj'}$ is an abstract set which corresponds to a quadrant of the object state shown in Fig. \ref{Figure28}(c), but which does not correspond to a realizable alternative. We can nevertheless define its formally as $\Sigma_{jj'}=u_{1}S_{O}(n+m'-1)\cap\tilde{\mathcal{P}}_{j}\cap\tilde{\mathcal{P}}_{j'}'$ and write

\begin{equation}
v_{jj'}=f(a_{jj'})=\frac{V[u_{1}S_{OO}(n+m'-1)\cap\vec{\mathcal{P}}_{j}\cap\vec{\mathcal{P}}_{j'}']}{V[u_{1}S_{OO}(n+m'-1)\cap\vec{\mathcal{P}}_{j}]}.
\label{vjk}
\end{equation}

Since $f(a)$ corresponds to a relative volume, we must require

\begin{equation}
0\leq f(a)\leq 1
\label{positivef}
\end{equation}
for all $a$ in the domain of $f$, in the addition to the conditions expressed in Eq. [\ref{acond}]. To determine $f$ from these conditions we make use of the assumption that the choice of $f$ should reflect the fact that the parts of the observational context $C$ that correspond to properties $P$ and $P'$ can be arranged independently from each other.

One way to express this fact is to say that the parameters describing the experimental setup that determine the mode of observation of property $P'$ are independent from the corresponding parameters that determine the mode of observation of property $P$. If both $P$ and $P'$ have knowability level 3, then this vague statement can be translated to a statement about relative volumes.

\begin{state}[\textbf{Relative volume independence}]Consider the set $\{C\}$ of all observational contexts $C$ with a given sequence $\ldots, P, P', \ldots $ of observed properties where $P$ and $P'$ have knowability level 3, and with given sets of possible values $\ldots,\{p_{j}\}, \{p_{j'}'\}, \ldots $. There are enough elements $C$ in in $\{C\}$ so that the relative volumes $\{v_{jj'}\}$ that describe the measurement of property $P'$ can be chosen independently from the relative volumes $\{v_{j}\}$ that describe a preceding measurement of $P$.
\label{volind}
\end{state}
This statement is quite trivial and follows from the fact that the only condition that the relative volumes has to fulfil \emph{a priori} is that they add to one: $1=\sum_{j}v_{j}$ and $1=\sum_{j'}v_{jj'}$. These relations do not mix relative volumes belonging to $\{v_{j}\}$ with those belonging to $\{v_{jj'}\}$. Therefore relative volumes associated with a property $P$ are independent from those associated with another property $P'$. 

If the values of property $P$ are unknowable, then we must generalize the above statement to account for the fact that the relative volumes $\{v_{jj'}\}$ may not be knowable either, since they do not correspond to a probability $q(p_{j'}'|p_{j})$ that is possible to determine by repeating the experiment a large number of times. Nevertheless, $\{v_{jj'}\}$ can still be formally defined in those cases, as shown in Eq. [\ref{vjk}].

However, in a strict epistemic approach we should not refer to potentially unknowable quantities in a physical statement about the independence of the properties $P$ and $P'$, just as we do not refer directly to the exact states $Z$ when we make statements about the evolution $u_{1}$ of the physical state $S$. A more general version of Statement \ref{volind} is then the following, which must now be seen as an assumption, since it cannot be motivated in the same straightforward way.

\begin{assu}[\textbf{Property independence}]Consider the set $\{C\}$ of all possible observational contexts $C$ with a given sequence $\ldots, P, P', \ldots$ of observed properties where $P'$ has knowability level 3, and with given sets of possible values $\ldots,\{p_{j}\}, \{p_{j'}'\}, \ldots $. Let $\{\alpha_{j}\}$ be a set of knowable property values that describe the part of the observational setup that is related to the observation of $P$, and let $\{\alpha_{j'}'\}$ be a corresponding set relating to $P'$. Suppose that these parameter sets are minimal in the sense that they determine the probability $q_{j'}'$ to get the value $p_{j'}'$ for each $j'$, so that we may write $\{q_{j'}'\}=f(\{\alpha_{j}\},\{\alpha_{j'}'\})$, but if we take away one parameter $\alpha_{j}$ or $\alpha_{j'}'$ this is no longer true. Then there are enough elements $C$ in $\{C\}$ so that $\{\alpha_{j}\}$ can be chosen independently from $\{\alpha_{j'}'\}$.
\label{propind}
\end{assu}

If property $P$ has knowability level 3, we can choose $\{\alpha_{j}\}=\{v_{j}\}$ and $\{\alpha_{j'}\}=\{v_{jj'}\}$, and we regain Statement \ref{volind}. These sets are minimal since $q_{j'}'=\sum_{j}v_{j}v_{jj'}$ for each $j'$, but we cannot take away any element from these sets and still determine all probabilities $q_{j'}'$.

We may ask how many elements are contained in the two minimal sets of independent parameters $\{\alpha_{j}\}$ and $\{\alpha_{j'}'\}$ that pertain to properties $P$ and $P'$, respectively. Suppose that there are $M$ and $M'$ possible values of these properties in the context $C$. There is then $M-1$ independent values of $v_{j}$ and $M(M'-1)$ independent values of $v_{jj'}$, taking into account the relations $1=\sum_{j}v_{j}$ and $1=\sum_{j'}v_{jj'}$. These numbers give the requested number of elements in $\{\alpha_{j}\}$ and $\{\alpha_{j'}'\}$ if both $P$ and $P'$ have knowability level 3, since then we can choose $\{\alpha_{j}\}=\{v_{j}\}$ and $\{\alpha_{j'}\}=\{v_{jj'}\}$. We may argue that the amount of freedom to choose the experimental setup should never be less than in this case. 

\begin{assu}[\textbf{Experimental freedom}]Consider the set $\{C\}$ of all possible observational contexts $C$ with a given sequence $\ldots, P, P', \ldots$ of observed properties where $P'$ has knowability level 3, and with given sets of possible values $\ldots,\{p_{j}\}, \{p_{j'}'\}, \ldots $. Suppose that $\{p_{j}\}$ and $\{p_{j'}'\}$ contain $M$ and $M'$ values, respectively. Then the sets of independent parameters $\{\alpha_{j}\}$ and $\{\alpha_{j'}'\}$, as defined in Assumption \ref{propind}, contain at least $M-1$ and $M(M'-1)$ values, respectively.
\label{expfree}
\end{assu}
If we add the numbers $M-1$ and $M(M'-1)$ we conclude that there is always at least $MM'-1$ free parameters to describe the observations of $P$ and $P'$.

We look for a function $f(a)$ such that the numbers $a_{j}$ and $a_{jj'}$ can always be used to parametrize the necessary level of experimental freedom, just as $v_{j}$ and $v_{jj'}$ can in the cases where both $P$ and $P'$ have knowability level 3. Otherwise the algebraic representation $\bar{S}_{C}$ of the contextual state becomes useless, since it cannot be used to calculate probabilities for the possible outcomes in all kinds of experiments. The whole point of the search for $f(a)$ is that we should find a function that makes the form of the representation [\ref{carep}] generally applicable, regardless the number of observed properties, their knowability level, and their sets of possible values. This means that we should be able to write

\begin{equation}\begin{array}{lll}
\{\alpha_{j}\} & = & F(\{a_{j}\})\\
\{\alpha_{j'}'\} & = & F'(\{a_{jj'}\}).
\end{array}
\label{bigf}
\end{equation}
It then follows from property independence that $f(a)$ is also such that the elements in $\{a_{j}\}$ and $\{a_{jj'}\}$ can be chosen independently. 

If both $P$ and $P'$ have knowability level 3 [Fig. \ref{Figure29}(a)] we could identify $\{\alpha_{j}\}=\{v_{j}\}$ and $\{\alpha_{j'}'\}=\{v_{jj'}\}$ with the independent parameter sets introduced in Assumption \ref{propind}. We cannot in general do the corresponding identifications $\{\alpha_{j}\}=\{a_{j}\}$ and $\{\alpha_{j'}'\}=\{a_{jj'}\}$. The parameters $\alpha_{j}$ and $\alpha_{j'}$ are defined in Assumption \ref{propind} as values of properties that define the design of the observational context $C$, which must be assumed to be known \emph{a priori}. This means that they should not only be knowable, but already known at the start of the experiment at time $n$. In contrast, the relative volumes $v_{j}$ and $v_{jj'}$, as well as the numbers $a_{j}$ and $a_{jj'}$ may be unknowable in principle when it is unknowable which value of $P$ is attained, when this property has knowability level 1. In that case the functions $F$ and $F'$ in Eq. [\ref{bigf}] are unknowable, even though it is known that they exist.

The assumptions \ref{propind} and \ref{expfree} make it possible to pinpoint a single acceptable function $f(a)$, given the other desiderata listed in Table \ref{desiderata}. We note first, however, that if properties $P$ and $P'$ both have knowability 3 within context, then any function $f(a)$ will do. In that case property independence and experimental freedom is automatically fulfilled. The fourth condition in Eq. [\ref{acond}] is replaced by $1=q_{1}'+q_{2}'= f(a_{1})f(a_{11})+f(a_{2})f(a_{21})+f(a_{1})f(a_{12})+f(a_{2})f(a_{22})$ according to the classical axioms of probability (see Eq. [\ref{classicprob}]). This condition follows from the first three conditions in Eq. [\ref{acond}], and is therefore not independent. This circumstance accounts for the fact that no restriction on $f(a)$ can be derived. 

In what follows, we therefore focus on the case where $P$ has knowablity level 1 and $P'$ has knowability level 3. We will see that the existence of the fourth condition in Eq. [\ref{acond}] is crucial, together with the requirement that the representation allows experimental freedom.

Let us first make a general observation that we must allow complex $a$. No function $f:\mathbb{R}\rightarrow\mathbb{R}$ such that $f(a)\neq a$ fulfils conditions [\ref{acond}], and can also be used to parametrize the necessary level of experimental freedom (Assumption \ref{expfree}). If all $a_{x}$ are restricted to be real and $f(a)\neq a$, we have four conditions in Eq. [\ref{acond}] that relate six real numbers $a_{1},a_{2},a_{11},a_{12},a_{21}$ and $a_{22}$. In that case two independent real parameters are sufficient determine the probabilities $q_{k}'$, whereas experimental freedom (Assumption \ref{expfree}) requires that at least $MM'-1=3$ are necessary, since $M=M'=2$.

One could try to identify the quantities $a_{x}$ with members of another collection (ring) of mathematical objects than the complex numbers (requiring that addition and multiplication is defined and yield another member of the same collection). We might, for example, consider vectors of three real numbers, or matrices. The discussion below will show, however, that complex numbers will do the job as arguments $a$ in a unique function $f(a)$, so that there is point in trying out more complicated mathematical objects.

Thus, we assume that $f: \mathbb{C}\rightarrow\mathbb{R}$. We may then write $v=f(a)=g(x,y)$, where $a=x+iy$ and $g: \mathbb{R}^{2}\rightarrow\mathbb{R}$. We then ask which functions $g(x,y)$ fulfil the requirements expressed in Eqs. [\ref{fanota}], [\ref{acond}], and [\ref{positivef}], as well as property independence and experimental freedom (Assumptions \ref{propind} and \ref{expfree}). We argue without proof that the only function that does the job is $g(x,y)=x^{2}+y^{2}$, that is, $f(a)=|a|^{2}$. This can be seen by inserting the Taylor expansion $g(x,y)=\sum_{m,n=0}^{\infty}d_{mn}x^{m}y^{n}$ into Eq. [\ref{acond}] and checking in what cases property independence and experimental freedom can be upheld. The general expressions become messy, but the lesson is that more terms and higher exponents make it impossible to comply with property independence and experimental freedom if we insist on fulfilling the other requirements. We take a shortcut through this mess by arguing that $f(a)$ should fulfil one additional condition, which makes the argument why $f(a)=|a|^{2}$ is the only possible choice much easier.

If we write the evolution of the contextual state representation $\bar{S}_{C}$ sequentially in the case where both $P$ and $P'$ have knowability level 3 [Fig. \ref{Figure29}(a)] we get the temporal sequence of contextual state representations starting with Eq. [\ref{initialsc}]. The probability $q_{jj'}$ to see the property values $p_{j}$ and $p_{j'}'$ becomes $q_{jj'}=f(a_{j})f(a_{jj'})$. However, we may also regard $(P,P')$ as one single property with values $(p_{j},p_{j'}')$, just like we did in the motivation of Eq. [\ref{recrep2}]. This vectorial value is decided at time $n+m'$. We may therefore write

\begin{equation}
\bar{u}_{C}\bar{S}_{C}(n+m)=a_{1}a_{11}\bar{S}_{PP'11}+a_{1}a_{12}\bar{S}_{PP'12}+a_{2}a_{21}\bar{S}_{PP'21}+a_{2}a_{22}\bar{S}_{PP'22}.
\end{equation}
In this way we see that $q_{jj'}=v_{jj'}=f(a_{j}a_{jj'})$. Thus the function $f(a)$ must also fulfil

\begin{equation}
f(a_{x}a_{y})=f(a_{x})f(a_{y}).
\end{equation}
There are only two operations on a pair of complex numbers that have this property, namely complex conjugation and exponentiation. That is, we must have $f(a)=(a^{*})^{\gamma}a^{\eta}$. Since $f(a)$ is real we hve to require $\gamma=\eta$, so that

\begin{equation}
f(a)=|a|^{2\gamma},
\label{exponentialf}
\end{equation}
where $\gamma$ is a positive integer.

Let us check first that the choice $f(a)=|a|^{2}$ is acceptable, as claimed. Consider again the case where property $P$ has knowability level 1, and $P'$ has knowability level 3. Inserting this ansatz in Eq. [\ref{acond}] we get

\begin{equation}\begin{array}{lll}
1 & = & |a_{1}|^{2}+|a_{2}|^{2}\\
1 & = & |a_{11}|^{2}+|a_{12}|^{2}\\
1 & = & |a_{21}|^{2}+|a_{22}|^{2}\\
0 & = & a_{1}a_{2}^{*}(a_{11}a_{21}^{*}+a_{12}a_{22}^{*})+\\
& & a_{1}^{*}a_{2}(a_{11}^{*}a_{21}+a_{12}^{*}a_{22}).
\end{array}
\label{a2cond}
\end{equation}
Suppose that a given choice $\{a_{j}\}$ and $\{a_{jj'}\}$ satisfies the last equation. Property independence (Assumption \ref{propind}) then means that we should be allowed to vary $\{a_{jj'}\}$ freely for the given choice of $\{a_{j}\}$, or vice versa, and the equality would still hold. To make this possible we must require that the following relation is always fulfilled.

\begin{equation}
0=a_{11}a_{21}^{*}+a_{12}a_{22}^{*}.
\label{protoinnerprod}
\end{equation}

This relation corresponds to two equations that relate the eight real parameters in the set $\{x_{jj'},y_{jj'}\}$, where $a_{jj'}=x_{jj'}+iy_{jj'}$. The second and third lines in Eq. [\ref{a2cond}] give two more conditions, so that we have four free parameters that are related to the setup to measure the value of $P'$. The necessary minimum number of free parameters is two according to Assumption \ref{expfree}. The first line in Eq. [\ref{a2cond}] gives one condition that relates the four real parameters $\{x_{j},y_{j}\}$, where $a_{j}=x_{j}+iy_{j}$. We have thus three free parameters that are related to the setup to measure the value of $P$, whereas the minimum number is just one in order to respect experimental freedom. We conclude that the choice $f(a)=|a|^{2}$ for complex $a$ is acceptable in experimental contexts involving two properties $P$ and $P'$ with two possible values each. 

Let us next try $f(a)=|a|^{4}$. If we require property independence, we get the following conditions that relate the numbers $a_{jj'}$, describing the setup to measure property $P'$:

\begin{equation}\begin{array}{lll}
0 & = & a_{11}a_{21}a_{21}^{*}a_{21}^{*}+a_{12}a_{22}a_{22}^{*}a_{22}^{*}\\
0 & = & a_{11}a_{11}a_{21}^{*}a_{21}^{*}+a_{12}a_{12}a_{22}^{*}a_{22}^{*}\\
0 & = & a_{11}a_{11}^{*}a_{21}a_{21}^{*}+a_{12}a_{12}^{*}a_{22}a_{22}^{*}\\
0 & = & a_{11}a_{11}a_{11}^{*}a_{21}^{*}+a_{12}a_{12}a_{12}^{*}a_{22}^{*}\\
\end{array}\label{messycond}
\end{equation}
These relations should be compared to the corresponding Eq. [\ref{protoinnerprod}] that holds for $f(a)=|a|^{2}$. They correspond to eight conditions that relate the eight real parameters in the set $\{x_{jj'},y_{jj'}\}$. The second and third lines in Eq. [\ref{a2cond}] give two more conditions, so that we have ten independent conditions, preventing any solution at all with property independence. Thus the choice $f(a)=|a|^{4}$ for complex $a$ is not acceptable.

It is easy too see that the higher exponent $\gamma$ is used in Eq. [\ref{exponentialf}], the more independent conditions like those in Eq. [\ref{messycond}] appear when we demand property independence. This fact spoils all chances to get any solution at all for $\gamma>1$.

Having concluded in this way that $f(a)=|a|^{2}$ is the only acceptable choice in the simple situation with two properties $P$ and $P'$ with two property values each, we must also check that it is acceptable in more complex situations with more than two properties that can take more than two values.

\begin{figure}[tp]
\begin{center}
\includegraphics[width=80mm,clip=true]{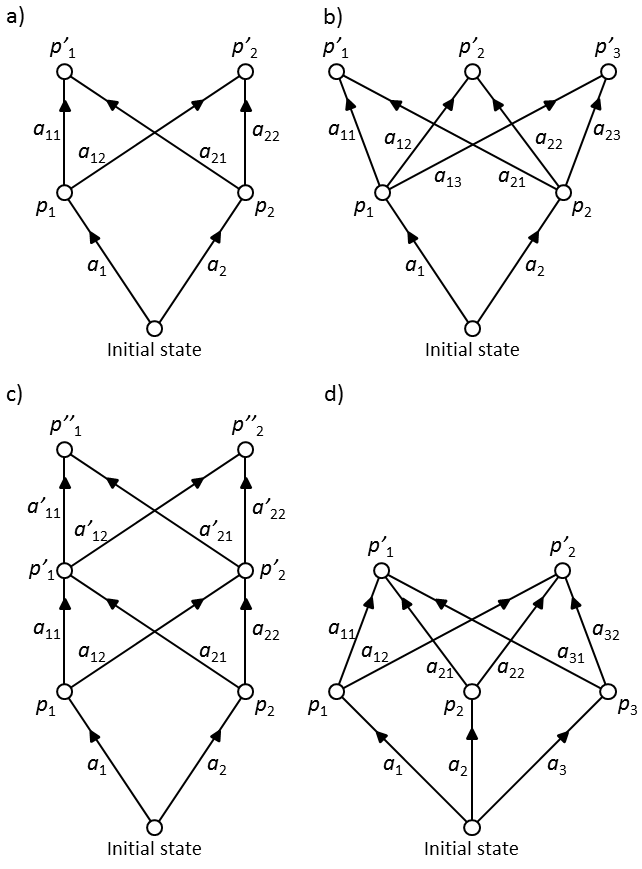}
\end{center}
\caption{Experimental contexts depicted as `networks of alternatives'. The nodes at each row correspond to the values of a given property that are possible in the given context. The numbers $a_{x}$ are seen as relations between the possible values of sequentially realized properties. Time flows upwards, as indicated by the directed edges. The context in panel a) is the one shown in \ref{Figure28}. In more complex contexts, the visual representation introduced here is easier to interpret. We do not distinguish graphically between different knowability levels in this figure, but this can, of course, be accomplished by marking the nodes or the edges in different ways.} 
\label{Figure30}
\end{figure}

Figure \ref{Figure30} shows some of the simplest of these more complex cases. Sequential time $n$ flows upwards in these diagrams. Each circle represents the `event' that a property value $p_{x}$ is attained. The contextual numbers $a_{x}$ are related to the relative volume of the event at the end point of the associated arrow given the event at the starting point. If this latter event is knowable (if the corresponding property has knowability level 3), this relative volume may correspond to a probability for the event at the end point of the arrow given the event at the starting point.

The context depicted in Fig. \ref{Figure30}(b) gives rise to the following relation for the numbers $a_{jj'}$ when $P$ has knowability level 1, and we require property independence (compare Eq. [\ref{protoinnerprod}]):

\begin{equation}
0=a_{11}a_{21}^{*}+a_{12}a_{22}^{*}+a_{13}a_{23}^{*}.
\label{protoinnerprod2}
\end{equation}
This relation corresponds to two conditions that relate twelve real parameters. The equations that correspond to the requirement $1=\sum_{j}v_{jj'}$ give three more conditions, so that we have left seven independent real parameters describing the setup to measure property $P'$. This is more than enough to fulfil experimental freedom.

If we have more than two properties that are observed in succession, we get a new set of relations of the type in Eqs. [\ref{protoinnerprod}] and [\ref{protoinnerprod2}] for each such property at knowability level 1 when we demand property independence. For the context in Fig. \ref{Figure30}(c) we get

\begin{equation}\begin{array}{lll}
0 & = & a_{11}a_{21}^{*}+a_{12}a_{22}^{*}\\
0 & = & a_{11}'(a_{21}')^{*}+a_{12}'(a_{22}')^{*}.
\end{array}\label{protoinnerprod3}
\end{equation}
It is easily understood that experimental freedom is respected in cases such as that as well. If we have more than two possible values of $P$, we also get more than one relation of this type when property independence is required. For the context in Fig. \ref{Figure30}(d) we get

\begin{equation}\begin{array}{lll}
0 & = & a_{11}a_{21}^{*}+a_{12}a_{22}^{*}\\
0 & = & a_{11}a_{31}^{*}+a_{12}a_{32}^{*}\\
0 & = & a_{21}a_{31}^{*}+a_{22}a_{32}^{*}.
\end{array}\label{protoinnerprod4}
\end{equation}
Here we have six conditions relating twelve real parameters. The requirement $1=\sum_{j}v_{jj'}$ gives three conditions more. Thus we are left with three real parameters which equals the minimum number $M(M'-1)$ needed to respect experimental freedom (Assumption \ref{expfree}). Here we are approaching a problem. If we increase the number of possible values $M$ of $P$ from three to four, and still have $M'=2$ possible values of $P'$, we get six relations of the type in Eq. [\ref{protoinnerprod4}]. This means twelve conditions relating sixteen real parameters. The requirement $1=\sum_{j}v_{jj'}$ gives four conditions more, so that we have at most a unique solution $\{a_{jj'}\}$ when we demand property independence. Experimental freedom is obviously not respected. This problem arises when $M>M'$, and gets worse as $M-M'$ gets bigger. 

Does this mean that there are contexts in which the choice $f(a)=|a|^{2}$ is not acceptable, so that we fail in the search for a function $f(a)$ that makes the representation [\ref{carep}] generally applicable? No, we are saved by the elusive relation between the numbers $a_{jj'}$ and the knowable, tangible physical properties that describe the experiment.

We may simply regard the set of contexts in which $M>M'$ as contexts in which $M=M'$ and the set of probabilities $\{q_{M'+1}', q_{M'+2}', \ldots, q_{M}'\}$ to see the last $M-M'$ values of property $P'$ are set to zero. In such contexts we regain the necessary experimental freedom (Assumption \ref{expfree}) that should be reflected in the representation. This is so since we add $M(M-M')$ numbers $a_{jj'}$, corresponding to $2M(M-M')$ new real parameters, while we only add $M-M'$ conditions that relate all real parameters that occur in the representation, corresponding to

\begin{equation}
q_{M'+1}'=q_{M'+2}'= \ldots= q_{M}'=0.
\label{virtualvalues}
\end{equation}

We may argue that we should also set all corresponding relative volumes $v_{j(M'+1)},v_{j(M'+2)},\ldots,v_{jM}$ and numbers $a_{j(M'+1)},a_{j(M'+2)},\ldots,a_{jM}$ to zero, in order to really erase the ghostly presence of the imagined extra $M-M'$ values of property $P'$. If we do this, the point of the trick is lost, since then we introduce as many new real parameters as we introdude conditions relating them. The degree of experimental freedom in the representation does not increase.

However, the procedure to add hypothetical values to $P'$ is needed only when the knowability level of $P$ is 1, when the value this property attains is forever unknowable. In that case $a_{jj'}$ and $v_{jj'}$ are also unknowable in principle. (To determine $v_{jj'}$ would mean to repeat the experiment many times until the conditional probability $q(p_{j'}'|p_{j})$ is determined. But this would require that we know in which repetitions the value $p_{j}$ of $P$ was attained.) Therefore we should not refer to $a_{jj'}$ or $v_{jj'}$ explicitly when we demand that the context is physically arranged so that imagined new values of $P'$ are not observable. That would go against explicit epistemic minimalism (Section \ref{explicitminimalism}). We should only refer to the knowable final probabilities $q(p_{j'}')=q_{j'}'$.

If, on the other hand, $P$ has knowability level 3, just as $P'$, then it is equivalent to demand that $q_{j'}'=0$ and to demand that $q(p_{j'}'|p_{j})=v_{jj'}=f(a_{jj'})=0$ for all $j$. We may introduce imagined values of $P'$ and erase them again by setting the relevant probablities or relative volumes to zero, but there is absolutely no point to it - nothing changes in the representation.

\begin{state}[\textbf{Born's rule}]
Consider the formal algebraic representation [\ref{carep}] of the contextual state $S_{C}$ of the experimental context $C$ (Definition \ref{observationalcontext}), where the relative volume $v_{j}$ of the corresponding future alternative $\vec{S}_{Oj}$ is given by $v_{j}=f(a_{j})$. An acceptable choice of function $f(a)$ is such that it makes this representation fulfil the four desiderata listed in Table \ref{desiderata}. An acceptable choice also allows property independence (Assumption \ref{propind}) as well as the necessary degree of experimental freedom (Assumption \ref{expfree}). The choice $f(a)=|a|^{2}$, where $f: \mathbb{C}\rightarrow\mathbb{R}$, is acceptable in this sense, and it is the only acceptable choice.
\label{acceptablef}
\end{state}

\subsection{The Hilbert space}

The form of relations [\ref{protoinnerprod}], [\ref{protoinnerprod2}], [\ref{protoinnerprod3}], and [\ref{protoinnerprod4}] resembles that of inner products in a vector space. We may treat them as actual Hermitian inner products in a complex vector space if we formally define the orthonormality relation

\begin{equation}
\delta_{ij}=\langle \bar{S}_{P'i},\bar{S}_{P'j}\rangle,
\label{ort1}
\end{equation}
where $\delta_{ij}$ is the Kronecker delta. Then the condition [\ref{protoinnerprod}] translates to

\begin{equation}\begin{array}{lll}
0 & = & a_{11}a_{21}^{*}+a_{12}a_{22}^{*}\\
& = & \langle a_{11}\bar{S}_{P'1}+a_{12}\bar{S}_{P'2},a_{21}\bar{S}_{P'1}+a_{22}\bar{S}_{P'2}\rangle\\
& = & \langle \bar{u}_{C}\bar{S}_{C1},\bar{u}_{C}\bar{S}_{C2}\rangle,
\end{array}
\label{ortscj}
\end{equation}
where the representations $\bar{S}_{C1}$ and $\bar{S}_{C2}$ of the corresponding `hypothetical contextual states' are introduced in Eq. [\ref{hypostates}].

Consider the general case where a property $P$ at knowability level 1, with an arbitrary number of possible values, is observed before another property $P'$, which also has an arbitrary number of possible values. Then the conditions corresponding to Eq. [\ref{protoinnerprod}] can be collapsed to the relation

\begin{equation}
\forall ij: \,\,\,\delta_{ij}=\langle \bar{u}_{C}\bar{S}_{Ci},\bar{u}_{C}\bar{S}_{Cj}\rangle.
\label{ort2}
\end{equation}

\begin{figure}[tp]
\begin{center}
\includegraphics[width=80mm,clip=true]{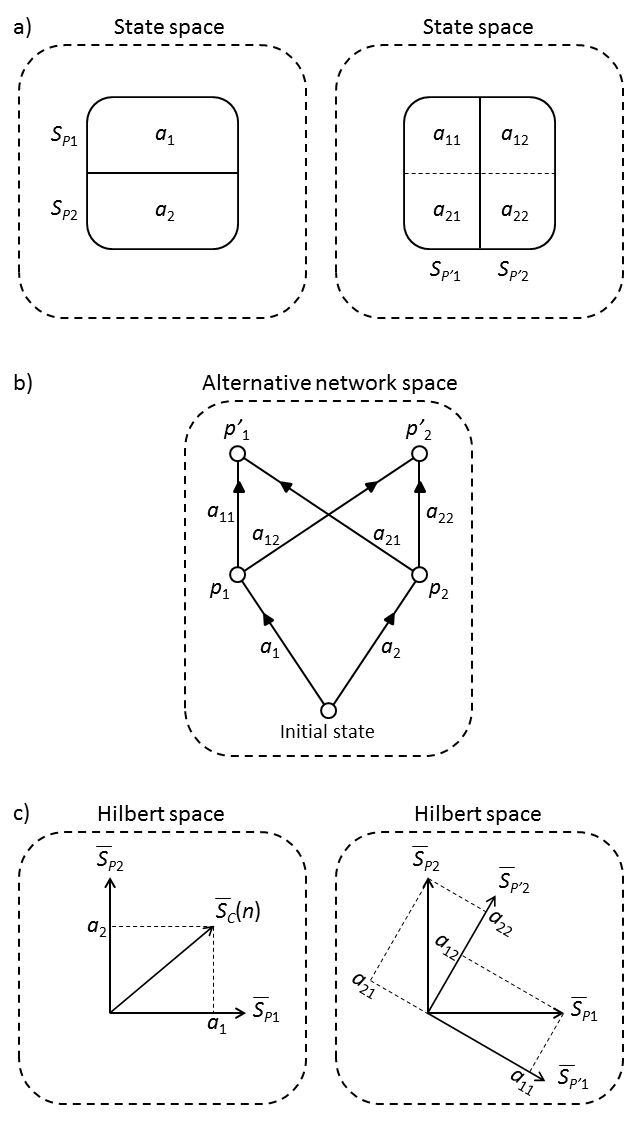}
\end{center}
\caption{Three kinds of representations of an experimental context which involves two properties $P$ and $P'$ at knowability levels 1 and 3, respectively, and in which $P$ and $P'$ have two possible values each. Compare Figs. \ref{Figure28} and \ref{Figure30}.} 
\label{Figure31}
\end{figure}

We see from Eqs. [\ref{scjpj}] and [\ref{uscjpj}] that $\bar{S}_{Cj}=\bar{u}_{C}\bar{S}_{Cj}=\bar{S}_{Pj}$. Therefore Eq. [\ref{ort2}] implies

\begin{equation}
\delta_{ij}=\langle \bar{S}_{Pi},\bar{S}_{Pj}\rangle.
\label{ort3}
\end{equation}
Further, Eqs. [\ref{scjpj}] and [\ref{umrep2}] mean that we may write

\begin{equation}
\bar{S}_{Pj}=a_{j1}\bar{S}_{P'1} + a_{j2}\bar{S}_{P'2},
\label{algebraicrel}
\end{equation}
so that Eq. [\ref{ort3}] implies

\begin{equation}
a_{jj'}=\langle \bar{S}_{Pj},\bar{S}_{P'j'}\rangle.
\label{inner}
\end{equation}
We should keep in mind, however, that the relations [\ref{algebraicrel}] and [\ref{inner}] are only defined contextually within $C$ via the hypothetical contextual states $S_{Cj}$.

The conditions of the form [\ref{protoinnerprod}] arose because we required that the representation [\ref{carep}] should be able to express property independence. Therefore the orthonormality relations [\ref{ort2}] and [\ref{ort3}] can be seen as consequences of this requirement. This goes for the inner product [\ref{inner}] as well, meaning that the sets $\{\bar{S}_{Pj}\}$ and $\{\bar{S}_{P'j'}\}$ can be seen as two different orthonormal bases in the same complex vector space $\mathcal{H}_{C}$.

To be able to illustrate these facts in a simple manner, let us focus again on the basic context in Fig. \ref{Figure30}(a). We see that we have now developed three ways to describe the same situation, as shown in Fig. \ref{Figure31}. We have the familiar state space description, the description as a network of alternatives, and now we also have a description in terms of a two-dimensional complex vector space, which we may call $\mathcal{H}_{C}$ [Fig. \ref{Figure31}(c)]. Since the numbers $a_{x}$ are contextual, so is the entire vector space $\mathcal{H}_{C}$. More precisely, it is defined only within an observational context $C$, as introduced in Defintion \ref{observationalcontext}.

\begin{table}
\caption{Types of experimental contexts $C$ involving two properties. The knowability levels are listed in Table \ref{levels}.}
\label{contexttypes}
\begin{tabular}{ll}
\hline\noalign{\smallskip}
		(a) & Property $P$ has knowability 1 and $P'$ has knowability level 3.\\
\noalign{\smallskip}
		(b) & Properties $P$ and $P'$ are simultaneously knowable. Both pro-\\
		    & perties have knowability level 3.\\
\noalign{\smallskip}
		(c) & Properties $P$ and $P'$ are not simultaneously knowable. Both\\
		    & properties have knowability level 3.\\
\noalign{\smallskip}\hline
\end{tabular}
\end{table}

The complex vector space representation arose naturally in experimental contexts $C$ in which two properties $P$ and $P'$ are involved, and the values of $P$ have knowability level 1 according to Table \ref{levels}. This is one of the three possible types of experimental contexts $C$ involving two properties that are listed in Table \ref{contexttypes}. Let us discuss whether contexts of types (b) and (c) in this list can be represented by a complex vector space $\mathcal{H}_{C}$ in an analogous manner.

We have seen that the principles listed in Table \ref{hprinciples} are fulfilled by the vector space $\mathcal{H}_{C}$ introduced for contexts of type (a) in Table \ref{contexttypes}, and we want them to be satisfied in contexts of the other two types as well.

Regarding principle 1, we identified the representations $\{\bar{S}_{Pj}\}$ of the property value states $\{S_{Pj}\}$ with orthogonal vectors in $\mathcal{H}_{C}$ in the above discussion about contexts of type (a) in Table \ref{contexttypes}. However, there are contexts in which we need to look at them more generally as subspaces, as we will see below.

Principle 2 should apply regardless whether the mutually exclusive states can actually be observed within context or not. For example, we introduced the hypothetical contextual states $S_{C1}$ and $S_{C2}$ that would have applied if we had known the value of property $P$ in a context of Mach-Zehnder type where $P$ has knowability level 1 (Eq. [\ref{hypostates}]). These states are represented as orthogonal vectors according to Eqs. [\ref{ortscj}] and [\ref{ort2}].

Turning to principle 3, let us denote by $M$ the number of elements in $\{p_{j}\}$ and let $M'$ be the number of elements in $\{p_{j'}'\}$. For Mach-Zehnder type contexts with two possible values each of properties $P$ and $P'$ this means that $M=M'=2$ (Fig. \ref{Figure28}). For the context shown in Fig. \ref{Figure28}(a) in which only $P$ is observed we get $D_{H}=M$, since there are two possible outcomes $p_{1}$ or $p_{2}$ of the experiment, which equals the maximum number of distinct states of knowledge that can be obtained. For the context shown in Fig. \ref{Figure28}(b) there are are two possible outcomes $p_{1}'$ or $p_{2}'$ of the experiment, so that we again have $D_{H}=M=M'=2$, as shown in Fig. \ref{Figure31}(c).

\begin{table}
\caption{Principles to be fulfilled by vector space representations of experimental contexts.}
\label{hprinciples}
\begin{tabular}{ll}
\hline\noalign{\smallskip}
		1 & To each value $p_{i}$ of any property $P$ that is involved in the context\\
		  & $C$ is associated a subspace $\bar{S}_{Pi}\subset\mathcal{H}_{C}$. If $p_{j}$ is another value of $P$,\\
		  & then $\delta_{ij}=\langle \bar{v}_{i},\bar{v}_{j}\rangle$, whenever $\bar{v}_{i}\in\bar{S}_{Pi}$ and $\bar{v}_{j}\in\bar{S}_{Pj}$\\
\noalign{\smallskip}
		2 & Any contextual state $S_{C}$ is represented as a unit vector:\\
		  & $\langle\bar{S}_{C},\bar{S}_{C}\rangle=1$. Any pair of mutually exclusive contextual states\\
		  & $S_{C}$ and $S_{C}'$ are represented as two orthonormal vectors:\\
			& $\langle\bar{S}_{C},\bar{S}_{C}'\rangle=0$. That such states are mutually exclusive means\\
		  & that $S_{C}\cap S_{C}'=\varnothing$. Conversely, if $S_{C}\cap S_{C}'\neq\varnothing$ then $\langle\bar{S}_{C},\bar{S}_{C}'\rangle\neq 0$.\\
\noalign{\smallskip}
		3 & Let $N(n')$ be the number of distinct states of potential knowledge\\
		  & about the specimen $OS$ that may be obtained at a given time $n'$.\\
		  & The dimension $D_{H}$ of $\mathcal{H}_{C}$ can be chosen as\\
		  & $D_{H}=\max\{N(n),N(n+1),\ldots,N(n+m^{(F)})\}$.\\
\noalign{\smallskip}
		4 & If it is possible to observe property value $p_{j}$ at time $n+m$, and if\\
		  & the corresponding probability $q_{j}$ exists, then \\
		  & $q_{j}=|\langle\bar{u}_{C}\bar{S}_{C}(n+m-1),\bar{S}_{Pj}\rangle|^{2}$.\\
\noalign{\smallskip}\hline
\end{tabular}
\end{table}

More generally, we should choose $D_{H}=\max\{M,M'\}$ in contexts of type (a) in Table \ref{contexttypes}. One might argue that if $M>M'$ it would suffice that $D_{H}=M'$, but in such cases we should add $M-M'$ `virtual' values of $P'$ to be able to uphold experimental freedom (Assumption \ref{expfree}) and to apply Born's rule, as discussed in relation to Eq. [\ref{virtualvalues}]. Only then is it proper to seek a vector space representation.

\begin{figure}[tp]
\begin{center}
\includegraphics[width=80mm,clip=true]{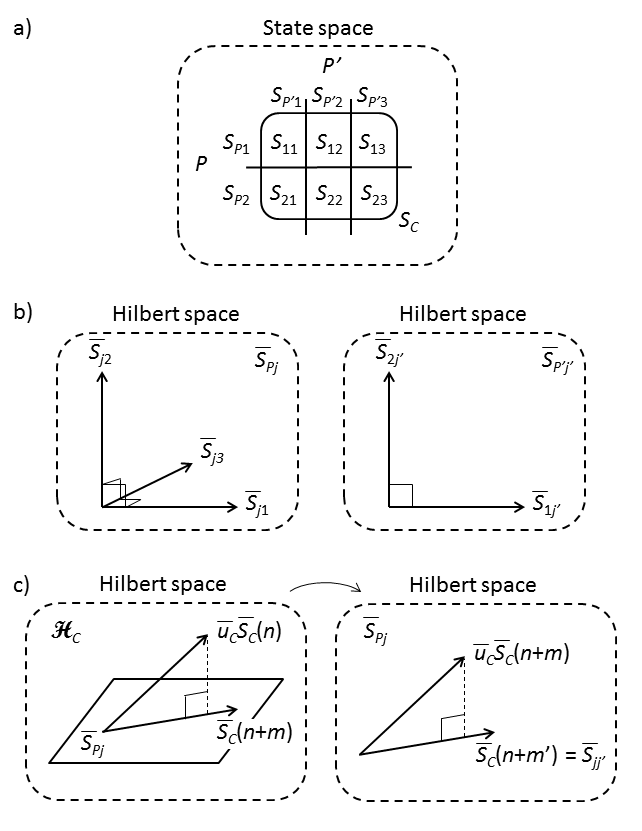}
\end{center}
\caption{A context involving two simultaneously knowable properties $P$ and $P'$ at knowability level 3, having $M=2$ and $M'=3$ possible values, respectively. We choose a vector space $\mathcal{H}_{C}$ with dimension $D_{H}=M\times M'=6$. a) We write $S_{jj'}$ as a short hand form of the specimen state $S_{PP'jj'}$ in object state space $\mathcal{S}_{O}$ (Eq. [\ref{sppjj}]). b) The property value spaces $S_{Pj}$ and $S_{P'j'}$ are represented in $\mathcal{H}_{C}$ by $3$- and $2$-dimensional subspaces, respectively. c) The observation of $P$ corresponds to a projection of the contextual state representation $\bar{S}_{C}$ down to $\bar{S}_{Pj}$. The subsequent observation of $P'$ corresponds to a further projection down to a vector $\bar{S}_{jj'}$, which corresponds to one of the six specimen states shown in panel a).} 
\label{Figure32}
\end{figure}

In contexts of type (b) in Table \ref{contexttypes} we should choose $D_{H}=M\times M'$. Then the potential knowledge about the specimen at time $n+m'$ contains knowledge about the values of both properties $P$ and $P'$, and there are $M\times M'$ pairs $(p_{j},p_{j'}')$ of possible such values. In this case the property value spaces $S_{Pj}$ are represented by $M'$-dimensional subspaces of $\mathcal{H}_{C}$, and the spaces $S_{P'j'}$ are represented by $M$-dimensional subspaces. Such a situation with $M=2$ and $M'=3$ is exemplified in Fig. \ref{Figure32}.

Principle 4 in Table \ref{hprinciples} expresses Born's rule in the vector space language. In the motivation of its original formulation in Statement \ref{acceptablef} we considered choices of functions $f(a)$ that are acceptable in contexts of types (a) or (b) in Table \ref{contexttypes}. Having sorted out the representation of property value spaces $S_{Pj}$ and the choice of vector space dimension $D_{H}$ in contexts of type (b) we conclude that all four general principles for vector space representations are fulfilled by contexts of type (b) apart from those of type (a).

What is left to do is to demonstrate that such a vector space representation is possible for contexts of type (c) as well. If we manage to do this we can safely say that all kinds of experimental contexts $C$ can be represented in a complex vector space $\mathcal{H}_{C}$ of the desired kind, since contexts of types (a)-(c) can be used as building blocks to construct all types of more complicated contexts involving more than two properties $P,P',\ldots,P^{(F)}$. At the other end of the spectrum, we note that an appropriate vector space $\mathcal{H}_{C}$ can trivially be constructed for contexts in which a single property $P$ is observed.

\begin{figure}[tp]
\begin{center}
\includegraphics[width=80mm,clip=true]{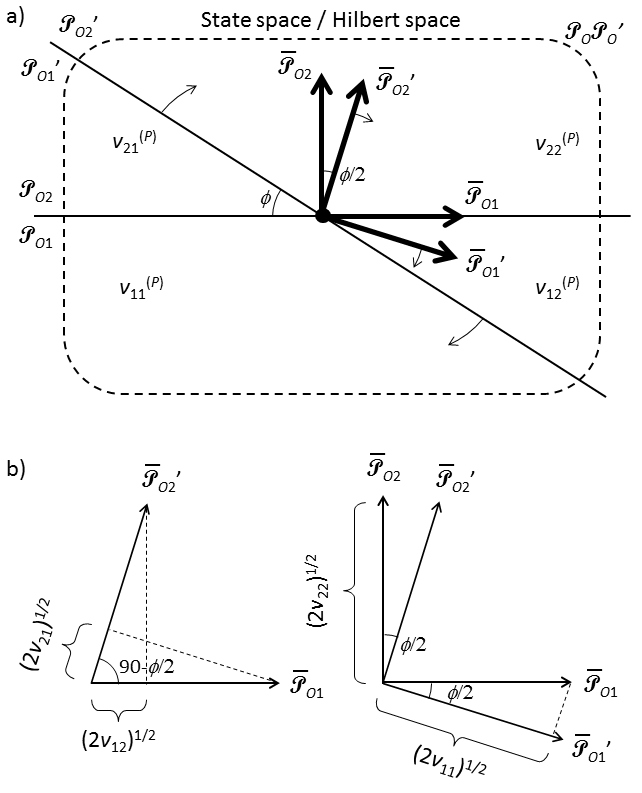}
\end{center}
\caption{Relation between the state space and vector space representation of two properties $P$ and $P'$ that are not simultaneously knowable. $P$ and $P'$ are assumed to have two possible values each. (a) The region of state space enclosed by the dashed curve is the set of object states for which both $P$ and $P'$ are defined. When the tilted line associated with $P'$ is rotated to the left, so that $P$ and $P'$ become `more different', the corresponding basis in the vector space is also rotated, but only half the angle. We assume that the properties $P$ and $P'$ are fundamental so that the state space volumes at each side of the horizontal line are equal, as well as the volumes at each side of the tilted line. (b) This fact implies that $v_{jj'}^{(P)}=v_{j'j}^{(P)}$ and $v_{jj}^{(P)}=v_{j'j'}^{(P)}$, making it possible to represent the property value spaces associated with $P$ and $P'$ as two orthonormal bases in a vector space.} 
\label{Figure33}
\end{figure}

A vector space representation cannot be constructed for contexts of type (c) in Table \ref{contexttypes} in the samy way as for contexts of type (a) or (b). The contextual state $S_{C}$ reduces to one of the property value states $S_{Pj}$ at time $n+m$ (Fig. \ref{Figure29}) in contrast to contexts of type (a). In terms of the algebraic representation $\bar{S}_{C}(n+m)$ of the contextual state, we get an expression of the form [\ref{reducedsc}] rather than of form [\ref{hypostates}]. 

In contrast to contexts of type (b) we cannot define future alternatives $\vec{S}_{Ojj'}$ according to Eq. [\ref{sojj}] and Fig. \ref{Figure28}, corresponding to the event that the value of property $P$ is found to be $p_{j}$ and the value of $P'$ is found to be $p_{j'}'$. A future alternative corresponds to a realizable object state according to Definition \ref{futurealt}, meaning that it should be possible to have $S_{OO}\subseteq \vec{S}_{Ojj'}$. However such an object state $S_{OO}$ would correspond to a situation where the values of both properties were simultaneously known, contrary to assumption. This means that we cannot define a sequence of projections down to one of $M\times M'$ orthonormal vectors $\{\bar{S}_{jj'}\}$ that spans $\mathcal{H}_{C}$ in Fig. \ref{Figure32}. 

According to principle 3 for vector space construction in Table \ref{hprinciples}, we should rather choose $D_{H}=\max\{M,M'\}=\max\{N(m+m),N(n+m')\}$ in contexts of type (c) in Table \ref{contexttypes}. For the sake of simplicity, let us focus first on the case $M=M'=2$, just as we did for contexts of type (a) and (b) when we considered context of the Mach-Zehnder type shown in Fig. \ref{Figure28}.

What we have to do is to relate the two bases in $\mathcal{H}_{C}$ that correspond to the two complete sets of future alternatives $\{\vec{S}_{O1},\vec{S}_{O2}\}$ and $\{\vec{S}_{O1}',\vec{S}_{O2}'\}$. To this end, we take a step back and focus first on the relation between the properties $P$ and $P'$ themselves, rather than the context $C$ in which they are observed.

If the property $P$ is fundamental (Section \ref{expcontext}), Definition \ref{ovoldef} implies that the property value spaces $\mathcal{P}_{Oj}$ of the object property space $\mathcal{P}$ fulfil the relation $V[\mathcal{P}_{Oi}]/V[\mathcal{P}_{O}]=V[\mathcal{P}_{Oj}]/V[\mathcal{P}_{O}]$ for any pair of indices $(i,j)$. If property $P'$ is also fundamental, then we may write

\begin{equation}\begin{array}{rcl}
v[\mathcal{P}_{Oi}] & = & v[\mathcal{P}_{Oj}]\\
v[\mathcal{P}_{Oi'}'] & = & v[\mathcal{P}_{Oj'}']
\end{array}
\label{equipropvol}
\end{equation}
for all pairs of indices $(i,j)$ and $(i',j')$ in terms of the relative volume $v[\mathcal{P}_{Oi}]\equiv V[\mathcal{P}_{Oi}]/V[\mathcal{P}_{O}]$.

Let us define the space $\mathcal{P}_{O}\mathcal{P}_{O}'\subseteq\mathcal{S}_{O}$ as the set of exact object states $Z_{O}$ for which \emph{both} properties $P$ and $P'$ are defined for the object $O$. Clearly, $\mathcal{P}_{O}\mathcal{P}_{O}'\subseteq\mathcal{P}_{O}$ and $\mathcal{P}_{O}\mathcal{P}_{O}'\subseteq\mathcal{P}_{O}'$. Let $v_{jj'}^{(P)}=v[\Sigma_{jj'},\mathcal{P}_{O}\mathcal{P}_{O}']$, where $\Sigma_{jj'}\subseteq\mathcal{P}_{O}\mathcal{P}_{O}'$ is the region inside which the value of $P$ is $p_{j}$ and the value of $P'$ is $p_{j'}'$. Explicitly,

\begin{equation}
v_{jj'}^{(P)}\equiv\frac{V[\mathcal{P}_{Oj}\cap\mathcal{P}_{Oj'}'\cap\mathcal{P}_{O}\mathcal{P}_{O}']}{V[\mathcal{P}_{O}\mathcal{P}_{O}']}.
\label{vjjp}
\end{equation}

Note that if $P$ and $P'$ is not simultaneously knowable, this region does not correspond to any realizable alternative; we cannot have $S_{OO}\subseteq\Sigma_{jj'}$. 

In any case, Eq. [\ref{equipropvol}] gives rise to the relations $v_{11}^{(P)}+v_{12}^{(P)}=v_{21}^{(P)}+v_{22}^{(P)}=v_{11}^{(P)}+v_{12}^{(P)}=v_{12}^{(P)}+v_{22}^{(P)}=1/2$ in the case $M=M'=2$, which imply that

\begin{equation}\begin{array}{rcl}
v_{jj'}^{(P)} & = & v_{j'j}^{(P)}\\
v_{jj}^{(P)} & = & v_{j'j'}^{(P)}
\end{array}
\label{vectorcond}
\end{equation}
for all pairs of indices $(j,j')$, as illustrated in Fig. \ref{Figure33}(a).

Equation [\ref{vectorcond}] makes it possible to represent the two sets of property value spaces $\{\mathcal{P}_{O1},\mathcal{P}_{O2}\}$ and $\{\mathcal{P}_{O1}',\mathcal{P}_{O2}'\}$ as two orthonormal bases in one vector space, as illustrated in Fig. \ref{Figure33}(b). These bases are specified by the following conditions.

\begin{equation}\begin{array}{rcl}
\langle \bar{\mathcal{P}}_{Oi}, \bar{\mathcal{P}}_{Oj}\rangle & = & \delta_{ij}\\
\langle \bar{\mathcal{P}}_{Oi'}', \bar{\mathcal{P}}_{Oj'}'\rangle & = & \delta_{i'j'}\\
\langle \bar{\mathcal{P}}_{Oj}, \bar{\mathcal{P}}_{Oj'}'\rangle & = & \sqrt{2v_{jj'}^{(P)}}\exp(i\theta_{jj'})
\end{array}
\label{vectorrepp}
\end{equation}
The relation between the inner product and the relative volume in the bottom row is chosen to conform with the following defining rule for inner products in vector space representations.

\begin{defi}[\textbf{Inner products for overlapping object states}]If the elements $\bar{v}$ and $\bar{w}$ in a vector space $\mathcal{H}$ represent two regions $\Sigma_{v}\subset\mathcal{S}_{O}$ and $\Sigma_{w}\subset\mathcal{S}_{O}$ such that $V[\Sigma_{v}]=V[\Sigma_{w}]$, then
$|\langle \bar{v},\bar{w}\rangle|^{2}=V[\Sigma_{v}\cap\Sigma_{w}]/V[\Sigma_{v}]$.
\label{generalborn}
\end{defi}

The requirement $V[\Sigma_{v}]=V[\Sigma_{w}]$ is necessary to make sense of the representation of $\Sigma_{v}$ and $\Sigma_{w}$ as vectors or subspaces in a vector space. This condition is fulfilled thanks to the relation [\ref{equipropvol}], which in turn follows from the fact that $P$ is assumed to be fundamental. Looking at the list in Table \ref{hprinciples} of four principles that govern the construction of vector space representation, we see that this definining rule conforms with and generalizes principle 2, and does not contradict any other princple.

We may look at the boundary between $\mathcal{P}_{O1}'$ and $\mathcal{P}_{O2}'$ in Fig. \ref{Figure33} as a line that is fixed at the centre of $\mathcal{P}_{O}\mathcal{P}_{O}'$ and may be rotated to alter the relative volumes $v_{jj'}^{(P)}$. If the line is vertical we have $v_{11}^{(P)}=v_{12}^{(P)}=v_{21}^{(P)}=v_{22}^{(P)}$. The corresponding basis $(\bar{\mathcal{P}}_{O1}',\bar{\mathcal{P}}_{O2}')$ is tilted $45^{\circ}$ in relation to the basis $(\bar{\mathcal{P}}_{O1},\bar{\mathcal{P}}_{O2})$. If the boundary line is horizontal the two properties $P$ and $P'$ are not independent and can be seen as two manifestations of the same property. We have $v_{11}^{(P)}=v_{22}^{(P)}=1/2$ and $v_{12}^{(P)}=v_{21}^{(P)}=0$. The corresponding bases in vector space coincide. In general, if we rotate the boundary line in state space between $\mathcal{P}_{O1}'$ and $\mathcal{P}_{O2}'$ the angle $\phi$, the corresponding basis in vector space is rotated the angle $\phi/2$.

The above abstract construction can be used in actual experimental contexts $C$ of type (c) in Table \ref{contexttypes} if we can choose two bases $\{\bar{S}_{P1},\bar{S}_{P2}\}$ and $\{\bar{S}_{P'1},\bar{S}_{P'2}\}$ for $\mathcal{H}_{C}$ with the same mutual relation as that between $\{\mathcal{P}_{O1},\mathcal{P}_{O2}\}$ and $\{\mathcal{P}_{O1}',\mathcal{P}_{O2}'\}$. This is not self-evident, as the practical details of the experiment can color the relative volumes that determine this relation so that they deviate from $\{v_{jj'}^{(P)}\}$ (Fig. \ref{Figure33}).

\begin{figure}[tp]
\begin{center}
\includegraphics[width=80mm,clip=true]{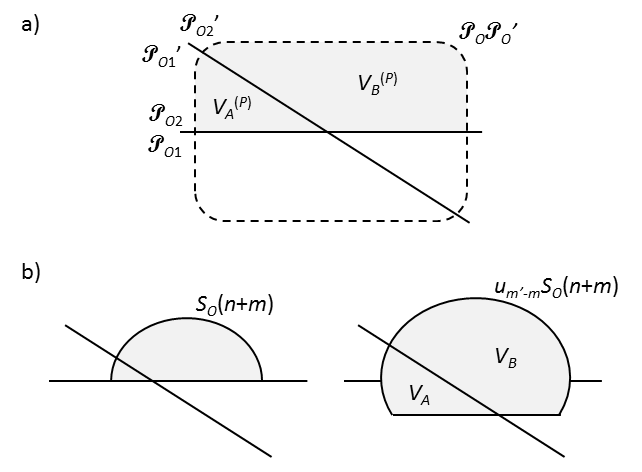}
\end{center}
\caption{Illustration of a neutral context. a) The same joint property space $\mathcal{P}_{O}\mathcal{P}_{O}'$ as shown in Fig. \ref{Figure33}, with two state space volumes $V_{A}^{(P)}$ and $V_{B}^{(P)}$ that correspond to the relative volumes $v_{21}^{(P)}$ and $v_{22}^{(P)}$, respectively. b) A context in which value $p_{2}$ of $P$ has been observed, and in which we are about to observe $P'$ at time $n+m'$. The two properties are not simultaneously knowable, just as in Fig. \ref{Figure24}(b). In a neutral context we have $V_{A}/V_{B}=V_{A}^{(P)}/V_{B}^{(P)}$, and correspondingly if $p_{1}$ would have been observed.} 
\label{Figure34}
\end{figure}

Regardless whether there is a vector space representation of the desired kind, we can always formulate the formal algebraic representation

\begin{equation}\begin{array}{rcl}
\bar{u}_{C}\bar{S}_{C}(n) & = & a_{1}\bar{S}_{P1}+a_{2}\bar{S}_{P2}\\
\bar{S}_{C}(n+m) & = & \bar{S}_{Pj}\\
\bar{u}_{C}\bar{S}_{C}(n+m) & = & a_{j1}\bar{S}_{P'1}+a_{j2}\bar{S}_{P'2},
\end{array}
\label{typecevol}
\end{equation}
where we have assumed in the second and third row that the value $p_{j}$ of property $P$ was observed at time $n+m$. We may say that the context does not affect or color the relative volumes if and only if

\begin{equation}
|a_{jj'}|^{2}=v_{jj'}^{(P)}
\label{transparenta}
\end{equation}
for each pair of indices $(j,j')$, where $v_{jj'}^{(P)}$ is defined in Eq. [\ref{vjjp}]. Such a context is illstrated in Fig. \ref{Figure34}, and it may be called \emph{neutral}. If the probabilities for all the different outcomes are defined in a neutral context we have

\begin{equation}\begin{array}{rcl}
q(p_{j}) & = & v_{j1}^{(P)}+v_{j2}^{(P)}\\
q(p_{j'}'|p_{j}) & = & 2v_{jj'}^{(P)},
\end{array}
\label{neutralprob}
\end{equation}
where $q(p_{j'}'|p_{j})$ is the conditional probability to observe the value $p_{j'}'$ of property $P'$ at time $n+m'$ given that the value $p_{j}$ of property $P$ was observed at time $n+m$.

If a context $C$ of type (c) Table \ref{contexttypes} is neutral so that Eq. [\ref{transparenta}] is fulfilled, and the underlying relative volumes $v_{jj'}^{(P)}$ defined in Eq. \ref{vjjp} fulfil Eq. [\ref{vectorcond}], then we can indeed represent $C$ of type (c) in a proper vector space $\mathcal{H}_{C}$. To do so we formally identify

\begin{equation}\begin{array}{rcl}
\bar{S}_{Pj}& = & \bar{\mathcal{P}}_{Oj}\\
\bar{S}_{P'j'} & = & \bar{\mathcal{P}}_{Oj'}'.
\end{array}
\end{equation}
The two sets of vectors $\{\bar{S}_{Pj}\}$ and $\{\bar{S}_{P'j'}\}$ become two bases that span the same vector space $\mathcal{H}_{C}$, and are related according to Eq. [\ref{vectorrepp}], as illustrated in the left panel of Fig. \ref{Figure36}(a).

\begin{figure}[tp]
\begin{center}
\includegraphics[width=80mm,clip=true]{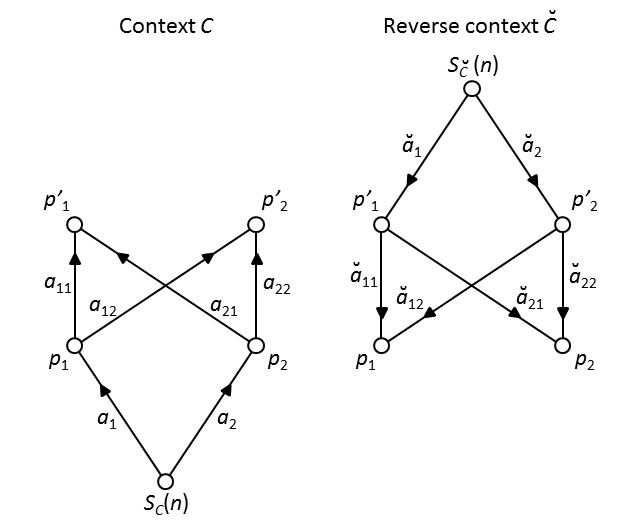}
\end{center}
\caption{A context $C$ in which two properties $P$ and $P'$ with two alternative values is observed, together with a reverse context $\breve{C}$ in which the order of observation is reversed. The contextual numbers $a_{x}$ and the reverse numbers $\breve{a}_{x}$ are not necessarily related in any particular way.} 
\label{Figure35}
\end{figure}

However, the two bases $\{\bar{S}_{P1},\bar{S}_{P2}\}$ and $\{\bar{S}_{P'1},\bar{S}_{P'2}\}$ of $\mathcal{H}_{C}$ are not on equal footing since the order in which two properties $P$ and $P'$ are observed is predefined within any experimental context $C$. We can restore the symmetry between the two bases if we consider a reverse context $\breve{C}$ in conjunction with $C$, as shown in Fig. \ref{Figure35}. In each such reverse context, the same set of properties $\{P,P',\ldots,P^{(F)}\}$ as in $C$ is observed, but in reverse order. Also, the same sets of possible values for each such property apply in both $C$ and $\breve{C}$.

\begin{figure}[tp]
\begin{center}
\includegraphics[width=80mm,clip=true]{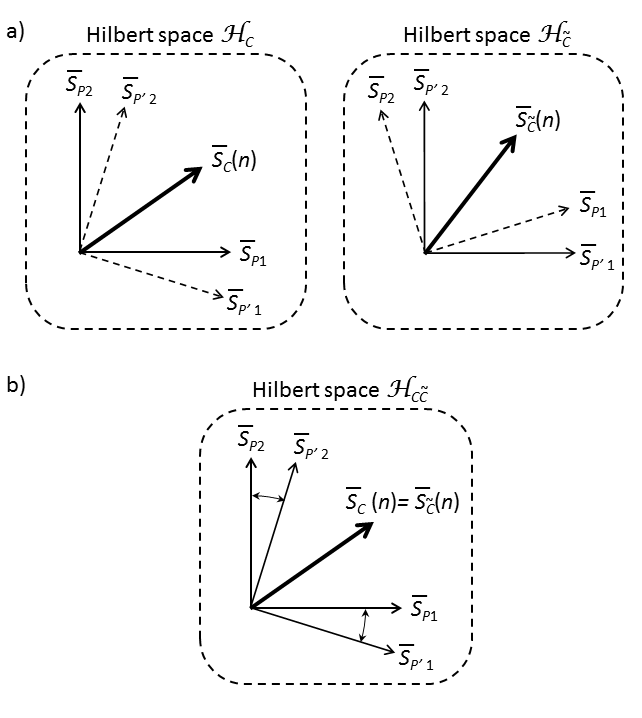}
\end{center}
\caption{Vector space representation of a context $C$ with a reciprocal $\tilde{C}$ where two properties $P$ and $P'$ that are not simultaneously knowable are observed. a) In $C$, property $P$ is observed first, and $\{\bar{S}_{P1},\bar{S}_{P2}\}$ is therefore the primary basis. In $\tilde{C}$ property $P'$ is observed first, and $\{\bar{S}_{P'1},\bar{S}_{P'2}\}$ beomes the primary basis. b) If we consider $C$ and $\tilde{C}$ together, we may identify $\bar{S}_{C}=\bar{S}_{\tilde{C}}$ and consider the two bases to be interchangeable in the common vector space.}
\label{Figure36}
\end{figure}

We may next define the \emph{reciprocal context} $\tilde{C}$ to $C$ as the reverse context for which

\begin{equation}\begin{array}{lcl}
\breve{a}_{j} & = & a_{1}a_{1j}+a_{2}a_{2j}\\
\breve{A} & = & A^{-1}
\end{array}
\end{equation}
for all $j$, where

\begin{equation}\begin{array}{rccccl}
\breve{A} & \equiv & \left(\begin{array}{cc}
\breve{a}_{11} & \breve{a}_{12}\\
\breve{a}_{21} & \breve{a}_{22}\end{array}\right), &
A & \equiv & \left(\begin{array}{cc}
a_{11} & a_{12}\\
a_{21} & a_{22}\end{array}\right). 
\end{array}
\end{equation}
Such a choice of reciprocal context $\tilde{C}$ is chosen since it gives rise to an evolution of $\bar{S}_{\tilde{C}}(n)$ as if we set $\bar{S}_{\tilde{C}}(n)=\bar{S}_{C}(n)$ in Eq. [\ref{typecevol}], and make an algebraic change of basis from $\{\bar{S}_{P1},\bar{S}_{P2}\}$ to $\{\bar{S}_{P'1},\bar{S}_{P'2}\}$, using the fact that contextually we have $\bar{S}_{Pj}=a_{j1}\bar{S}_{P'1}+a_{j2}\bar{S}_{P'2}$ according to Eq. [\ref{typecevol}]. The vector space representation $\mathcal{H}_{\tilde{C}}$ is shown in the right panel of Fig. \ref{Figure36}(a). The reciprocal context clearly exists if and only if $M=M'$

At this stage we may introduce the common vector space $\mathcal{H}_{C\tilde{C}}$ according to Fig. \ref{Figure36}(b). Looking at a context $C$ and its reciprocal $\tilde{C}$ together in this way, we have restored the symmetry between the two bases. They are at equal footing in the sense that we can express the initial contextual state at time $n$ in either basis. A change of basis in $\mathcal{H}_{C\tilde{C}}$ correponds to at change of focus from $C$ to $\tilde{C}$ or vice versa.  

In this way we give epistemic meaning to the operation in which we change basis in a quantum mechanical Hilbert space. It is no longer seen as an abstract operation on a given Hilbert space that is `out there' in some transcendent sense, but as an operation that corresponds to a change of focus to a well-defined experimental context in which we reverse the order in which two properties are observed.

Up until now we have discussed the case where $M=M'=2$. We argued that if the two properties $P$ and $P'$ are fundamental, meaning that physical law allows exactly two possible values of each property, then Eq. [\ref{vectorcond}] is fulfilled. These are the relations that enable a vector space representation, the skeleton of which is defined by Eq. [\ref{vectorrepp}].

What about other cases? Can we say anything more general about the conditions for the existence of vector space representations of contexts of type (c) in Table \ref{contexttypes}? First, we note that in the case $M=M'=2$ it is a sufficient but not a necessary condition that the two properties and the entire experimental context $C$ are fundamental. Second, we note that for $M=M'>2$ this condition is no longer sufficient. Then Eq. \ref{vectorcond}] does not follow from Eq. [\ref{equipropvol}]. Third, we recall the necessary condition $M=M'$ for the existence of the reciprocal context $\tilde{C}$, ensuring that both sets of property value state representations $\{\bar{S}_{Pj}\}$ and $\{\bar{S}_{P'j'}\}$ span $\mathcal{H}_{C}$, and that we can change basis at will if we consider the common vector space $\mathcal{H}_{C\tilde{C}}$.

\begin{figure}[tp]
\begin{center}
\includegraphics[width=80mm,clip=true]{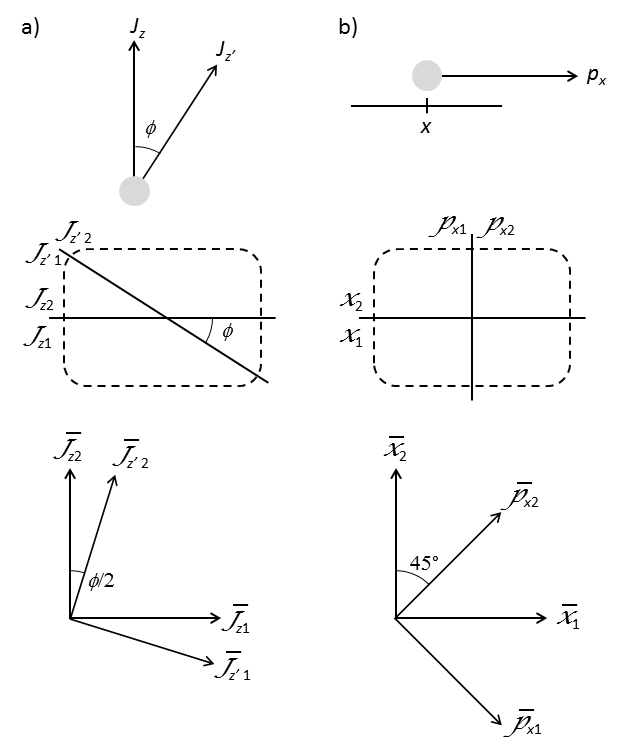}
\end{center}
\caption{Representation of pairs of properties $(P,P')$ that are not simultaneously knowable. a) A mutually defined pair of properties, exemplified by the angular momenta $(J_{z},J_{z'})$ of a given object along two axes. b) A pair of independent properties, exemplified by the position $x$ and momentum $p_{x}$ of a given object. Compare Fig. \ref{Figure33}.} 
\label{Figure37}
\end{figure}

Let us discuss two kinds of property pairs $(P,P')$ for which we can have $M=M'>2$ and still be sure that a vector space representation exists. We denote these two kinds of property pairs \emph{mutually defined} and \emph{independent}, respectively.

As an example of a mutually defined property pair we may consider the angular momenta $(J_{z},J_{z'})$ of a given object, observed along two axes $\bar{e}_{z}$ and $\bar{e}_{z'}$. The relation between these properties are defined by the angle $\phi_{zz'}$ between these two axes. When we speak of the relation between two properties in their own right, we are not allowed to refer to any particular experimental context containing objects that can be used to define an external coordinate system in which we can assign coordinates to the two axes. The only possible relation between $J_{z}$ and $J_{z'}$ is thus given by the angle $\phi_{zz'}$. An angle has no direction, meaning that

\begin{equation}
\phi_{zz'}=\phi_{z'z}.
\label{angleredundancy}
\end{equation}
The permutation of the indices $z$ and $z'$ has no epistemic meaning if we consider the properties stripped from any context. Therefore, in a symbolic representation where the permutation nevertheless makes a graphical difference, like the one above, we must compensate for this redundancy in the representation by invoking a symmetry relation like Eq. [\ref{angleredundancy}]. We conclude immediately that for the property pair $(J_{z},J_{z'})$ Eq. [\ref{vectorcond}] must be fulfilled because of such redundancies in the representation. We can therefore represent $(P,P')$ in the joint property space $\mathcal{P}_{O}\mathcal{P}_{O}'$, and also as two orthonormal bases in a vector space, as shown in Fig. \ref{Figure37}(a).

More generally, mutually defined property pairs can be defined as two properties $(P,P')$ with a difference that is completely specified by one or several relational attributes or properties that lack direction, such as angles or spatial distances. It follows that mutually defined property pairs must have the same number $M=M'$ of possible values, since a difference in this number defines a direction - from the property with the smaller number of possible values to the one with the larger number of values. In the same way we see that two properties that are related according to a temporal difference cannot be mutually defined, since time is directed (Definition \ref{directededvalues}).

As an example of an independent property pair $(P,P')$ we can take the position and momentum of the same object. There is no inherent relation between them at all, directed or not, that allows the observation of a given position $x_{i}$ to make it more probable to observe some momentum $p_{xi'}$ along the same $x$-axis rather than $p_{xj'}$. This may be so in a particular experimental context, but not in the entire property space $\mathcal{PP}'$. This becomes even more clear if we consider the fact that the numerical value of $x_{i}$ is just a matter of choice of a coordinate system, which requires a context of external objects to define. The same line of reasoning applies to $p_{xi'}$, since velocity is also defined in relation to an external coordinate system. For these reasons we must set $v_{ii'}^{(P)}=v_{jj'}^{(P)}$ for all index pairs $(i,i')$ and $(j,j')$ because we cannot define any quantity that relates $x_{i}$ and $p_{xi'}$ on which any difference could depend. Specifically we see that Eq. [\ref{vectorcond}] holds, so that a vector space representation is possible regardless the value of $M=M'$. The situation is illustrated in Fig. \ref{Figure37}(b).

More generally, we may say that a property pair $(P,P')$ is independent if and only if there is no set of relational attributes that can be used to specify the difference between them. If both $P$ and $P'$ are fundamental, then we can be sure that $v_{ii'}^{(P)}=v_{jj'}^{(P)}$ for all index pairs $(i,i')$ and $(j,j')$, so that Eq. \ref{vectorcond}] holds and a vector space representation exists such that

\begin{equation}
|\langle\bar{\mathcal{P}}_{i},\bar{\mathcal{P}}_{i'}'\rangle|^{2}=|\langle\bar{\mathcal{P}}_{j},\bar{\mathcal{P}}_{j'}'\rangle|^{2}.
\label{contspace}
\end{equation}

The reason why we add the condition that both properties should be fundamental is that otherwise we could define one observed property value to consist of two property values of the corresponding fundamental property. Then the symmetry between two property value spaces $\mathcal{P}_{j}$ and $\mathcal{P}_{j'}'$ could be broken, and we could motivate an assignment $v_{ii'}^{(P)}\neq v_{jj'}^{(P)}$.

This is not possible in the case of the property pair $(x_{i},p_{xi'})$, since we cannot exclude that both these properties are continuous. In such a case each property value $p_{i}$ that is observed within a context corresponds to a bin of continuous values within some interval $\Delta p_{i}=[p_{i},p_{i}+\delta p_{i})$ which cannot be excluded by an observation with limited resolution. In other words, continuous properties are never fundamental within an experimental context $C$. We cannot count the number of values in such a bin $\Delta p_{i}$, and therefore we cannot compare the state space volumes $V[\Delta p_{i}]$ and $V[\Delta p_{j}]$ of two such bins. Each element in one bin can be put into one-to-one correspondence with each element in another bin. The condition $V[\mathcal{S}_{OO}(A,\upsilon)]=V[\mathcal{S}_{OO}(A,\upsilon')]$ in the specification of the object state space volume (Definition \ref{ovoldef}) implies that $V[\Delta p_{i}]=V[\Delta p_{j}]$ because of this one-to-one correspondence. Therefore $v_{ii'}^{(P)}=v_{jj'}^{(P)}$ for all index pairs $(i,i')$ and $(j,j')$ when properties $P$ and $P'$ are continuous, just like in the case where they are fundamental. Consequently, a vector space representation according to Eq. \ref{contspace} can be constructed.

To exemplify the procedure, imagine that we have a double slit towards which we shoot a specimen. There is a detector at each slit. There are three alternative outcomes in this context. The specimen may be detected at detector 1 or 2, or it may not be detected at all. We may then discretize the positions on the plane that is defined by the double slit screen into three compartments: those points that define slit 1 may be called $\Delta x_{1}$, those that define slit 2 may be called $\Delta x_{2}$, and the set of all other points are placed in the trashbin $\Delta x_{3}$. Then we may make a three-dimensional vector space representation. We may also combine the position measurement with a measurement of the momentum, which we should then discretize into three bins to enable a common vector space representation $\mathcal{H}_{C}$ with $D_{H}=M=M'=3$.

The conclusion that $V[\Delta p_{i}]=V[\Delta p_{j}]$ for a continuous property $P$ regardless the length of the intervals $\Delta p_{i}$ and $\Delta p_{j}$ corresponds to coordinate system independence. It is crucial that the possibility to make vector space representations of discretized continuous properties is coordinate system independent. Since the choice of coordinate system is arbitrary, basic physical distinctions, such as that between contexts that can be represented in vector spaces or not, should not depend on such arbitrary choices. Furthermore, when we take relativity into account, different subjects may assign different coordinate values to parts of \emph{the same} context. We have chosen to build the algebraic representation upon the measure $V[S]$ on state space. We conclude that the coordinate independence of this measure is essential.

It may seem counter-intuitive that the choice of bin widths is irrelevant when we construct the vector space representation. Consider, for example, a context in which we observe position and momentum in succession, and in which there are just two alternative values for each property. This means that the bins may be `very big'. Even more, we can make one of the position bins, say $x_{2}$, very wide compared to the other. Say that we indeed find value $x_{2}$ of the discretized position property. Even so, the outcome of the momentum measurement is completely undetermined, since the position and momentum bases are tilted $45^{\circ}$ in relation to one another regardless the details of the discretization. One might think that the finding of $x_{2}$ correspond to such a wide interval $\Delta x_{2}$ that Heisenberg would have allowed us to determine momentum well enough to be sure whether we would find $p_{x1}$ or $p_{x2}$, corresponding to the intervals $\Delta p_{x1}$ and $\Delta p_{x2}$, which may also be wide.  

Recall, however, that the vector space representation is only possible for neutral contexts, as defined by the condition [\ref{transparenta}]. It may be harder to obtain such neutrality if we use a context with large bins. Recall also that the state that we represent in vector space is the memoryless contextual state $S_{C}$ of the specimen $OS$ (Definition \ref{contextualstate}). Whenever $OS$ is not a quasiobject like an electron, we may have more knowledge about its state than is encoded in $S_{C}$. In other words, we may have $S_{OS}\subset S_{C}$. That is, the vector space description does not always represent all knowledge about a specimen. In particular, if the specimen is a directly perceived object like a ball, we obviously can keep track of position $x_{i}$ and momentum $p_{xi'}$ simultaneously if we choose large enough bins $\Delta x_{i}$ and $\Delta p_{xi'}$.

The discussion in this long section has been winding, so it might be worthwhile to provide a brief overview. We have been seeking a vector space representation that applies to as many different type of experimental contexts $C$ as possible, and to as many different arrangements of contexts of a given type as possible.

Contexts of type (a) in Table \ref{contexttypes} made it necessary to make use of complex numbers $a_{x}$ in the algebraic representation [\ref{carep}] and utilize Born's law (Statement \ref{acceptablef}) to calculate probabilities. The requirement of property independence (Definition \ref{propind}) naturally gave rise to relations of the form [\ref{ortscj}] that can be interpreted as inner products. Such relations were used to construct a complex contextual vector space $\mathcal{H}_{C}$ that fulfils the four principles in Table \ref{hprinciples}.

It was seen that contexts of types (a) and (b) in Table \ref{contexttypes} can be always represented in this kind of vector space, as well as contexts of type (c), under some mild conditions. First, contexts of type (c) have to be neutral; the practical setup should not color the inherent relation between the observed properties $P$ and $P'$, as expressed in Eq. [\ref{transparenta}]. Second, these two properties should be either mutually defined or independent, as exemplified in Fig. \ref{Figure37}. Third, these two properties should either be fundamental, meaning that all values that are allowed by physical law can also be observed within context, or they should be continuous. Fourth, there should be equally many values of $P$ that can be observed within context as there are values of $P'$.

Simple contexts in which just one property is observed can trivially be represented in the desired kind of vector space, and contexts of type (a), (b) and (c) can be combined to more complex contexts which can also be represented in this way. We see that the vector space representation is indeed quite general.

\begin{state}[\textbf{Representable experimental contexts} $C$]
An experimental context $C$ (Definition \ref{observationalcontext}) can be represented in a finite-dimensional complex vector space $\mathcal{H}_{C}$ that fulfils the principles in Table \ref{hprinciples}, provided the following sufficient conditions are met in the case when $C$ involves pairs of properties $P$ and $P'$ which are not simultaneously knowable (Fig. \ref{Figure24}). All such pairs should have the same number of observable values $M=M'$. Further, the context must be neutral with regard to their observation, in the sense of Eq. [\ref{transparenta}]. Also, the two properties $P$ and $P$ should be either mutually defined or independent (Fig. \ref{Figure37}). Finally, all such property pairs should be either be fundamental, or be allowed by physical law to take continuous values.
\label{representablec}
\end{state}

The statement that it is sufficient to consider finite-dimensional contextual vector spaces $\mathcal{H}_{C}$ follows from the fact that the number $M$ of observable values of any property $P$ is always finite within any experimental context $C$. This is so since these values correspond to one element in a complete set of future alternatives (Definition \ref{setfuturealt}), and since the number of alternatives in such a set is always finite (Statement \ref{finitenoalt}). Further, in any context $C$ we observe no more than a finite number $P,P',\ldots,P^{(F)}$ of properties. Therefore the dimension $D_{H}$ of $\mathcal{H}_{C}$ fulfils $D_{H}\leq M\times M'\times\ldots\times M^{(F)}$, where we have equality if all observed properties have knowability level 3 (Table \ref{levels}).

\subsection{Properties and operators}

We may ask whether there is a general way to represent properties and their values in the vector space language. We have discussed the subject to some extent in relation to the observation of properties that are not simultaneously knowable. Here we continue that discussion.

A property transcends the object for which it is defined, as well as the context in which it can be observed. Therefore we should consider the entire property space $\mathcal{P}$ or $\mathcal{P}_{O}$, which contain the states of all such objects and contexts. Also, we should consider the set $\{p_{j}\}$ of all possible values of the property allowed by physical law, not just the set of possible values that are possible to see in a particular object observed in a particular context.

For simplicity, in the following discussion we use indices $i$ and $j$ to distinguish different property values and corresponding entities, even if we do not exclude the possiblity that they form a continuous set. That is, we do not pressuppose that we can count them according to $\{p_{j}\}=\{p_{1},p_{2},p_{3},\ldots\}$.

\begin{figure}[tp]
\begin{center}
\includegraphics[width=80mm,clip=true]{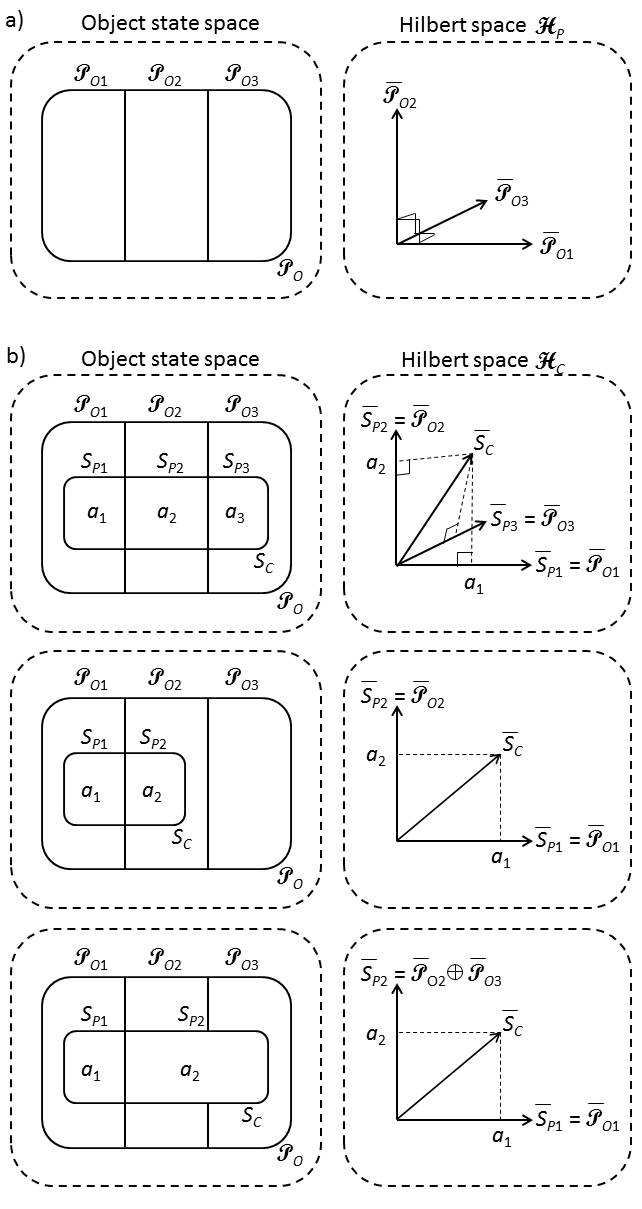}
\end{center}
\caption{a) The property space $\mathcal{P}_{O}$ as a vector space $\mathcal{H}_{P}$ spanned by the eigenvectors $\bar{\mathcal{P}}_{Oj}$ of property operator $\bar{P}$. b) Relation between the property space $\mathcal{P}_{O}$ and the contextual state $S_{C}$ and the corresponding relation between the vector spaces $\mathcal{H}_{P}$ and $\mathcal{H}_{C}$. Three cases are shown. Top panel: all possible property values can be observed within context. Middle panel: Some property values cannot be observed. Bottom panel: some future alternatives ($\bar{S}_{O2}$ in this example) correspond to several property values - the resolution of the observation is limited.} 
\label{Figure38}
\end{figure}

We can formally construct a vector space $\mathcal{H}_{P}$ such that each property value space $\mathcal{P}_{Oj}\subset\mathcal{P}_{O}$ corresponds to a vector $\bar{\mathcal{P}}_{Oj}$ such that $\langle\bar{\mathcal{P}}_{Oi},\bar{\mathcal{P}}_{Oj}\rangle=\delta_{ij}$ [Fig. \ref{Figure38}(a)]. This vector space is analogous to $\mathcal{H}_{C}$ except for the fact that the coordinates $a_{j}$ are not determined by any experimental context in the case of $\mathcal{H}_{P}$. In the vector space language, each property $P$ is uniquely specified by the complete basis $\{\bar{\mathcal{P}}_{Oj}\}$ with associated real numbers $\{p_{j}\}$. Conversely, each such pair of sets $(\{\bar{\mathcal{P}}_{Oj}\},\{p_{j}\})$ in $\mathcal{H}_{P}$ defines at least one property $P$.

\begin{equation}
P\leftrightarrow\left\{\begin{array}{ll}
\mathcal{H}_{P}\\
\{\bar{\mathcal{P}}_{Oj}\}, & \mathrm{a\:complete\:basis\:for}\:\mathcal{H}_{P}\\
\{p_{j}\}, & p_{j}\in\mathbb{R}\:\:\mathrm{for\:all}\:j
\end{array}\right.
\label{propcorr}
\end{equation}
A linear operator is uniquely defined by its eigenvectors and eigenvalues if the basis of eigenvectors is complete. Therefore we can associate $P$ with exactly one linear operator $\bar{P}$ with domain $\mathcal{H}_{P}$, with a complete basis of eigenvectors $\bar{\mathcal{P}}_{Oj}$, and with real eigenvalues $p_{j}$. Any such operator $\bar{P}$ is necessarily self-adjoint: $\langle \bar{P}\bar{v},\bar{w}\rangle=\langle\bar{v},\bar{P}\bar{w}\rangle$ whenever $\bar{v},\bar{w}\in\mathcal{H}_{P}$. Furthermore, the eigenvalues of $\bar{P}$ are non-degenerate, since different property values $p_{i}$ and $p_{j}$ are different by definition.

For any observational context $C$ in which property $P$ is observed, we have $S_{C}\subseteq\mathcal{P}_{O}$. It would therefore be natural to define $\mathcal{H}_{C}$ so that $\mathcal{H}_{C}\subseteq\mathcal{H}_{P}$. To relate the two vector spaces, we define the action of $\bar{P}$ on a vector $\bar{S}_{C}\in\mathcal{H}_{C}$ so that

\begin{equation}
S_{C}\subseteq\mathcal{P}_{Oj}\Leftrightarrow\bar{P}\bar{S}_{C}\equiv p_{j}\bar{S}_{C}.
\label{eigenstate}
\end{equation}
If the property value state fulfils $S_{Pj}\subseteq\mathcal{P}_{Oj}$, then we may identify $\bar{S}_{Pj}=\bar{\mathcal{P}}_{Oj}$ since both quantities are defined as an eigenspace of the same operator with the same eigenvalue. Since $\{\bar{\mathcal{P}}_{Oj}\}$ is a complete basis for   $\mathcal{H}_{P}$ we indeed get $\mathcal{H}_{C}\subseteq\mathcal{H}_{P}$, as desired.

The last statement is not necessarily true when more than one property is observed within $C$. If two such properties $P$ and $P'$ are simultaneously knowable and have $M$ and $M'$ possible values each, we argued in relation to Fig. \ref{Figure32} that we should choose the dimension $D_{HC}=M\times M'$ of $\mathcal{H}_{C}$. Since we have $D_{HP}=M$, where $D_{HP}$ is the dimension of $\mathcal{H}_{P}$, we see that $\mathcal{H}_{C}\not\subseteq\mathcal{H}_{P}$.

In such a case we should compare $\mathcal{H}_{C}$ with a formal vector space $\mathcal{H}_{PP'}$, constructed from the common property space $\mathcal{P}_{O}\mathcal{P'}_{O}$. This space can be decomposed as $\mathcal{P}_{O}\mathcal{P'}_{O}=\bigcup_{jj'}\mathcal{P}_{jj'}$, where each property value space $\mathcal{P}_{jj'}=\mathcal{P}_{Oj}\cap \mathcal{P}_{Oj'}'$ corresponds to the pair of values $(p_{j},p_{j'}')$ of the combined property $(P,P')$. To each property value space $\mathcal{P}_{jj'}$ we associate a vector $\bar{\mathcal{P}}_{jj'}$ such that $\langle\bar{\mathcal{P}}_{ii'},\bar{\mathcal{P}}_{jj'}\rangle=\delta_{(ii'),(jj')}$. Then, in anaology with Eq. [\ref{eigenstate}] we may say that 

\begin{equation}
S_{C}\subseteq\mathcal{P}_{jj'}\Leftrightarrow\left\{\begin{array}{lll}
\bar{P}\bar{S}_{C} & \equiv & p_{j}\bar{S}_{C}\\
\bar{P}'\bar{S}_{C} & \equiv & p_{j'}\bar{S}_{C}
\end{array}\right.,
\label{eigenstate2}
\end{equation}
while Eq. [\ref{eigenstate}] still holds true, of course, meaning that 

\begin{equation}
S_{C}\subseteq\mathcal{P}_{Oj}=\bigcup_{j'}\mathcal{P}_{jj'}\Leftrightarrow\bar{P}\bar{S}_{C}\equiv p_{j}\bar{S}_{C}
\label{eigenstate3}
\end{equation}
and

\begin{equation}
S_{C}\subseteq\mathcal{P}_{Oj'}'=\bigcup_{j}\mathcal{P}_{jj'}\Leftrightarrow\bar{P}'\bar{S}_{C}\equiv p_{j'}'\bar{S}_{C}.
\label{eigenstate4}
\end{equation}
At this stage we can identify $\bar{S}_{jj'}=\bar{\mathcal{P}}_{jj'}$ as orthonormal basis vectors in both $\mathcal{H}_{C}$ and $\mathcal{H}_{PP'}$, and it makes sense to write $\mathcal{H}_{C}\subseteq\mathcal{H}_{PP'}$, since both vector spaces now have dimension $D_{HC}=D_{HPP'}=M\times M'$.
Here $\bar{S}_{jj'}$ is the vector space representation of the set defined in Eq. [\ref{sppjj}].

We argued in relation to Eq. [\ref{propcorr}] that the eigenvalues of the self-adjoint operator $\bar{P}$ are always non-degenerate when we let it act on the elements of $\mathcal{H}_{P}$. This is no longer true when we enlarge the domain of $\bar{P}$ to the vector space $\mathcal{H}_{PP'}$, as illustrated by Eqs. [\ref{eigenstate3}] and [\ref{eigenstate4}]. The entire set of eigenvectors $\{\bar{\mathcal{P}}_{jj'}\}_{j'}$ is now associated with the same eigenvalue $p_{j}$, and the entire set of eigenvectors $\{\bar{\mathcal{P}}_{jj'}\}_{j}$ becomes associated with the same eigenvalue $p_{j'}'$.

Figure \ref{Figure38}(b) shows the relation between  $\mathcal{H}_{C}$ and $\mathcal{H}_{P}$ in three basic cases in which a single property $P$ is observed within the experimental context $C$. In the top panel $C$ is fundamental, meaning that each value $p_{j}$ allowed by physical law can also be observed within context. In that case we may write $\mathcal{H}_{C}=\mathcal{H}_{P}$.

In the middle panel only two of the three values $p_{j}$ allowed by physical law can be observed within $C$. The context is not fundamental, and we get $\mathcal{H}_{C}\subset \mathcal{H}_{P}$.

In the bottom panel the resolution of the observation of $P$ is limited within $C$. There is a future alternative $\vec{S}_{Ok}$ that corresponds to more than one of the property values $p_{j}$ allowed by physical law. In such a case we may define a \emph{contextual property} $P_{C}$ in which some values corresponds to several values of the fundamental property $P$. In the present example we have $p_{C1}=p_{1}$ and $p_{C2}=\{p_{2},p_{3}\}$. This situation occurs for all properties which possibly are continuous, like position or momentum, as discussed in the previous section. If \begin{equation}
p_{Ck}=\{p_{j},p_{j+1},\ldots,p_{j+l}\}
\label{pck}
\end{equation}
we set

\begin{equation}
\bar{S}_{Pk}=\bar{P}_{Oj}\oplus \bar{P}_{O(j+1)}\oplus\ldots\oplus\bar{P}_{O(j+l)},
\label{valuecontraction}
\end{equation}
where we identify the unit vectors in $\mathcal{H}_{C}$ and $\mathcal{H}_{P}$ with the subspaces they span. We get $\mathcal{H}_{C}=\mathcal{H}_{P}$.

In general we state that the number of possible values $p_{Ck}$ of a contextual property $P_{C}$ is always finite, even if physical law allows an infinite number of values $p_{j}$. Keeping Eq. [\ref{valuecontraction}] in mind, this means that the dimension of $\mathcal{H}_{C}$ can always be considered to be finite, in contrast to that of $\mathcal{H}_{P}$.

To make the notation less cluttered in the following, we write $P$ instead of $P_{C}$ when we speak about a contextual property, and we write $p_{j}$ instead of $p_{Ck}$ even though this contextual value corresponds to several values of the corresponding fundamental property, according to Eq. [\ref{pck}].

\begin{state}[\textbf{Properties as self-adjoint operators}]
To each contextual property $P$, which may or may not be fundamental, there corresponds exactly one self-adjoint linear operator $\bar{P}$ with domain $\mathcal{H}_{C}$, and with a finite complete set of eigenspaces $\{\bar{\mathcal{P}}_{Oj}\}=\{\bar{S}_{Pj}\}$ that span $\mathcal{H}_{C}$, and a corresponding set of non-degenerate eigenvalues $\{p_{j}\}$. The number $M$ of such non-degenerate eigenvalues and eigenspaces filfils $M=D_{HC}$ if and only if the experimental context $C$ does not involve a pair of simultaneously knowable properties $P$ and $P'$. Otherwise $M<D_{HC}$. We have $\bar{P}\bar{v}=p_{j}\bar{v}$ whenever $v\in\bar{S}_{Pj}$. If value $p_{j}$ is observed at time $n+m$, then $S_{C}(n+m)\subseteq\mathcal{P}_{Oj}$ and $\bar{S}_{C}(n+m)=\bar{\mathcal{P}}_{Oj}=\bar{S}_{Pj}$. 
\label{propopnofun}
\end{state} 

Let us check that the operators $\bar{P}$ fulfil the familiar commutation rules. Consider a context in which two properties $P$ and $P'$ are observed in succession. If they are simultaneously knowable we have $S_{C}(n+m')=S_{Pj}\cap S_{P'j'}=S_{jj'}$ at the time $n+m'$ when we have just observed the value of $P'$, according to Fig. \ref{Figure32}. We have $S_{Pj}\subseteq \mathcal{P}_{Oj}$ and $S_{P'j'}\subseteq \mathcal{P}_{Oj'}'$. According to Eq. [\ref{eigenstate}] this means that $\bar{P}S_{C}(n+m')=p_{j}S_{C}(n+m')$ and $\bar{P}'S_{C}(n+m')=p_{j'}'S_{C}(n+m')$, so that $\bar{P}\bar{P}'\bar{S}_{C}(n+m')=\bar{P}'\bar{P}\bar{S}_{C}(n+m')=p_{j}p_{j'}'\bar{S}_{C}(n+m')$. This holds true for any final contextual state. These final states are the same as those called $S_{jj'}$ in Fig. \ref{Figure32}, and the corresponding subspaces $\bar{S}_{jj'}$ can be identified with a complete basis for $\mathcal{H}_{C}$. Thus, for any vector $\bar{v}\in\mathcal{H}_{C}$ we get

\begin{equation}
\bar{P}\bar{P}'\bar{v}=\bar{P}'\bar{P}\bar{v}.
\label{simcom}
\end{equation}

The situation is different if $P$ and $P'$ are not simultaneously knowable. Then $S_{C}(n+m')=S_{P'j'}$. We have $\bar{P}\bar{P}'\bar{S}_{C}(n+m')=p_{j'}'\bar{P}\bar{S}_{C}(n+m')$, but $\bar{P}\bar{S}_{C}(n+m')\neq p_{j}\bar{S}_{C}(n+m')$ for all $j$, as implied by Eq. [\ref{eigenstate}] and Fig. \ref{Figure24}. Therefore we cannot arrive at the conclusion that $\bar{P}\bar{P}'\bar{S}_{C}(n+m')=\bar{P}'\bar{P}\bar{S}_{C}(n+m')$ in the same way as before.

We could nevertheless try to use the same kind of basis $\{\bar{S}_{jj'}\}$ for $\mathcal{H}_{C}$ as in the case when $P$ and $P'$ are simultaneously knowable. Then we could try to apply $\bar{P}$ to $\bar{S}_{C}(n+m')$ written as a linear combination of these basis vectors. We would get the same commutation relation [\ref{simcom}] as for simultaneously knowable properties.

Why do we not accept this approach? The reason is that the corresponding sets $\Sigma_{jj'}\equiv S_{Pj}\cap S_{P'j'}$ in state space do not correspond to realizable future alternatives (Definition \ref{futurealt}) like the sets $S_{jj'}$ do in Fig. \ref{Figure32}(a). Referring again to Fig. \ref{Figure24}, we understand that $S_{C}$ can never be squeezed into any of the individual compartments $S_{jj'}$ in Fig. \ref{Figure32}(a). The principle of explicit epistemic minimalism implies that we would get an incorrect physical theory if we tried to operate on enitites that are not knowable. The same argument was used to infer that the evolution operator $u_{1}$ or $u_{O1}$ cannot be applied to exact states $Z$ or $Z_{O}$. In fact the domain $\mathcal{D}_{\bar{P}}$ of each contextual property operator $\bar{P}$ must be the same as that of $\bar{u}_{C}$, which acts on all contextual state representations $\bar{S}_{C}$. That is,

\begin{equation}
\mathcal{D}_{\bar{P}}=\mathcal{D}_{\bar{u}_{C}}=\mathcal{H}_{C}.
\end{equation}
For this reason it does not help to write $\bar{S}_{C}(n+m')$ as a linear combination of vectors $\bar{\Sigma}_{jj'}$, since we cannot apply $\bar{P}$ to the outcome. We have $\bar{P}\bar{S}_{C}(n+m')=\bar{P}S_{P'j'}$, but we cannot write $\bar{P}\bar{S}_{P'j'}=\bar{P}\bigcup_{j}\bar{\Sigma}_{jj'}=\bigcup_{j}\bar{P}\bar{\Sigma}_{jj'}$ even though $S_{P'j'}=\bigcup_{j}\Sigma_{jj'}$, since $\bar{\Sigma}_{jj'}$ is not an element of $\mathcal{H}_{C}$ when $P$ and $P'$ are not simultaneously knowable.

These considerations block the road for any attempt to derive the result that $\bar{P}$ and $\bar{P}'$ commute when $P$ and $P'$ are not simultaneously knowable. As we discussed above, in this case we should choose a Hilbert space $\mathcal{H}_{C}$ with dimension $M=M'$, rather than $M\times M'$. This reduction of dimensionality means that $\mathcal{H}_{C}$ will be spanned by two `competing' orthogonal bases, one associated with $P$ and one with $P'$, as explained in connection with Fig. \ref{Figure33}, rather than a combined basis that spans a larger space. Since the two competing bases are not orthogonal to each other, we immediately see that the corresponding two operators cannot commute.

\begin{state}[\textbf{Commutation rules for property operators}]Assume that we have a context $C$ in which two contextual properties $P$ and $P'$ are observed in succession. Then the pair of operators $\bar{P}$ and $\bar{P}'$ defined according to Statement \ref{propopnofun} fulfil the following commutation rules: $[\bar{P},\bar{P}']\equiv\bar{0}$ if $P$ and $P'$ are simultaneously knowable, and $[\bar{P},\bar{P}']\not\equiv\bar{0}$ if they are not.
\label{commutationrules}
\end{state}

Here $\bar{0}$ is the zero operator that maps any vector to the origin.

\begin{figure}[tp]
\begin{center}
\includegraphics[width=80mm,clip=true]{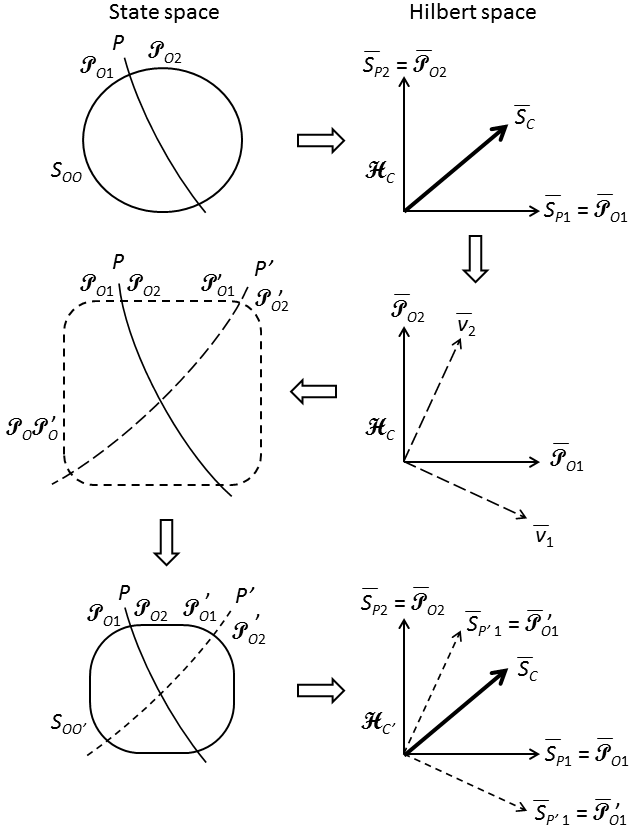}
\end{center}
\caption{In a context $C$, described by the object state $S_{OO}$, in which property $P$ is observed with possible values $\{p_{j}\}$, we may define a vector space $\mathcal{H}_{C}$ with basis $\{\bar{S}_{Pj}\}$. We may define another orthonormal basis $\{\bar{v}_{j'}\}$ in $\mathcal{H}_{C}$. Such a basis can always be associated with another property $P'$, where we can identify $\{\bar{v}_{j'}\}=\bar{S}_{P'j'}$ in another context $C'$ in which $P'$ is observed after $P$. This context $C'$ is described by the state $S_{O'O'}$. Compare Figs. \ref{Figure33} and \ref{Figure36}.}
\label{Figure39}
\end{figure}

The direction of the reasoning behind Statement \ref{propopnofun} can be reversed - just like a contextual property corresponds to a self-adjoint operator in $\mathcal{H}_{C}$, a self-adjoint operator in $\mathcal{H}_{C}$ corresponds to a contextual property.

\begin{state}[\textbf{Self-adjoint operators as properties}]
Consider a context $C$ in which $P$ is the last contextual property to be observed. To each linear, self-adjoint operator $\bar{P}'$ that acts upon any vector $\bar{v}\in\mathcal{H}_{C}$ and has a complete set of $D_{H}$ orthonormal eigenvectors $\bar{\mathcal{P}}_{Oj}'$ with non-degenerate eigenvalues, there corresponds at least one contextual property $P'$ with $D_{HC}$ possible values in the following sense: there exists another context $C'$ that is the same as $C$ except that another property $P'$ is observed after $P$, and this property corresponds to the operator $\bar{P}'$, in the sense expressed in Statement \ref{propopnofun}. $P'$ and $P$ are not simultaneously knowable whenever $\bar{P}'\neq\bar{P}$.
\label{opisprop}
\end{state}

To see why this statement is reasonable, consider Fig. \ref{Figure39}. In the top panel we show the state space and the vector space representation in $\mathcal{H}_{C}$ of the original context $C$. Choose an arbitrary self-adjoint operator $\bar{P}'$ that acts in $\mathcal{H}_{C}$. It will have a set of orthogonal eigenvectors $\{\bar{v}_{j'}\}$ that spans $\mathcal{H}_{C}$ (middle panel, right). To define the operator uniquely, we also have to fix a set of eigenvalues $\{\epsilon_{j'}\}$. We may write

\begin{equation}
\bar{P}'\leftrightarrow\left\{\begin{array}{l}
\{\bar{v}_{j'}\}\\
\{\epsilon_{j'}\}
\end{array}\right.
\end{equation}

To interpret the basis $\{\bar{v}_{j'}\}$ as the vector space representation of the property value spaces $\mathcal{P}_{Oj'}'$ of another contextual property $P'$, we have to make a partition of a presumed joint property space $\mathcal{P}_{O}\mathcal{P}_{O}'$ that conforms with such an interpretation. To this end we choose

\begin{equation}\begin{array}{rcl}
\mathcal{P}_{O}\mathcal{P}_{O}' & = & \mathcal{P}_{O} \cap \left(\bigcup_{j'}\mathcal{P}_{Oj'}'\right)\\
\mathcal{P}_{Oi'}'\cap\mathcal{P}_{Oj'}' & = & \varnothing,
\end{array}\end{equation}
with 

\begin{equation}\begin{array}{rcl}
v[\mathcal{P}_{Oi'}'] & = & v[\mathcal{P}_{Oj'}']\\
v_{jj'}^{(P)} & = &  |\langle \bar{\mathcal{P}}_{Oj}, \bar{v}_{j'}\rangle|^{2}/2,
\end{array}
\end{equation}
for all triplets of indices $(j,i',j')$, where the involved relative volumes are defined in Eqs. [\ref{equipropvol}] and [\ref{vjjp}]. The inner product is chosen to conform with the third row of Eq. [\ref{vectorrepp}].

It is clear that sets $\mathcal{P}_{Oj'}'\subset\mathcal{S}_{O}$ of the desired kind always exist, but they may not be uniquely determined by the operator $\bar{P}'$ and its eigenvectors $\{\bar{v}_{j'}\}$. The set of relative volumes $\{v_{jj'}^{(P)}\}$ does not uniquely determine $\{\mathcal{P}_{Oj'}'\}$. The boundaries $\partial\mathcal{P}_{Oj'}'$ may wiggle around some mean position without changing the relative volumes.

As a last step, we have to assign values $p_{j'}'$ to the new contextual property $P'$. Since the property value spaces $\mathcal{P}_{Oj'}'$ are disjoint by construction, they correspond to states of the observed object $O$ that are subjectively distinguishable. As such, they can be encoded by any set of values $\{p_{j'}'\}$, provided all of them are different. We may set $p_{j'}'=\epsilon_{j'}$ without loss of generality, since this is just a matter of choosing a coordinate system for the new property $P'$.

\subsection{State reduction by epistemic consistency}

We have tried to argue that quantum mechanics arises as the only algebraic representation that applies to most experimental contexts $C$ (Defintion \ref{observationalcontext}) and respects the guiding principles for physical law discussed in Section \ref{guiding}. But can all scientific experiments be described as such a context $C$, or is the definition too strict?

We required that all properties $P, P',\ldots,P^{(F)}$ that are involved in $C$ have knowability level 1 or 3 according to Table \ref{levels}. However, it is indeed possible to construct well-defined experiments in which it is not known at the start of the experiment at time $n$ whether one of the predefined values $\{p\}$ of some property $P\neq P^{(F)}$ will ever become known or not, meaning that it has knowability level 2.

A state reduction $u_{O1}S_{OO}(n+m-1)\rightarrow S_{OO}(n+m)\subset u_{O1}S_{OO}(n+m-1)$ occurs in a context $C$ if and only if such a value $p$ becomes potentially known to some subject at time $n+m$. We may call such an event a \emph{state reduction by observation}. If we allow properties $P$ at the contingent knowability level 2, it is necessary to introduce another kind of state reduction, which we may call \emph{state reduction by epistemic consistency}.

\begin{figure}[tp]
\begin{center}
\includegraphics[width=80mm,clip=true]{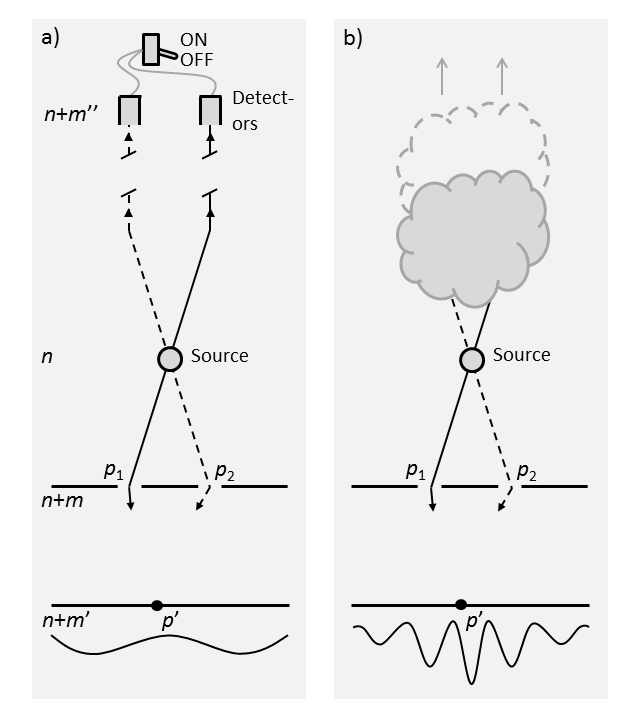}
\end{center}
\caption{Gedankenexperiment to illustrate state reduction by epistemic consistency. The source emits two particles in opposite directions at time $n$. One of these particles passes a slit at time $n+m$. Property $P$ is the slit that particle passes. Property $P'$ is the position at the screen hit at time $n+m'$ by the same particle. a) If there are two detectors that can detect the backward-moving twin particle at a still later time $n+m''$, there may or may not appear path knowledge depending on whether the detectors are turned on or off. This may not be decided until after time $n+m'$. The state reduces at time $n+m'-1$ in either case to preserve epistemic consistency. b) If path information is erased in a `cloud', the values of $P$ are degraded to knowability level 1 at time $n$, and there is no state reduction at time $n+m'-1$. The cloud has to be close enough, so that information is erased before the particle hits the screen.}
\label{Figure40}
\end{figure}

Imagine a double-slit experiment where the source emits two (entangled) particles in opposite directions, as illustrated in Fig. \ref{Figure40}. Property $P$ represents the slit one of the particles passes. Property $P'$ is the position on the detector screen behind the slits that is hit by the same particle.

Suppose first that the other particle is deflected backwards into a detector [Fig. \ref{Figure40}(a)]. There is one detector for each slit, so that detection of this twin particle determines which slit the original particle passed. The two detectors may be placed far away, so that the twin particle reaches them at a moment in time $n+m''$ a long time after the original particle hits the screen at time $n+m'$. Until time $n+m''$, the detectors can be turned on and off at will. Therefore the alternatives $S_{O1}$ and $S_{O2}$ corresponding to property values $p_{1}$ or $p_{2}$ may have knowability level 2 during the entire course of the experiment.

We may write $S_{OO}(n')=\vec{S}_{O1}(n')\cup \vec{S}_{O2}(n')$ for $n'\geq n$. If there would be no state reduction $u_{O1}S_{OO}(n')\rightarrow S_{OO}(n'+1)=\vec{S}_{Oj}(n'+1)\subset u_{O1}S_{OO}(n')$ at some time $n'+1<n+m'$, then an interference pattern would appear at the screen (if the experiment were repeated many times). However, if we choose to gain path knowledge afterwards, at time $n+m''$, then this knowledge would refer backwards to the time $n+m$ at which the particle passed the slit, and this knowledge would contradict the appearance of the interference pattern. Therefore a state reduction must occur no later than at time $n+m'-1$ to preserve epistemic consistency in the sense expressed in Fig. \ref{Figure16} in Section \ref{consistency}. Note that the mere \emph{risk} of inconsistency is enough to trigger a state reduction.

In contrast, if it is known \emph{a priori} a the start of the experiment at time $n$ that path knowledge is erased before time $n+m'$, then there is no risk of inconsistency. Such an eraser is represented in Fig. \ref{Figure40}(b) as a cloud in which the twin particle disappears before the original particle passes through one of the slits. The alternatives corresponding to property values $p_{1}$ or $p_{2}$ are degraded from knwability level 2 to knowability level 1.

However, if the eraser is successively moved away, the possibility to gain path path knowledge finally survives after time $n+m'$, and we are back in a situation where Nature has to choose path no later than at time $n+m'-1$ to make sure that no contradiction will occur. Nature is about to paint itself into a logical corner, so to say, and must jump out of it to preserve epistemic consistency.

We conclude that if we allow the algebraic representation $\bar{S}_{C}$ of the contextual state to `collapse' from $\bar{S}_{C}(n+m'-2)=a_{1}\bar{S}_{P1}+a_{2}\bar{S}_{P2}+\ldots$ to $\bar{S}_{C}(n+m'-1)=\bar{S}_{Pj}$ in order to uphold epistemic consistency, just as $\bar{S}_{C}$ collapses according to Eqs. [\ref{initialsc}] and [\ref{reducedsc}] as a result of an observation, then quantum mechanics can indeed be used to represent a more general kind of experiment that involves properties at knowability level 2. 

\section{Discussion}
\label{discussion}

\subsection{A bird's eye perspective}

The aim of this study has been to use epistemic principles and concepts to construct the formalism of quantum mechanics (QM). We have tried to loosen the feathers, to make the discussion as general and abstract as possible without making it empty.

The fact that it seems to be possible to motivate QM from such a general epistemic discussion conforms with the fact that QM is the general framework which gives form to all modern physical theories. In this sense, QM can be seen as a metaphysical theory. From that perspective, it calls for a general philosophical motivation of the kind suggested here.

As an example of the generality of the present approach, we do not even address the question whether space-time or other attributes are fundamentally continuous or discrete. Instead, we use only the more primitive notions of ordered and directed attributes. Nonetheless, the approach does suggest that it is always sufficient to consider finite-dimensional quantum mechanical Hilbert spaces, since the number of alternative values of an attribute or property that can be detected in an experimental context is always finite.

This unconventional conclusion is enabled by the viewpoint that Hilbert space is not the fundamental physical state space. Rather, it can be defined on top of the epistemic state space $\mathcal{S}$ in certain well-defined experimental contexts, as summarized in Statement \ref{representablec}. In such contexts, QM can be seen as an algebraic representation of the relations between the events that occur during the course of the experiment. As such, it applies only to the specimen that is studied in the experiment, never to the entire world. From this perspective, the conceptual troubles we run into when we try to apply QM to the entire universe arise because we try to apply a formalism outside its domain of validity. 

Probability and probability amplitudes are central to QM. The strict epistemic approach adopted here means, however, that the probability for an event is not always defined, just like the Hilbert space is not always defined. Nevertheless, the underlying epistemic state space $\mathcal{S}$ resembles the sample space $\Omega$ in probability theory in several respects. A physical state $S$ is defined as a subset of $\mathcal{S}$ just like an event is defined as a subset of $\Omega$. Also, we define a measure $V$ on $\mathcal{S}$ just like a probability measure is defined on $\Omega$.

However, since we want the same state space $\mathcal{S}$ to describe all physical situations we need to make it more general than $\Omega$, provide a deeper interpretation, and give it more structure. For example, we need to define carefully the evolution in time of the state $S$.

The shift of perspective offered in this paper can thus be described as follows. We downgrade the Hilbert space so that it is no longer seen as a fundamental physical state space. Rather, we define different Hilbert spaces in different experimental contexts. Each of them applies in its own setting, but nowhere else. In contrast, we upgrade and develop the sample space, so that can now be regarded as the fundamental physical state space. Traditionally, the sample space is only defined in certain probabilistic experiments, and we define different sample spaces for different such experiments.

In this paper we have not tried to derive differential evolution equations analogous to the Shr\"{o}dinger equation. That requires a more elaborate treatment of time than that provided here. This will be the subject of an upcoming paper. A preliminary account of the matter can be found in Ref. \cite{epistemic}. We will argue that such evolution equations involving a continuous wave function are useful even though we have argued in this paper that a finite-dimensional Hilbert space is sufficient to describe any given experimental context. 

\subsection{A comparison to other points of view}

In this final section we try to position the present approach to QM among the wealth of approaches currently pursued within the field of quantum foundations. There are similarities to some of them, and differences to all.

Interpretations of the wave function are often classified as either $\psi$-ontic or $\psi$-epistemic. The former class of interpretation attach an element of objective reality to the wave function $\psi$, meaning that it represents a physical entity that does not presuppose a relation to an observer. Such a relation is supposed in the latter class of interpretations, where the wave function represents a state of knowledge.

In the former class we find, for example, the many worlds interpretation \cite{manyworlds,dewitt} and the de Broglie-Bohm pilot wave theory \cite{bohm}. Theories in which the wave function collapse is considered to be an objective physical process do also belong to this class \cite{GRW,OR}. A recent $\psi$-ontic theory is suggested by Gao, who argues that the wave function reflects the random, discontinuous motion of particles \cite{gao}.

The perspective put forward in this paper is $\psi$-epistemic, of course. However, we have to distinguish between two types of such interpretations. In the first type the knowledge represented by the wave function is incomplete in the sense that there is an underlying physical state that represents objective reality. This state is specified by the precise values of a set of hidden variables. These values cannot be extracted from the wave function itself. This was the viewpoint of Einstein \cite{hs}.

After Einstein's death, a series of theorems have been proven that rule out the more natural theories with hidden variables that conform with the predictions of QM. Bell's theorem excludes local hidden variables \cite{bell1,bell2}, and the Kochen-Specker theorem rules out non-contextual hidden variables \cite{ks} that belong to the objects themselves, regardless the device used to observe them. More recently, Pusey, Barrett and Rudolph ruled out $\psi$-epistemic models with hidden variables in which the states of two systems that are described as independent by QM are truly independent \cite{pbr}. In other words, a $\psi$-epistemic model with hidden variables must be such that even if we manage to prepare two systems independently to the best of our knowledge, the hidden variables that specify the underlying physical states of these two systems are nevertheless \emph{not} independent \cite{ljbr}. A comprehensive review of restrictive theorems of this kind is provided by Leifer \cite{leifer}.

Despite the obstacles presented by these no-go theorems when it comes to the construction of simple and natural $\psi$-epistemic models with hidden variables, some researchers, such as Matt Leifer and Robert Spekkens, regard them as the most promising path towards an understanding of the meaning of QM. Spekkens has constructed a toy model that restricts the amount of knowledge an observer can have about reality \cite{spekkens}. This model reproduces several characteristic features of QM, and Spekkens argues that this fact together with the simplicity of the model is a sign that points in the $\psi$-epistemic direction. Leifer and Spekkens find it striking that QM can be seen as a certain kind of generalization of classical probability theory \cite{ls}, a circumstance that also points in the epistemic direction. The same goes for the fact that several reconstructions of QM from probabilistic or information-theoretic principles have been successfully carried out in recent years \cite{axiomatic,dakic,masanes,infoderivation,goyal,hoehn}.

The assumption that there are hidden variables is not essential in all of these $\psi$-epistemic models or reconstructions. The wave of information-theoretic approaches to QM can be traced back to John Wheeler's ideas, summarized in his phrase "it from bit" \cite{wheeler}. Wheeler was quite explicit that he associated this phrase with a "participatory universe" in which the subjective act of observation is crucial: "Physics gives rise to observer-participancy; observer-participancy gives rise to information; and information gives rise to physics." This self-referential loop is very similar to the view advocated in the present approach, as illustrated in Fig. \ref{Figure1}. If this view is indeed true, there is no observer-independent underlying layer of physical reality, the state of which has to be described by a set of hidden variables. Anton Zeilinger adheres to a similar view in his attempt to implement Wheeler's ideas \cite{zeilinger,zeilinger2}. So do the adherents of the Qbist interpretation \cite{fuchs}. Of course, Bohr, Heisenberg, Pauli and several other pioneers of QM also rejected the notion of an observer-independent layer of physical reality in their Copenhagen interpretation. In modern language, they advocated the second type of $\psi$-epistemic model, where the knowledge represented in the wave function is complete in the sense that no additional hidden variables are needed to specify the fundamental physical state. The present approach belongs to this tradition.

To throw away the hidden variables is a relief since we can forget about all the previously mentioned no-go theorems. To avoid them, we do not have to choose among more or less far-fetched $\psi$-epistemic models, if we still aim at a theory that conforms with QM. Why are several researchers that promote $\psi$-epistemic models unwilling to take this step? First and foremost they seem to consider realism too precious to abandon, and therefore cling to this metaphysical idea as a template for scientific models. In addition, Spekkens argues that a proper scientific explanation of statistical correlations between two perceived events require a causal relation between them, or a common cause \cite{allen,wood}. Such well-defined causal relations are not provided by QM itself \cite{ls}.
 
Let me relate the present approach to QM to some of the concepts and ideas discussed above, and do so in more detail than just to say that it arrives at a $\psi$-epistemic model of the second type, without any hidden variables. We have tried to reconstruct the basic formalism of QM from a set of epistemic principles. Therefore the approach follows in the footsteps of other recent reconstructions of QM \cite{axiomatic,dakic,masanes,infoderivation,goyal,hoehn}. The present attempt is more ambitious in the sense that the assumed principles are chosen to be so general that they, as I see it, have power to shed new light not only on the structure of QM, but also on other aspects of physical law \cite{epistemic}.

I agree with those researchers who stress that any proper interpretation of QM must have a well-defined ontology. However, I disagree with the idea that such an ontology must contain observer-independent objects, which motivate the introduction of hidden variables. Very generally, we might say that the ontological content of a physical theory corresponds to those aspects of that theory that transcend the contents of subjective perceptions and cannot be manipulated at will. In the present epistemic approach we have to assume a distinction between true and false interpretations of perceptions, as illustrated in Fig. \ref{Figure2}. Knowledge and belief are not the same thing. The assumed existence of \emph{truth} is therefore part of the ontology. This also goes for the \emph{structure} of truth, or the structure of knowledge. For example, we assume the possibility to divide perceptions into objects and to arrange them hierarchically so that some objects may be part of another object as subsets. These forms of perception are given once and for all and are therefore also part of the ontology. The same goes for physical law. Also, we assume the existence of potential knowledge that transcends the knowledge we happen to be consciously aware of at a given moment. We also assume that there is collective knowledge that transcends our own personal knowledge. The existence of collective knowledge reflects the assumed existence of several observers, meaning that the present approach is not solipsistic. We may say that we assign objective existence to a set of subjects rather than to a set of objects.

There is a subtlety to be noted in this connection. The world view sketched in Fig. \ref{Figure1} presumes the ability to distinguish objects in our own body from external objects. However, both kinds of objects lose their meaning if they are not perceived by anyone. That is, we do not consider external objects to be the same thing as observer-independent objects.

Some would say that the above set of ontological `beables' in the present approach contains mystical elements. But every physical theory must contain elements of the world that enter the construction as assumptions, just as a formal mathematical system must contain axioms. Any such set of elements must be considered `mystical' in the sense that they cannot be explained within the theory in terms of other, more primitive elements. All we can do to minimize the amount of mysticism in our world view is to look for a set of assumptions that is as small as possible and account for as much as possible of the structure and behavior of the physical world.

Such a set of assumptions can obviously be too small. The wave of informational approaches to physics explores the idea that the bit or the qubit is the only necessary fundamental building block of the world. It is true that virtually all the knowledge about the world that we receive via our senses can be encoded as strings of bits, but it does not follow that the perceived world can be described as a collection of bits. It is true that the simplest quantum mechanical systems have two possible states, and therefore may be called qubits. But it does not follow that all other quantum mechanical systems can described as collections of qubits.

If that would indeed be the case, then the world would correspond to a structureless heap of bits or qubits. To assume spatial or other relations between these binary units would be the same as to introduce additional structure of the world external to the binary units themselves, admitting that this description is too narrow. If we insist that everything can nevertheless be described in terms of binary units we might want to encode this additional structure in terms of an additional set of binary units. But then we must mark the members of this additional set in a particular way to avoid that they are thrown into the structureless heap of binary units that we started out with. Such markings would correspond to a key to the code that is again external to the binary units themselves. We may want to encode this key with still another set of binary units, but we see that we end up in infinite regress if we continue along this road.

The crucial mistake is to conflate the concepts of \emph{encoding} and \emph{description}. An encoding presupposes three distinct proper subsets of the world: the object to be encoded, the code and the key. Therefore it is misguided to describe the world as a whole as a code made up of binary units. It would amount to a description of the structure of the world in terms of a proper subset of its own structure. To take a trivial example, if we are given just a heap of bits, it is impossible to account for the fundamental distinction between different binary attributes of the same object and binary attributes of different objects.

The exclusion of the subjective aspect of the world in classical physics can be seen as a mistake of the same kind. (This amounts to an exclusion of the lower half of Fig. \ref{Figure1}). If the description of the world in terms of a physical model contains observed objects but no observers, then it becomes impossible to account for the appearance of conscious awareness within the model. This is often called the `hard problem of consciousness' \cite{chalmers}, but I would rather call it a consequence of a misguided attempt to describe all fundamental aspects of the world in terms of just one of these aspects. 

In the present approach we try instead to identify all fundamental aspects or `degrees of freedom' of the world and include them explicitly in the physical model. This goes for the vertical degree of freedom in Fig. \ref{Figure1} where the objective and subjective aspects corresponds to its binary values. Also, it goes for the treatment of attributes and attribute values in Section \ref{structure}, where we do not try to squeeze them all into a binary paradigm, but allow attributes with any number of discrete values, as well as attributes with continuous values.

Another difference between the present epistemic reconstruction of QM and epistemic reconstructions within the informational paradigm is that the present attempt aims at the postulates of QM formulated in terms of state vectors in Hilbert space, whereas other reconstructions often aim at the formulation of QM in terms of density operators \cite{hoehn}. In this connection, I agree with Roger Penrose \cite{interpretations} that it is misleading to focus on density operators in discussions about the foundations of QM. They should be seen exclusively as an elegant practical tool to do calculations in the case the quantum state is not exactly known to the experimenter.

Density operators mix classical and quantum probabilities in a way that is impossible to decompose without additional information external to the density operator itself. Many different ensembles of state vectors are represented by the same density operator. This may cause confusion in foundational discussions. For example, given a diagonal density matrix, it is impossible to tell whether it corresponds to a state of decoherence or de-superposition. This makes it possible to argue erroneously that decoherence solves the measurement problem, since de-superposition corresponds to the collapse of the wave function. As I see it, the fact that the transition from state vectors to density operators is not one-to-one means that a reconstruction that arrives at the correct rules that govern density operators has not gone all the way to a reconstruction of the fundamental formulation of QM in terms of state vectors.

In the language used in the present study, a description of an experimental context in terms of a density operator corresponds to the best possible description given the aware knowledge $AK$ of the experimenter, whereas the description in terms of a state vector corresponds to a description given all the potential knowledge $PK$ (Fig. \ref{Figure7}). Most often the aware knowledge is smaller than the potential knowledge, meaning that the experimenter cannot pinpoint the state vector that describes the context precisely. Instead, there will be an ensemble of state vectors that cannot be excluded. If her aware knowledge allows it, the experimenter may assign a classical probability to each member of this ensemble and define the density operator. However, we have argued that the only well-defined state of knowledge that every observer may agree upon is the collective potential knowledge $PK$. To arrive at an epistemic approach to physics in which the physical state $S$ is well-defined and associated with the world as a whole rather than the fuzzy knowledge of any particular observer, we must let it correspond to $PK$ rather than $AK$. The aware knowledge $AK$ and the associated density operator becomes secondary.

Similar considerations apply to to the treatment of probability in the present approach. Probabilities do not exist in all physical states $S$, but when they do, they are considered to be well-defined quantities associated with the state of the world as a whole, but applying to a particular object, being a functional of its state. More precisely, a probability is defined as a measure on a \emph{future alternative}, which is a subset $\vec{S}_{Oj}$ of the physical state $S_{OO}$ of an experimental context $C$, defined at the start of the experiment (Section \ref{probability}). As such it transcends the beliefs and the personal knowledge of individual experimenters and acquire a certain objectivity. All observers who make proper interpretations of their perceptions of the experimental setup at the start of the experiment, and are in a position to perceive all relevant parts of it, must also agree on the probability of a certain outcome.

This view is very different from that in \emph{quantum Bayesianism} or \emph{qbism} \cite{fuchs}. This interpretation of QM corresponds to a $\psi$-epistemic model without hidden variables, just like the present approach. However, the qbists adhere to the Bayesian view that all probabilities reflect personal beliefs. When an individual observer makes a measurement the new information makes it possible for her to update her beliefs, corresponding to an update from a prior to a posterior probability. As far as I understand, this persepctive makes no clear distinction between proper and improper beliefs, so that different observers may assign different probabilities to the same experimental outcome, in which case they assign different quantum states to the same experimental setup. Therefore it seems to me that the statement that two observers look at \emph{the same} object becomes ill-defined, just like the statement that they live in the same world.

Apart from giving them a more objective role, the present approach to probability differs from that of qbism in another respect. In the present construction it is not possible to define any posterior probabilities at all after the final measurement in the experimental context $C$. The Bayesian approach breaks down. At that stage there is no longer any wave function, and no probabilities. (The experimenter may of course design another experimental context $C'$ at a later time, but the probabilities associated with $C'$ has no direct relation to those associated with $C$.)

The physical state $S$ that corresponds to the state of collective potential knowledge $PK$ is still defined after the disappearance of the wave function. Since we need this more basic layer of physical description, we may borrow a term from those who promote $\psi$-epistemic models with hidden variables and say that QM is considered incomplete in the present approach, in contrast to the Copenhagen and the qbist interpretations. However, it is not considered incomplete in the sense that the wave function cannot describe all relevant aspects of the physical state, but rather in the sense that it cannot describe the physical state in all situations.

The introduction of the state $S$ and the state space $\mathcal{S}$ as a more basic layer of physical description than the wave function might nevertheless please those $\psi$-epistemicists like Robert Spekkens who sense the need for a clearer concept of causality than that provided by conventional QM \cite{allen,wood}. The physical state $S(n)$ at sequential time $n$ always has a boundary $\partial S(n)$, meaning that $S(n)\subset \mathcal{S}$ for each $n$, and the same goes for the evolved state $u_{1}S(n)$. This means that given the present state we can always exclude some future states as incompatible with physical law. This circumstance is reflected in the strict implication $S(n)\Rightarrow \{S(n+1)\cap[\mathcal{S}\setminus u_{1}S(n)]=\varnothing\}$, and may be called `negative causality'. If I grab a trumpet and start blowing, it will not remain silent. Of course, such a clear causal implication may occur in conventional QM as well, and corresponds to a situation in which the wave function has limited support, whose boundary changes with time. However, the connection to causality in such situations is rarely spelled out, and wave functions with unlimited support are seemingly treated as the typical case.

\end{document}